\documentclass{article}
\usepackage[margin=1.15in]{geometry}
\usepackage{amsmath}
\usepackage{amssymb}
\usepackage{hyperref}
\usepackage{natbib}
\usepackage{mathtools}
\usepackage{booktabs}
\usepackage{graphicx}
\usepackage{caption}
\usepackage{color}
\usepackage{subcaption}
\usepackage[section]{placeins}
\usepackage{color} 
\definecolor{mygreen}{RGB}{28,172,0} 
\definecolor{mylilas}{RGB}{170,55,241}

\usepackage{listings}
\title{\textbf{On 1-D traveltime tomography and linear inhomogeneity
}}
\author{
Md Abu Sayed\footnote{%
Department of Earth Sciences, Memorial University of Newfoundland,
{\tt m.abusayed.stu@gmail.com}}
}
\date{}
\begin{document}
\maketitle
\begin{abstract}
In this article, 
we develop a 1-D traveltime tomography method to calculate the seismic $P$-wave velocity of a medium.
We use the results of 1-D tomography to obtain linear inhomogeneity parameters in a specific layer. 
To get the trustworthiness of the method, we perform several synthetic experiments. 
We show that the inverted model parameters are reasonably accurate and stable. 
To examine the results of linear inhomogeneity parameters using a different method, we also develop an inversion method based on a two-parameter velocity model.  Finally, we apply both the methods to Vertical Seismic Profile (VSP) data and do a study comparing their results. 
\end{abstract} 

\section{Introduction}
\label{sec:Introduction}
We examine linear inhomogeneity of a medium by applying two inversion methods on seismic traveltime.
In the first method, we derive an analytical expression for the solution of Hamilton's ray equation in vertically inhomogeneous and isotropic media.
Considering the analytical solution as a forward model, we construct an inversion method based on the Levenberg-Marquardt damped least square solution (\citet{Levenberg1944}, \citet{Marquardt1963}).
In the second inversion method, we use the traveltime expression based on a two-parameter velocity model as the forward model. 
We perform several synthetic experiments on the first method based on a linear velocity model.
While we study the linear velocity in synthetic studies to reduce the model parameters to two, the inversion method can be used to construct a velocity model that varies with depth in any order.

The synthetic experiments show that the traveltime convergence occurs even with a significant change in the start-up values; however, as the discrepancy gets higher, the inverted velocity diverges more from the reference velocity model. 
In comparison to the start-up values, the inversion method is less sensitive to the number of data points and the noise.   

We provide the source codes in the appendices~\ref{sec:code_syn}, \ref{sec:code_real}, and \ref{sec:code_ab}. The checkshot (VSP) data is provided by the Canada-Newfoundland \& Labrador Offshore Petroleum Board~\citep{CNLPOB2009}.
Therein, the traveltime data corresponds to a single source and multiple receivers. 
The source is placed at a 26.50~m offset, and the receivers are located along a the vertical axis, starting at a depth of 1865~m and ending at 2650~m\,. 

\section{Method developement}
\subsection{Solution of the ray equation in vertically inhomogeneous media}
\label{sec:raytracing}
In a medium, the velocity of seismic waves can vary in any direction. 
However, assuming the velocity only a function of vertical depth, we present an analytical solution for Hamilton's ray equation.  
In the derivation, we apply the method of characteristics, similar to the approaches described by~\citet{Slawinski2015} and~\citet{Cerveny2001}.
To parameterize the ray equation, \citet{Slawinski2015} used arc length as opposed to traveltime, and \citet{Cerveny2001} used the level set equation $p^{2}-v^{-2}=0$ as opposed to $p^{2}v^{2}=1$, where $p$ is the slowness parameter, and $v$ is the wave velocity.
In our derivation, we use traveltime for the parametrization and $p^{2}v^{2}=1$ as the level set equation.

In this section, we present the solution of Hamilton's ray equation for a vertically inhomogeneous and isotropic medium.
We start with a 3-D inhomogeneous medium and then move into a 1-D medium by considering velocity as a function of depth.
In a smoothly inhomogeneous isotropic medium, the high-frequency seismic wave field can be separated into two independent waves, $P$ and $S$~\citep[p.~277]{Slawinski2015}. 
Both waves satisfy the eikonal equation 
\begin{equation}	
p^{2}=\dfrac{1}{v^{2}\left(\mathbf{x,p}\right)}\,,
\label{EikEqnRef}
\end{equation}
where, $p^{2}=\mathbf{p}\cdot\mathbf{p}$\,, $\mathbf{p}$ is the slowness and
$p_{i}:=\dfrac{\partial\psi}{\partial x_{i}}\,, \,i\in\;\left\{
1,2,3\right\}$, $\psi$ is the phase function.
Equation~\eqref{EikEqnRef} is a set of first order partial differential equations that depends on the variables $\mathbf{x}$ and $\mathbf{p}\left(\mathbf{x}\right)$\,. It relates the magnitude of phase slowness of the wave to the medium properties~\citep{Slawinski2015}.
The method of characteristics is commonly applied in the eikonal equation to get a system of six first-order ordinary differential equations~\citep[p.~343]{Slawinski2015}
\begin{equation}	
\begin{array}{c}
\dfrac{\mathrm{d}x_{i}}{\mathrm{d}s}=\zeta\dfrac{\partial F}{\partial
p_{i}}\\ \\\dfrac{\mathrm{d}p_{i}}{\mathrm{d}s}=-\zeta\dfrac{\partial
F}{\partial x_{i}}
\end{array}
\,,\qquad i\in\;\left\{
1,2,3\right\} \,,
\label{preHam}
\end{equation}
where $\zeta$ is a scaling factor and $s$ is the parameter along the curve. 
The choice
of $s$ determines the parametrization.
As discussed in~\citet[p.~343]{Slawinski2015}, the solution of the eikonal equation is a surface in the $\bf xp$-space.
This surface can be described as level sets of a function, which we denote by $F(\mathbf{x},\mathbf{p})$. It is a Hamiltonian with a factor of $\frac{1}{2}$.
A relationship for the scaling factor $\zeta$ in equation~\eqref{preHam} to the flow parameter $s$ are provided in~\citet{Cerveny2001}. They consider three cases of $s$ along the curve: the arclength, the traveltime and the parameter $\sigma$.

For a vertically inhomogeneous isotropic medium, we solve Hamilton's ray equation by parametrizing the characteristic equations in terms of time and scaling factor as a constant number, so that expression~\eqref{preHam} becomes 
\begin{equation}	
\begin{array}{c} \dot{x}_{i}=\dfrac{\partial}{\partial
p_{i}}\left(\dfrac{F}{2}\right)=\dfrac{\partial\mathcal{H}}{\partial p_{i}}\\
\smallskip\\
\dot{p}_{i}=-\dfrac{\partial}{\partial
x_{i}}\left(\dfrac{F}{2}\right)=-\dfrac{\partial\mathcal{H}}{\partial
x_{i}}
\end{array}
\,,\qquad i\in\;\left\{
1,2,3\right\} \,,
\label{HamRayEqn}
\end{equation}

where $\mathcal{H}:=\dfrac{F}{2}\,,$ known as the ray-theory Hamiltonian.
We choose $F\left(\mathbf{x,p}\right)=p^{2}v^{2}\left(\mathbf{x,p}\right)$ as the level sets,
which leads to a Hamiltonian 
\begin{equation}	
\mathcal{H}\left(\mathbf{x,p}\right)=\dfrac{1}{2}p^{2}v^{2}
\left(\mathbf{x}\right)=\dfrac{1}{2}\left[p_{1},p_{3}\right]\cdot
\left[p_{1},p_{3}\right]v^{2}\left(x_{1},x_{3}\right)\,,
\label{FourEik}
\end{equation}

and the corresponding ray equations 
\begin{subequations}
\begin{equation}	
\dfrac{dx_{1}}{dt}
=p_{1}v^{2}\,,
\label{FourCharA} 
\end{equation}
\begin{equation}
\dfrac{dx_{3}}{dt}
=p_{3}v^{2}\,,
\label{FourCharB}
\end{equation}
\begin{equation}
\dfrac{dp_{1}}{dt}
=-p^{2}v\dfrac{\partial v}{\partial x_{1}} = 0\,,
\label{FourCharC}
\end{equation}
\begin{equation}
\dfrac{dp_{3}}{dt}
=-p^{2}v\dfrac{\partial v}{\partial x_{3}}\,.
\label{FourCharD}
\end{equation}
\end{subequations}

Dividing expression~\eqref{FourCharA} by \eqref{FourCharB}
\begin{equation}
\label{FourCharE}
\dfrac{dx_{1}}{dx_{3}} = \dfrac{p_{1}}{p_{3}}\,.
\end{equation}

Using the eikonal equation $p_{1}^{2}+p_{3}^{2} = v^{-2}$ in expression~\eqref{FourCharE}
\begin{equation}
\label{FourCharF}
\dfrac{dx_{1}}{dx_{3}} = 
\dfrac{p_{1}}{\sqrt{ v^{-2}-p_{1}^{2}}} =
\dfrac{p_{1}v}{\sqrt{1-p_{1}^{2}v^{2}}}\,.
\end{equation}

Using expression~\eqref{FourCharF} in expression~\eqref{FourCharA}
\begin{equation}
\label{FourCharG}
dt = \dfrac{dx_{1}}{p_{1}v^{2}} =
\dfrac{\dfrac{p_{1}v}{\sqrt{1-p_{1}^{2}v^{2}}}dx_{3}}{p_{1}v^{2}} =
\dfrac{dx_{3}}{v\sqrt{1-p_{1}^{2}v^{2}}}\,.
\end{equation}

Expression~\eqref{FourCharC} shows that the slowness parameter, $p_{1}$, is constant along the whole ray path.  
For a vertically inhomogeneous medium $p_{1}$ is a conserved quantity, which is known as the ray parameter.
We express the ray parameter by $\mathfrak{p}$. 
We obtain the solution of ray equation by integrating expressions~\eqref{FourCharE} and \eqref{FourCharG} for $x_{3}$ to get
\begin{equation}
\label{HamSolA}
x_{1}\left(x_{3}\right)=\int\limits_{z_{0}}^{z}\dfrac{\mathfrak{p}v
\left(x_{3}\right)}{\sqrt{1-\mathfrak{p}^{2}v^{2}
\left(x_{3}\right)}}\,\mathrm{d}x_{3}\,,
\end{equation} 
and 
\begin{equation}
\label{HamSolB}
t\left(x_{3}\right)=\int\limits_{z_{0}}^{z}\dfrac{1}{v\sqrt{1-\mathfrak{p}^{2}v^{2}
\left(x_{3}\right)}}\,\mathrm{d}x_{3}\,,
\end{equation} 
where $x_{3}$ is the vertical depth.
Equations~\eqref{HamSolA} and \eqref{HamSolB} are in agreement with~\citet{Cerveny2001}.
To trace a ray, we need to solve expressions~\eqref{HamSolA} and \eqref{HamSolB} simultaneously. 

If the velocity changes linearly with depth, i.e., $v(x_{3})=a+b x_{3}$, using expressions~\eqref{HamSolA} and \eqref{HamSolB}, the ray parameter and the traveltime expressions can be written as~\citep{SlawinskiSlawinski1999}

\begin{equation}
\label{lin_p}
\mathfrak{p} = \dfrac{2bx_{3}}{
\sqrt{\left(b^{2}x_{3}^{2}+a^{2}+(a+bx_{3})^{2}\right)^{2} - 4a^{2}(a+bx_{3})^{2}}
}\,,
\end{equation}
 
\begin{equation}
\label{lin_t}
t = \dfrac{1}{b}
\left| \log\left(\dfrac{a+bx_{3}}{a}\dfrac{1+\sqrt{1-a^2\mathfrak{p}^2}}{1+\sqrt{1-\mathfrak{p}^{2}(a+bx_{3})^{2}}}\right)\right|\,.
\end{equation} 

We use expression~\eqref{lin_t} as the forward model to the $ab$-model inversion. 
\subsection{Discretizing the forward model for 1-D tomography}
\label{sec:forward}
In this section, we discretize the expressions~\eqref{HamSolA} and \eqref{HamSolB} to solve the ray equation numerically.
To perform the integration for multiple source-receiver pairs, we consider the medium to be composed of N layers; $H$ is the layer thickness, where the layers are equally thin, homogeneous, and isotropic.
Using expressions~\eqref{HamSolA} and \eqref{HamSolB}, the ray tracing equations from the $i$-th source to the $k$-th receiver are

\begin{equation}	
\label{offset}
x_{k,i}= x_{1k,i} +x_{2k,i} +........+x_{jk,i} 
= \sum_{j=1}^{m} \frac{H_{j}B_{kj,i}}{\sqrt{1-B_{kj,i}^2}}
\,,\qquad j\in\;\left\{ 1,2,3 ....., m\right\}\,,
\end{equation}

\begin{equation}	
\label{traveltime}
t_{k,i}= t_{1k,i} +t_{2k,i} +........+t_{jk,i} 
=  \sum_{j=1}^{m}  \frac{H_{j}}{v_{j}\sqrt{1-B_{kj,i}^2}} 
\,,\qquad j\in\;\left\{ 1,2,3 ....., m\right\}\,,
\end{equation}

where $\theta_{1k}$ is the take-off angle, $B_{kj,i} = p_{k,i}v_j$, $i$ and $k$ denote the indices of sources and receivers.
Traveltime in the $j$-th segment is $t_{jk}$.
The total number of model parameters is $m$, which is equal to the number of layers.
We consider the sources to be located at the surface and the receivers to be set along the vertical axis. 
To calculate the total traveltime and the offset for a given source-receiver pair, we modify the upper limit of the summation by replacing $m$ to $L(k)$.
For a given source-receiver pair, we modify the upper limit of the summation by replacing $m$ by $L(k)$ to calculate the total traveltime and the offset.
This is because, the Geophone locations may not be related to the layering, therefore, an index $L(k)$ is introduced that indicates in which layer the $k$-th geophone is located. If the geophone locations $k$ and $k + 1$ are in the same layer, then $L(k) = L(k + 1)$.

\subsection{Development of the inversion method for 1-D tomography}
Using the analytical solution as a forward model, we develop an inversion method based on Levenberg-Marquardt (L-M) damped least-squares solution.
The L-M method is a powerful tool for the iterative solution for both linear and nonlinear problems \citep{Pujol2007}. 
\citet{Levenberg1944} used the technique for the first time, and about twenty years later, \citet{Marquardt1963} independently rediscovered the method utilizing an independent approach. 

In this section, we develop the L-M method for a vertically inhomogeneous and isotropic medium. 
As the forward model, we use expressions~\eqref{offset} and \eqref{traveltime} from section~\ref{sec:forward}. 
In the case of $t = t(v_{j})$, the traveltime residual can be written as
\begin{equation}	
\label{dt}
dt_{k}= \sum_{j=1}^{L(k)} \frac{\partial t_{k}}{\partial v_{j}}  dv_{j} 
\,,\qquad j\in\;\left\{ 1,2,3 ....., L(k)\right\}\,.
\end{equation}
Where, we neglect the higher order terms in Taylor series expansion.
Taking the derivative of expression~\eqref{traveltime} with respect to $v_{j}$,

\begin{equation}	
\label{ddv_t}
\frac{\partial t_{k}}{\partial v_{j}} =  \frac{p_{k}^2h_{j}}{\sqrt{1-B_{kj}^2}} \,  - \, \frac{h_{j}}{v_{j}^2\sqrt{1-B_{kj}^2}}
\,,\qquad j\in\;\left\{ 1,2,3 ....., L(k)\right\}\,.
\end{equation}

Also, the system of linear equations~\eqref{dt} may be written in the matrix form,
\begin{equation}
\label{matrix}
\bf{
C =  t_{obs} - t_{mod} = AX \,,
}
\end{equation}

where~\footnote{Throughout the Chapter~\ref{ch:tomo}, we present vectors and matrices in bold letters.} 
\begin{equation}
\label{matrix_C}
\mathbf{C}= (dt_{1},dt_{2}.......dt_{M})^{T}  \qquad {\rm and} \qquad \mathbf{X}= (dv_{1},dv_{2}.......dt_{N})^{T} \,.
\end{equation}

In expression~\eqref{matrix}, $\mathbf{A}$ is an $(M\times N)$ matrix of partial derivatives,
$M$ and $N$ are the total number of receivers and layers, respectively.
$\bf{X}$ represents the model parameter adjustment vector, and $\bf{C}$ is the traveltime residual vector. 
We calculate both the traveltime residual vector and the partial derivative matrix in each iteration.

For a particular source-receiver pair, the basic algorithm is as follows---we apply the Newton-Raphson method to calculate the take-off angle from equation~\eqref{offset} by assuming we have the velocities in each layer.
The corrected take-off angle is used to calculate the model traveltime.
The parameter adjustment vector is calculated from expression~\eqref{matrix}, which allows us to update the velocity in each iteration. 
We repeat the process until we achieve a satisfactory agreement between the model and observed data.  

To solve equation \eqref{matrix} for ${\bf X}$, \cite{Pujol2007} stated that the convergence is not assured when $\bf{X}$ is computed using ordinary least squares.
The assumption behind linearizing the problem no longer remains valid if the initial model is far from the real solution. 
One of the ways to overcome this problem is the application of Levenberg-Marquardt method.

\subsection{A review of Levenberg-Marquardt Method}

In this section, we review the basic steps of Levenberg-Marquardt iteration scheme.
We follow the description of~\citet{Pujol2007}.
Let us consider the higher order terms in Taylor series expansion that we ignored in equation \eqref{matrix} 

\begin{equation}
\label{L_M_1}
\bf{R = C - A X}\,.
\end{equation}

The elements of $\bf{C}$ represent the residuals of traveltime for each source-receiver pair.
The problem is to calculate the elements of $\bf{X}$'s which minimize $\bf{R}$. 
The misfit function is defined as follows,
\begin{equation}
\label{obj}
S = \sum_{i=1}^{n} R_{i}^{2} = {\bf{R^{T}R}} \qquad \left\{ 1,2,3,......,n\right\}\,,
\end{equation}

where $n$ is the number of data points. Substituting equation \eqref{L_M_1} into \eqref{obj}, we get
\begin{equation}
\label{obj_exp}
S = \bf{(C^{T} - X^{T}A^{T})(C- A X) 
= C^{T}C- 2C^{T}AX+X^{T}A^{T}AX}\,.
\end{equation}

Instead of minimizing the misfit function $S$, \citet{Levenberg1944} proposes to minimize the following function
\begin{equation}
\label{obj_up}
\bar{S} = wS + Q \,,
\end{equation}
where $w$ is known as Levenberg damping parameter, $Q= {\bf{X^{T}DX}}$ with $\bf{D=I}$, the identity matrix.
Using equation~\eqref{obj_exp} in equation~\eqref{obj_up}
\begin{equation}
\label{obj_up_exp}
\bar{S} = w\left({\bf C^{T}C-2C^{T}AX+X^{T}(A^{T}A}+\frac{1}{w}{\bf I}){\bf X}\right)\,.
\end{equation}
Minimizing Equation \eqref{obj_up_exp} 
\begin{equation*}
\dfrac{d\bar{S}}{d{\bf X}} = \left(\dfrac{d\bar{S}}{dX_{1}}, \dfrac{d\bar{S}}{dX_{2}}, \dots, \dfrac{d\bar{S}}{dX_{N}}\right)^{T} = \, {\bf 0} \,,
\end{equation*}

The iteration scheme becomes
\begin{equation}
\label{L_M}
({\bf A^{T}A}+\lambda {\bf I}) {\bf X} = {\bf A^{T}c}\,,
\end{equation}

where $\lambda = \frac{1}{w}$. 
Using the method of \cite{PujolEtAl1985},
we assign a constant value to $\lambda$ and in each iteration we reduce it by a factor of 10.
At $p$-th iteration, we solve 
\begin{equation}
\label{L_M_it}
\left( \left({\bf A^{T}A}\right)^{(p)}+\lambda^{(p)} {\bf I}\right) \left({\bf X}\right)^{(p)} = \left({\bf A^{T}c}\right)^{(p)}\,.
\end{equation}

To otherwise improve the numerical aspects of the method, we use the scaled version of equation~\eqref{L_M_it}, which is suggested by~\cite{Marquardt1963}.
Instead of using ${\bf A^{T}A}$ and ${\bf A^{T}c}$ in expression~\eqref{L_M_it}, we use the scaled forms $\left[{\bf A^{T}A}\right]^{\ast}$ and $\left[{\bf A^{T}c}\right]^{\ast}$, The components of the scaled matrix are~\citep{Pujol2007}
\begin{equation}
\label{L_M_sc_1}
\left(\left[{\bf A^{T}A}\right]^{\ast}\right)_{ij} = S_{ii}S_{jj} \left({\bf A^{T}A}\right)_{ij}
\end{equation}
and 
\begin{equation}
\label{L_M_sc_2}
\left(\left[{\bf A^{T}c}\right]^{\ast}\right)_{i} = S_{ii} \left({\bf A^{T}c}\right)_{i}\,,
\end{equation}

where 
\begin{equation}
\label{L_M_sc_3}
S_{ii} = \dfrac{1}{\sqrt{\left(\left[{\bf A^{T}A}\right]^{\ast}\right)_{ii}}}\,.
\end{equation}

The scaled Levenberg-Marquardt equation is

\begin{equation}
\label{L_M_sc}
(\left[{\bf A^{T}A}\right]^{\ast(p)}+\lambda^{(p)} {\bf I}) {\bf X^{\ast(p)}} = \left[{\bf A^{T}c}\right]^{\ast(p)}\,.
\end{equation}

In each iteration step, we solve equation~\eqref{L_M_sc} for ${\bf X^{\ast}}$ and then calculate the components of ${\bf X^{\ast}}$ based on ${\bf X}$,
\begin{equation}
\label{L_M_sc_x}
X_{i} = S_{ii}X_{i}^{\ast}\,.
\end{equation}

The vector form of expression~\eqref{L_M_sc_x} is
\begin{equation}
\label{L_M_sc_x_v}
{\bf X} = {\bf S} {\bf X^{\ast}}\,,
\end{equation}
where ${\bf S}$ is a diagonal matrix with diagonal elements $S_{ii}$.
In each iteration, we update the velocity as 
\begin{equation}
\label{update}
{\bf V}^{(p+1)} = {\bf V}^{(p)} + {\bf X}^{(p)}\,.
\end{equation}

The iteration process continues until we reach a specific value of the misfit functional. 
Under the assumption of uncorrelated data with equal variances, $\sigma_{0}^{2}$, at $p$-th iteration, the misfit functional is defined as~\citep[p. 73]{Zhdanov2002} 
\begin{equation}
\label{misfit}
f({\bf X}^{(p)}) = \frac{1}{\sigma_{0}^{2}} \left(t_{obs} - t_{mod}^{(p)}\right)^{2}\,.
\end{equation}

In synthetic cases, we add normally distributed noise to the traveltime data, and following equation \eqref{misfit}, we set the iteration to stop while $f({\bf X}^{(p)}) \approx N$, where $N$ is the number of data points.

In each iteration of the Levenberg-Marquardt method, for a given set of velocities in layers, we use equation~\eqref{offset} to update the take-off angle. 
We apply a root-finding algorithm known as the Newton-Raphson method~\citep{Heath2002} to calculate $p_{k,i}$.  
It produces successively better approximations to the roots of a real-valued function.
To optimize the computation time, we terminate the iteration once we reach to the value of $10^{-6}$ for the $dx_{1}$, which is the difference between the horizontal distance of the shooting ray and the offset given from the data.

The updated take-off angle is used to calculate the velocity in the next iteration of the Levenberg-Marquardt method.
The process of calculation makes the method two-step as opposed to the one-step approach described by~\citet{PujolEtAl1985}.
The two-step approach provides us with a better initial model for the traveltime since it calculates only the take-off angle in first and the velocity in the second.
It also allows us to use a single unit for model parameters, which reduces the work of nondimensionalization to define misfit functional.   

In contrast to the other local optimization method, such as Gauss-Newton or steepest descent method, the Levenberg-Marquardt method minimizes both model parameters and the data residuals~\citep{Pujol2007}. As a result, the chances of convergence increases. 
 
\section{Synthetic experiments}
In the synthetic experiments, we consider multiple sources at the surface, many receivers along the vertical depth and assign a reference velocity which changes linearly with depth. 
The linear velocity is described by two parameters, i.e., the velocity at the surface and the velocity gradient. 
The variations of both parameters in the startup model allow us to observe the influence of the initial model to the inversion result.
We also study the effects of the noise on the data and the number of data points. 
In the synthetic study, the forward traveltime is calculated based on the analytic solution, the observed traveltime is calculated based on the variations in the reference velocity model by changing the startup model and the amount of noise in the data. 

\subsection{Test of the noise and the number of data points}
In Table~\ref{tab:syn_n_d_p}, we consider the reference velocity model as a linear function of depth, $v = a+bz$, with $a = 1000 \, {\rm ms^{-1}}$ and $b = 0.12 \, s^{-1}$. 
We choose $a$ based on the typical value of the $P$-wave velocity at the surface in the offshore. 
To have more options in choosing the number of layers in the synthetic experiments, we decide to consider the velocity gradient in the lower side, such as 0.12. 
If the velocity gradient is higher, with the increase of layers, the ray hits the critical angle in a relatively lower take-off angle. 
For the first six cases, the startup velocity for inversion is considered as $v_{ref} \pm 20 \, {\rm ms^{-1}}$ and for the last six cases, the startup velocity is considered as $v_{ref} \pm 40 \, {\rm ms^{-1}}$.  
Following~\citet{PujolEtAl1985}, we choose the value of the parameter $\lambda$ in the Levenberg-Marquardt algorithm.
We start at $10^{4}$, and in each iteration, it reduces by a factor of 10.
We consider the number of traveltime data and the number of model parameters to be equal. 
However, the inversion method can be applied to both underdetermined and overdetermined cases.

\begin{table}[h]
\centering
\begin{tabular}{c*{9}{c}}
\toprule
Test &
Noise (\%)&
Source &
Geophone &
Layer &
$f({\bf M})$ &
$a_{inv}$ &
$b_{inv}$ &
Figure
\\[1pt]
\toprule
1 & 1 & 101 & 1 & 101 & 97.74 & 1002.15 & 0.1179 &  \ref{t_t_n_1_d_p_101},\ref{vel_n_1_d_p_101} \\
2 & 1 & 101 & 2 & 202 & 199.90 & 1001.46 & 0.1184 &  \ref{t_t_n_1_d_p_202},\ref{vel_n_1_d_p_202} \\
3 & 5 & 101 & 1 & 101 & 100.70 & 1002.17 & 0.1173 &  \ref{t_t_n_5_d_p_101},\ref{vel_n_5_d_p_101} \\
4 & 5 & 101 & 2 & 202 & 200.95 & 1001.73 & 0.1187 & \ref{t_t_n_5_d_p_202},\ref{vel_n_5_d_p_202} \\
5 & 10 & 101 & 1 & 101 & 100.53 & 1002.40 & 0.1178 & \ref{t_t_n_10_d_p_101},\ref{vel_n_10_d_p_101} \\
6 & 10 & 101 & 2 & 202 & 199.87 & 1002.22 & 0.1171 & \ref{t_t_n_10_d_p_202},\ref{vel_n_10_d_p_202} \\
\midrule
7 & 1 & 101 & 1 & 101 & 99.91 & 1003.42 & 0.1157 &  \ref{t_t_n2_1_d_p_101},\ref{vel_n2_1_d_p_101} \\
8 & 1 & 101 & 2 & 202 & 201.90 & 1003.15 & 0.1166 &  \ref{t_t_n2_1_d_p_202},\ref{vel_n2_1_d_p_202} \\
9 & 5 & 101 & 1 & 101 & 100.23 & 1002.62 & 0.1159 & \ref{t_t_n2_5_d_p_101},\ref{vel_n2_5_d_p_101} \\
10 & 5 & 101 & 2 & 202 & 201.32 & 1001.91 & 0.1166 & \ref{t_t_n2_5_d_p_202},\ref{vel_n2_5_d_p_202} \\
11 & 10 & 101 & 1 & 101 & 100.23 & 1004.20 & 0.1153 &  \ref{t_t_n2_10_d_p_101},\ref{vel_n2_10_d_p_101}\\
12 & 10 & 101 & 2 & 202 & 200.12 & 1002.48 & 0.1170 &  \ref{t_t_n2_10_d_p_202},\ref{vel_n2_10_d_p_202} \\
\bottomrule
\end{tabular}
\caption[Varying noise and number of data points]{
Model set-up : test of the first six, $a_{true}=1000 {\rm ms^{-1}}$, $b_{true}=0.12 {\rm s^{-1}}$, 
$a_{in}=a_{true}\pm 20 {\rm ms^{-1}}$, $b_{in}=b_{true}$;
test of the last six, $a_{true}=1000 {\rm ms^{-1}}$, $b_{true}=0.12 {\rm s^{-1}}$, 
$a_{in}=a_{true}\pm 40 {\rm ms^{-1}}$, $b_{in}=b_{true}$ 
}
\label{tab:syn_n_d_p}
\end{table}

\begin{figure}[ht]
        \captionsetup[subfigure]{aboveskip=-.1cm}
        \centering
        \begin{subfigure}[b]{.45\textwidth}
                \includegraphics[width=\textwidth]{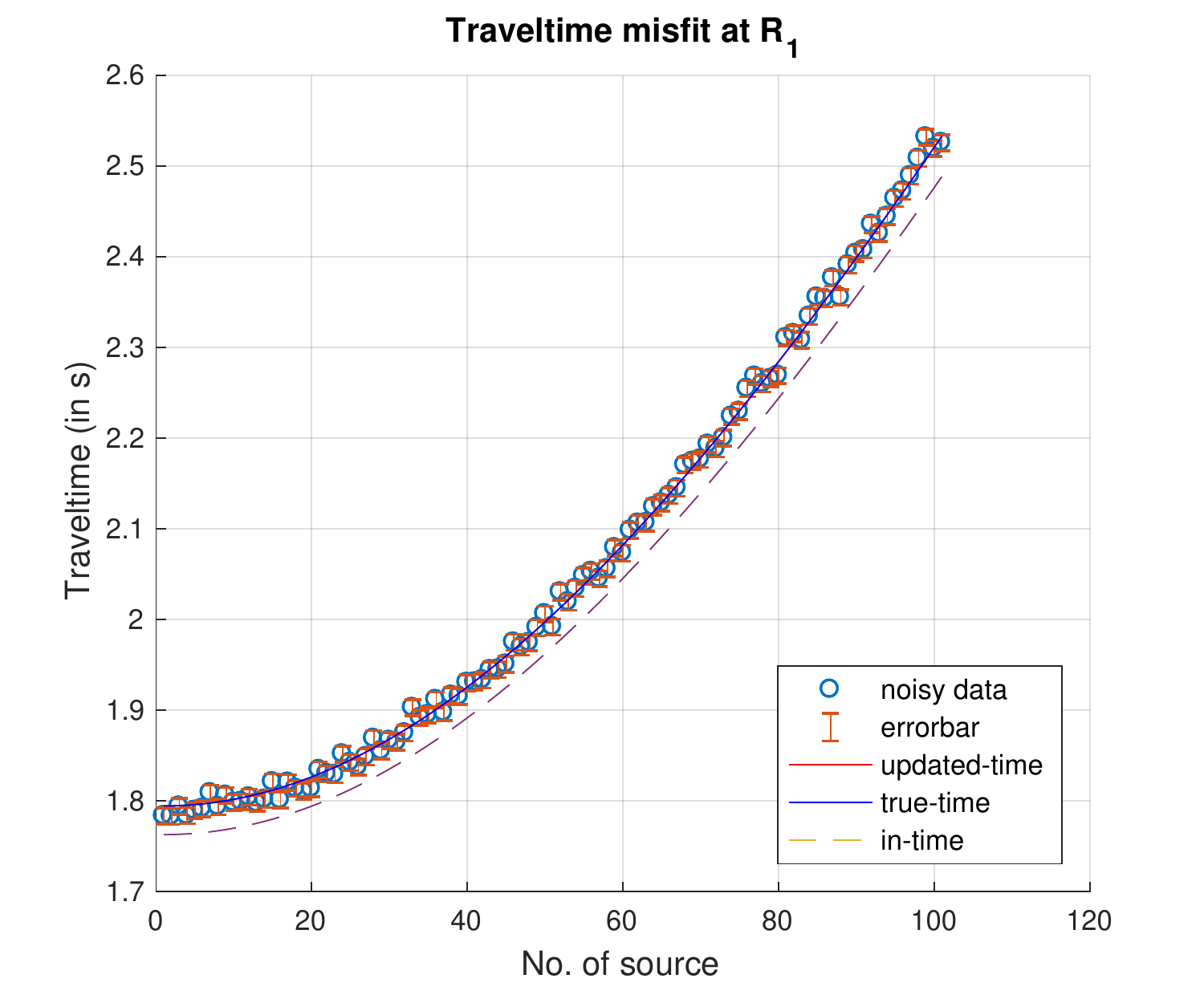}
                \caption{}
                \label{t_t_n_1_d_p_101}
        \end{subfigure}
        \begin{subfigure}[b]{.45\textwidth}
                \includegraphics[width=\textwidth]{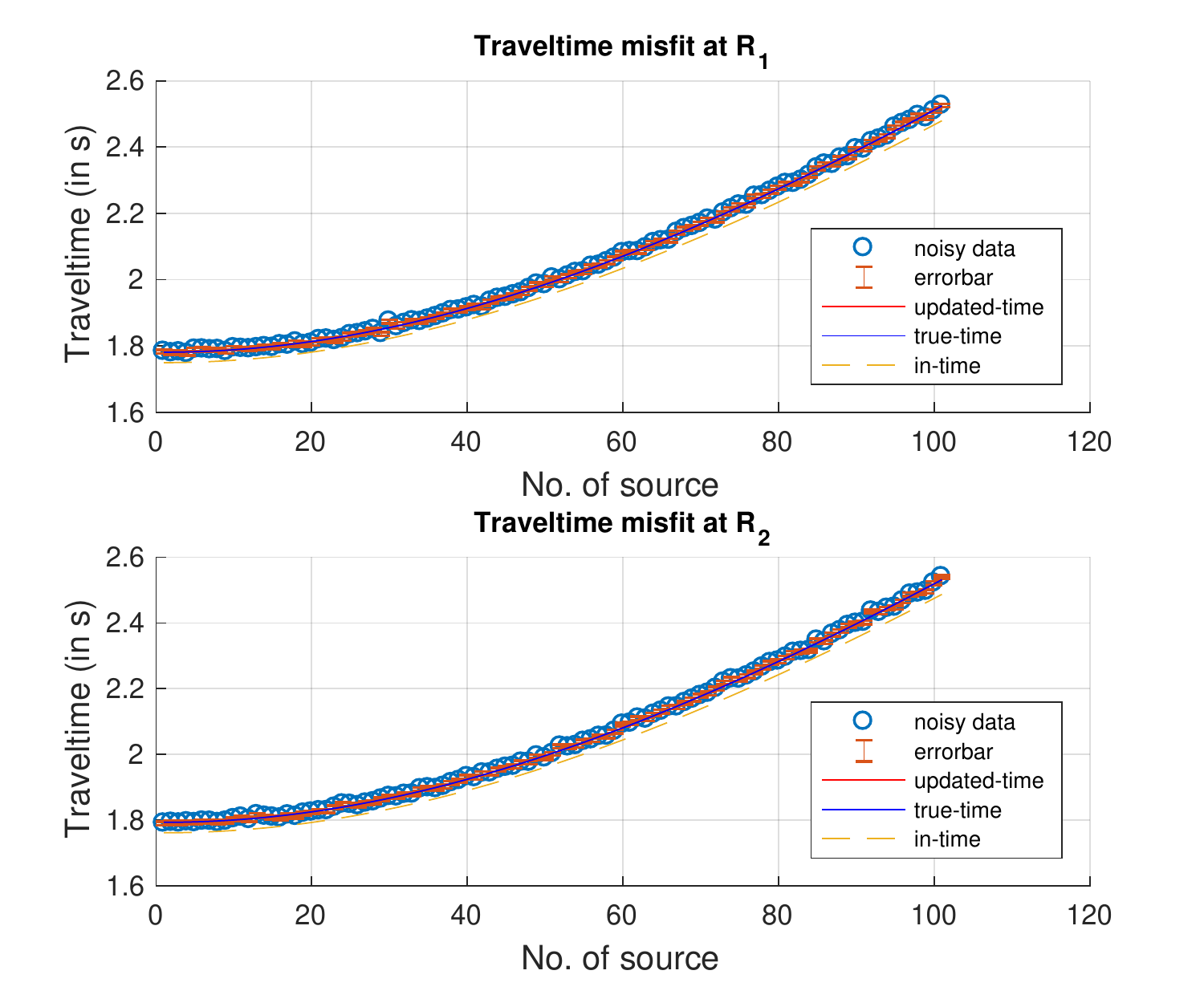}
                \caption{}
                \label{t_t_n_1_d_p_202}
        \end{subfigure}
        \begin{subfigure}[b]{.45\textwidth}
                \includegraphics[width=\textwidth]{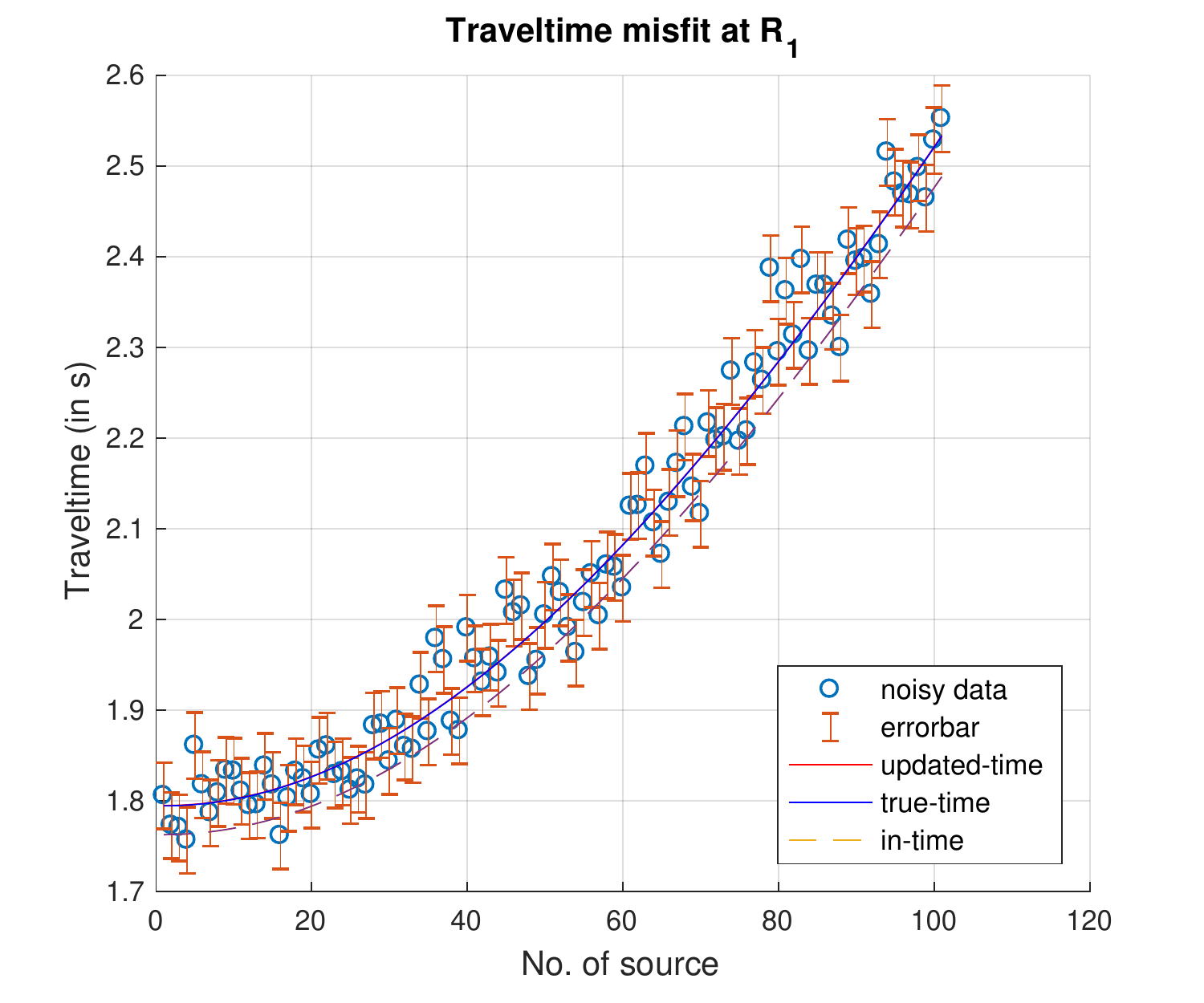}
                \caption{}
                \label{t_t_n_5_d_p_101}
        \end{subfigure}       
        \begin{subfigure}[b]{.45\textwidth}
                \includegraphics[width=\textwidth]{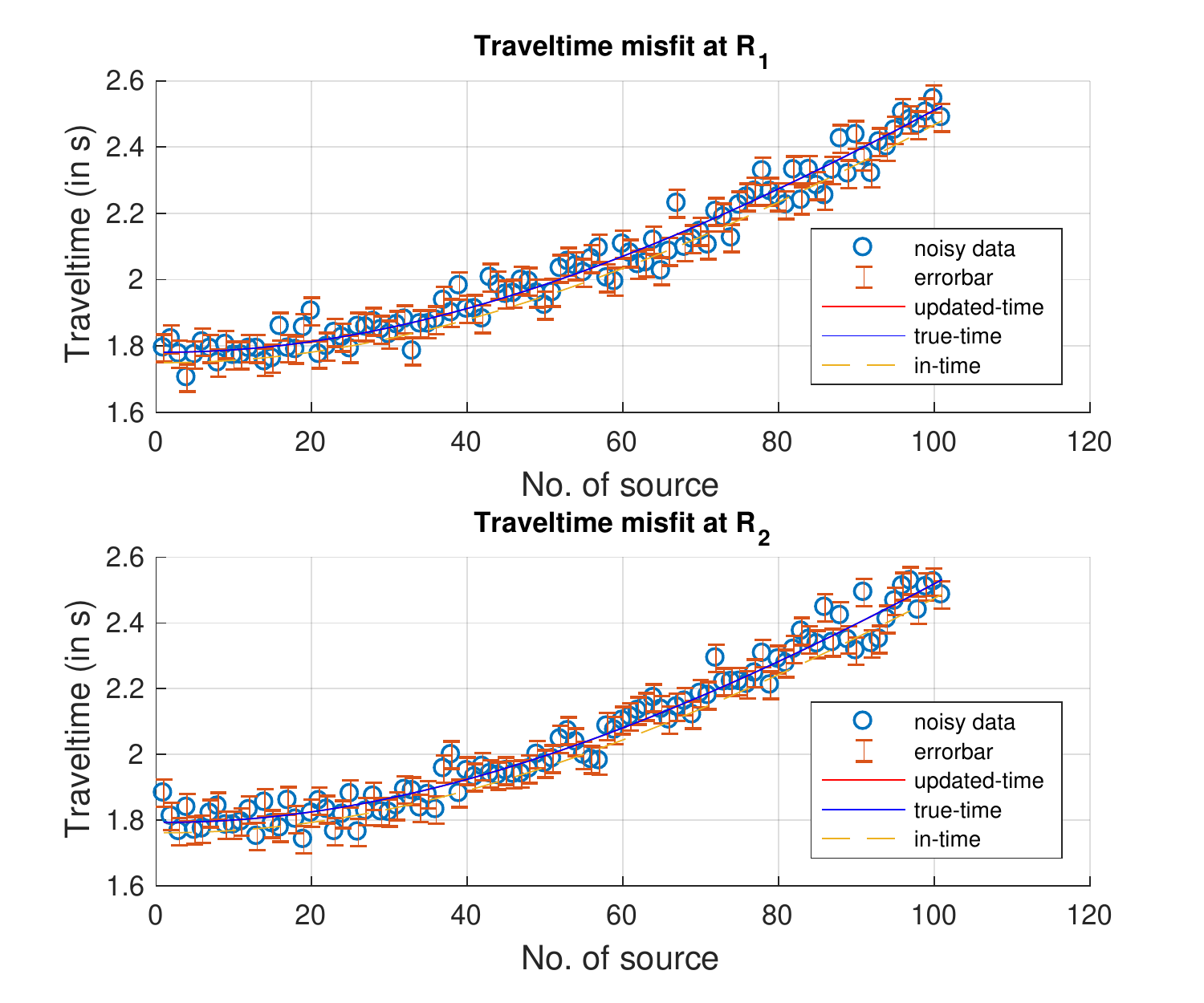}
                \caption{}
                \label{t_t_n_5_d_p_202}
        \end{subfigure}
        \begin{subfigure}[b]{.45\textwidth}
                \includegraphics[width=\textwidth]{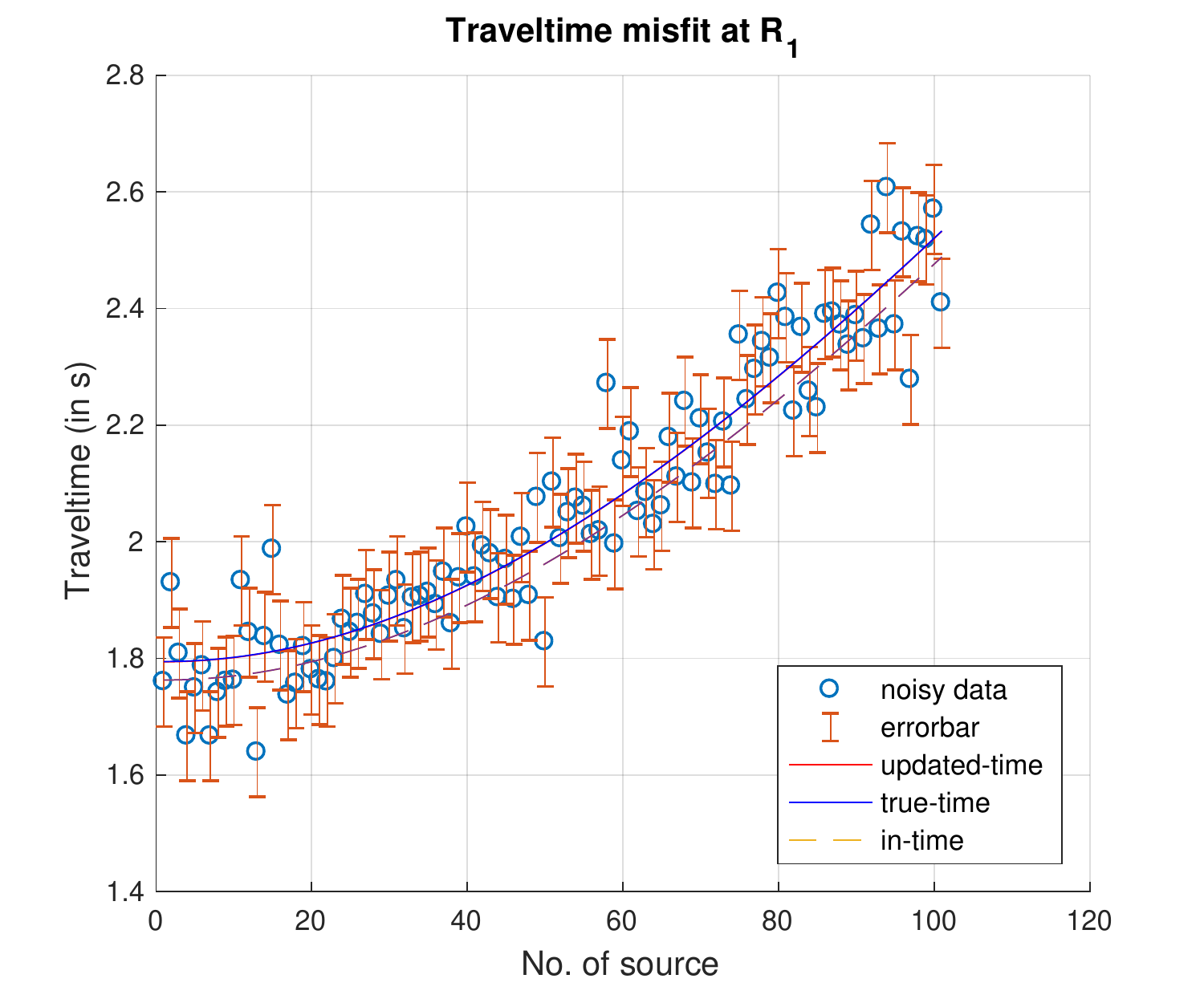}
                \caption{}
                \label{t_t_n_10_d_p_101}
        \end{subfigure}
                \begin{subfigure}[b]{.45\textwidth}
                \includegraphics[width=\textwidth]{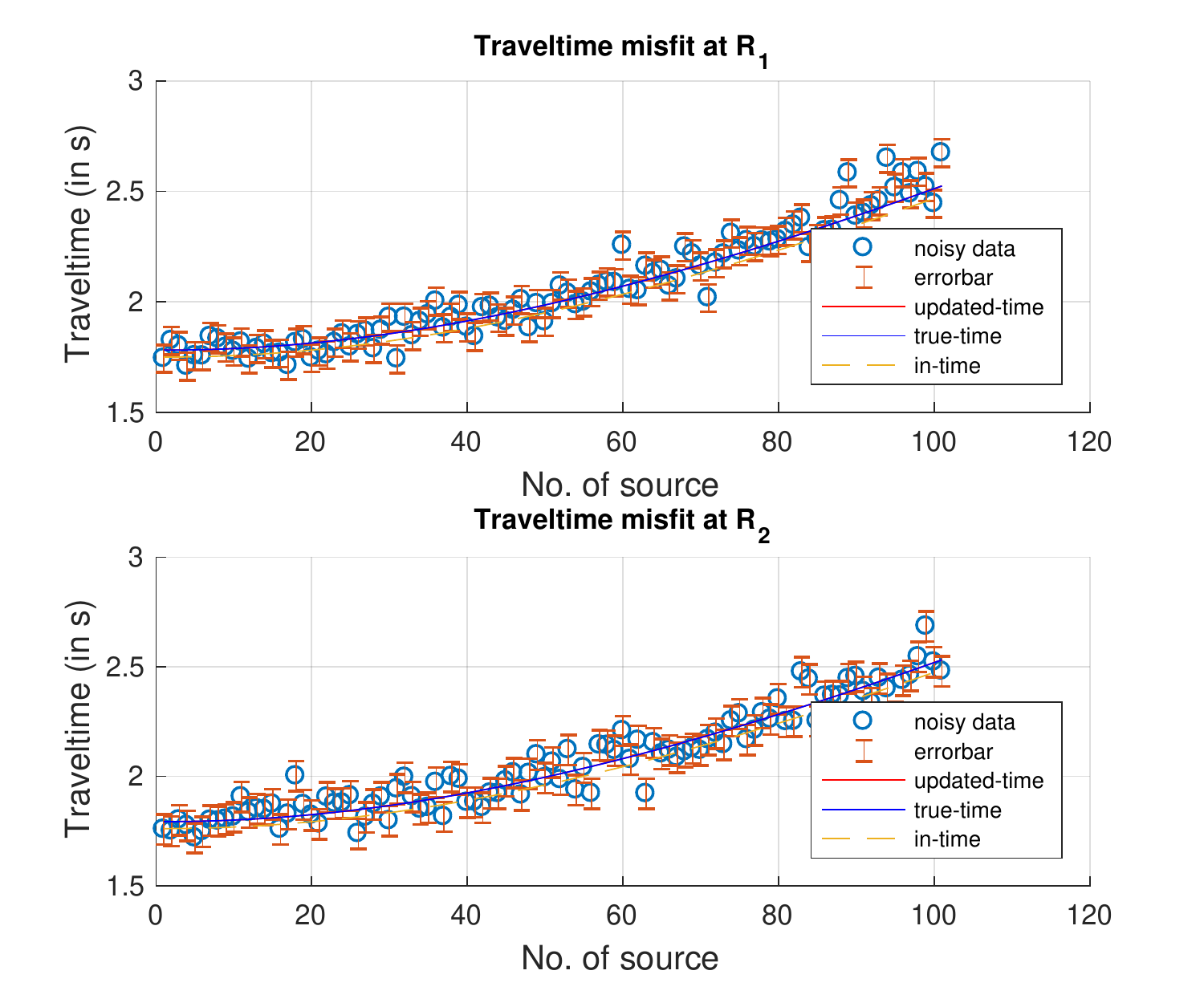}
                \caption{}
                \label{t_t_n_10_d_p_202}
        \end{subfigure}       
        \caption{Travetime inversion: variation of noise and number of data points, $v_{in} = v_{true}\pm20$}
        \label{fig:n_d_p_tt_20} 
\end{figure}

\begin{figure}[ht]
        \captionsetup[subfigure]{aboveskip=-.1cm}
        \centering
        \begin{subfigure}[b]{.45\textwidth}
                \includegraphics[width=\textwidth]{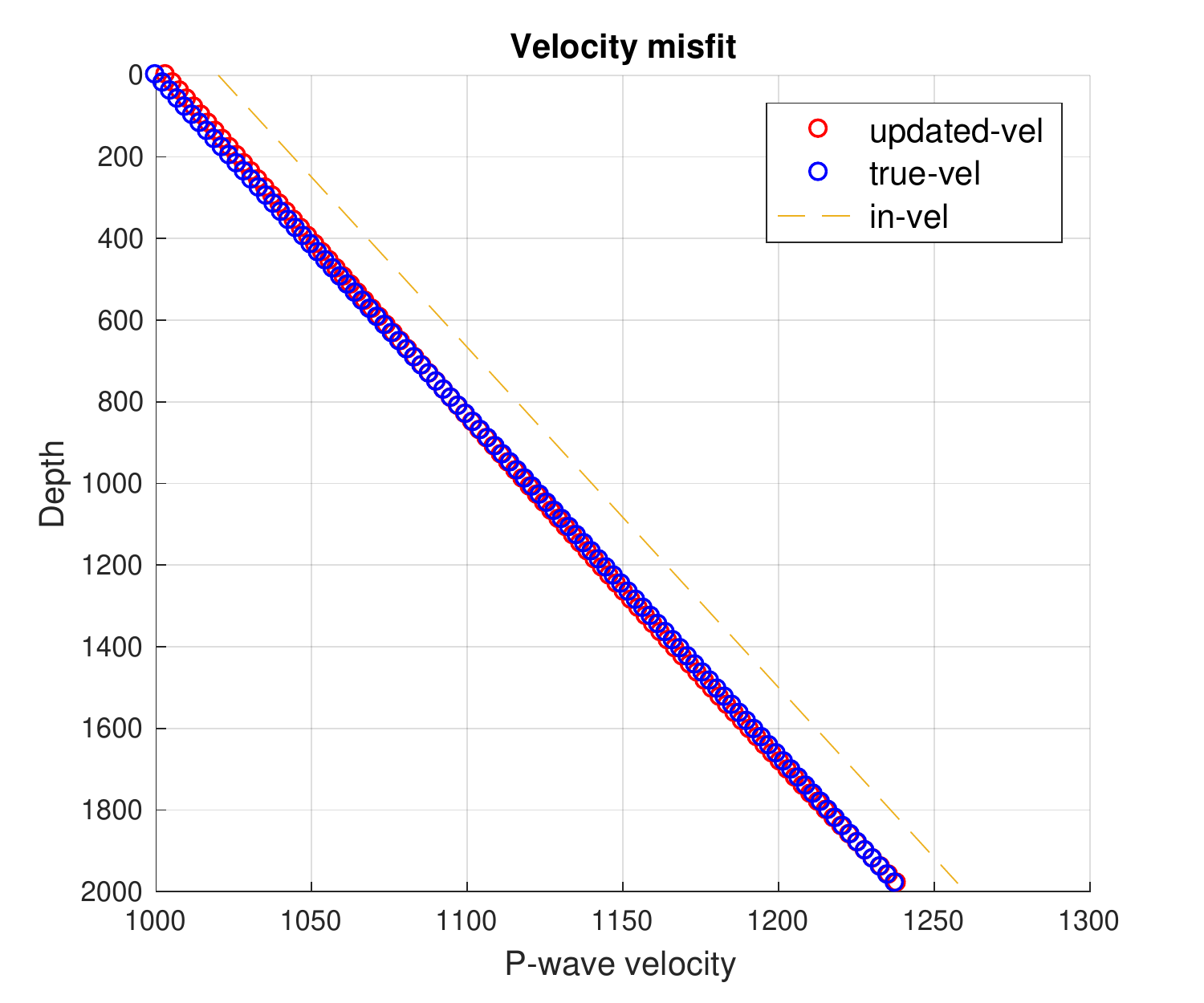}
                \caption{}
                \label{vel_n_1_d_p_101}
        \end{subfigure}
        \begin{subfigure}[b]{.45\textwidth}
                \includegraphics[width=\textwidth]{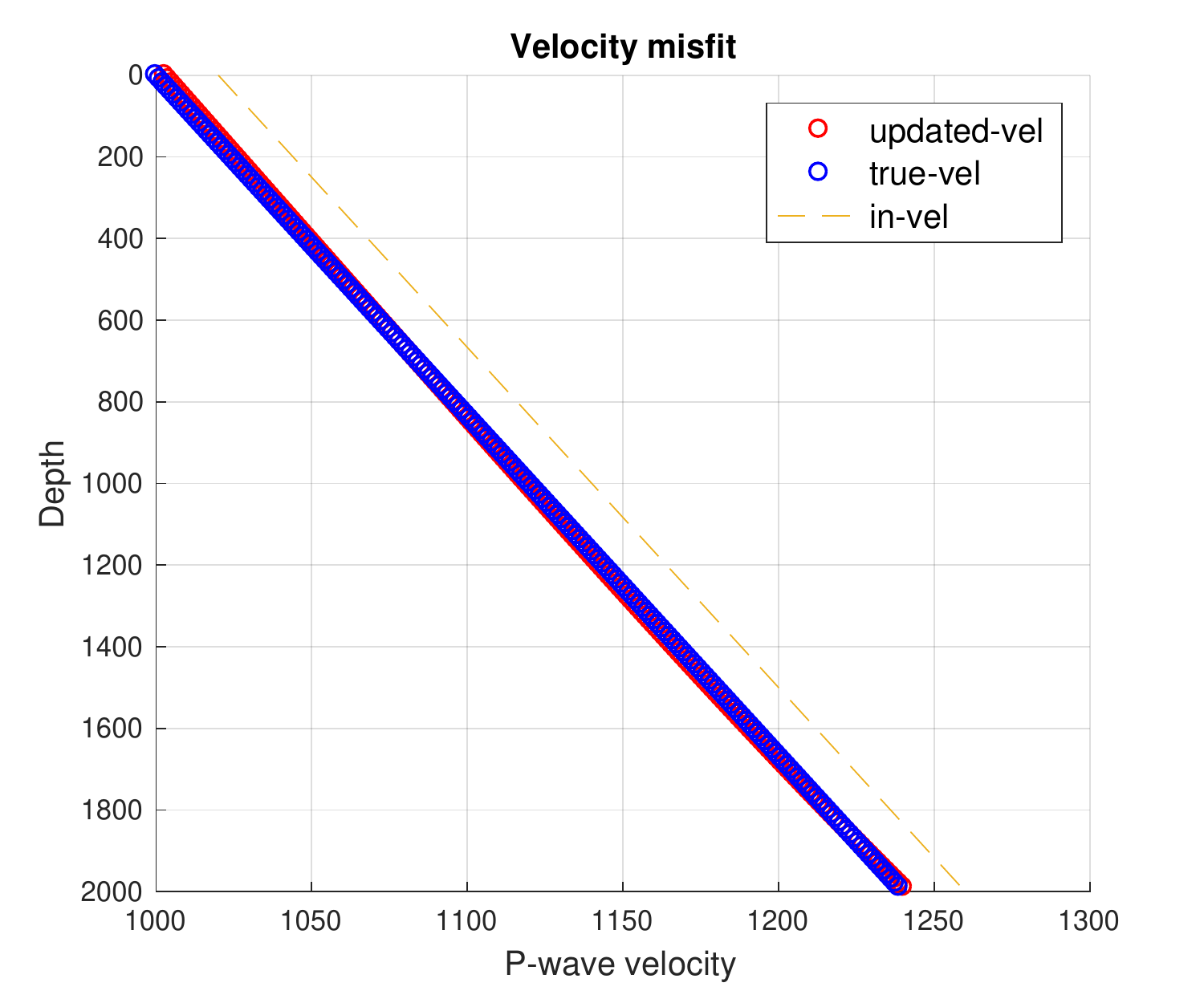}
                \caption{}
                \label{vel_n_1_d_p_202}
        \end{subfigure}
        \begin{subfigure}[b]{.45\textwidth}
                \includegraphics[width=\textwidth]{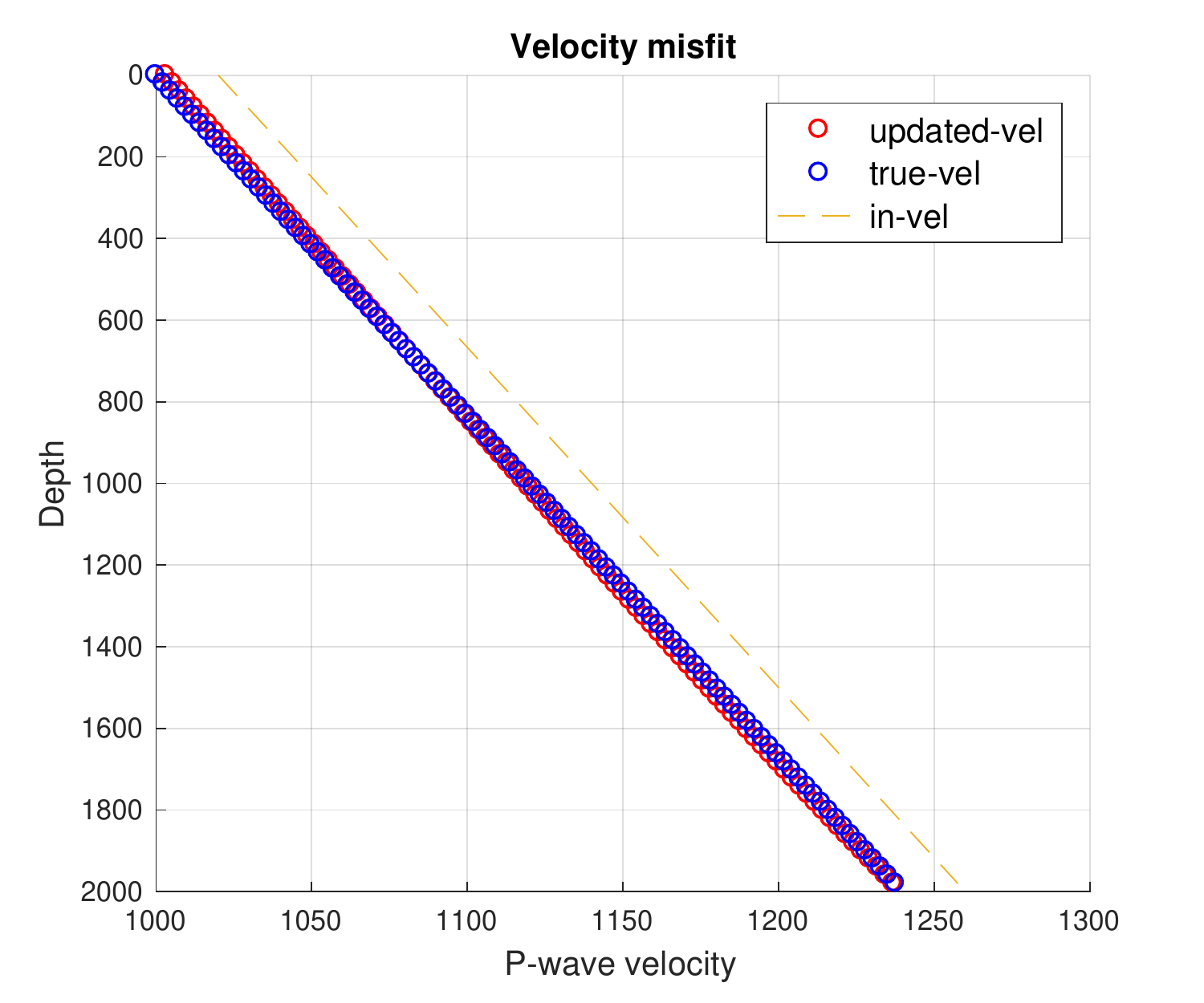}
                \caption{}
                \label{vel_n_5_d_p_101}
        \end{subfigure}       
        \begin{subfigure}[b]{.45\textwidth}
                \includegraphics[width=\textwidth]{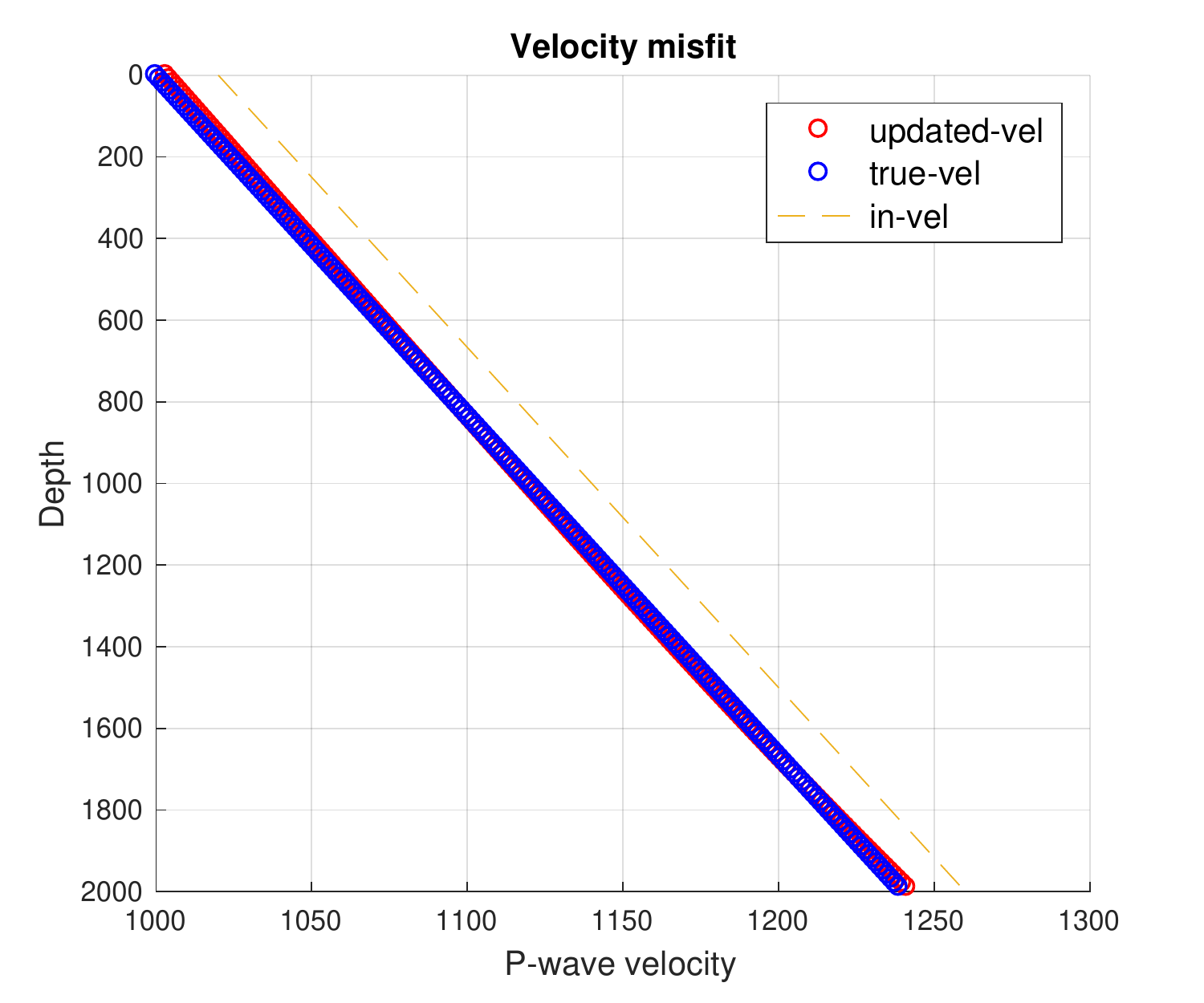}
                \caption{}
                \label{vel_n_5_d_p_202}
        \end{subfigure}
        \begin{subfigure}[b]{.45\textwidth}
                \includegraphics[width=\textwidth]{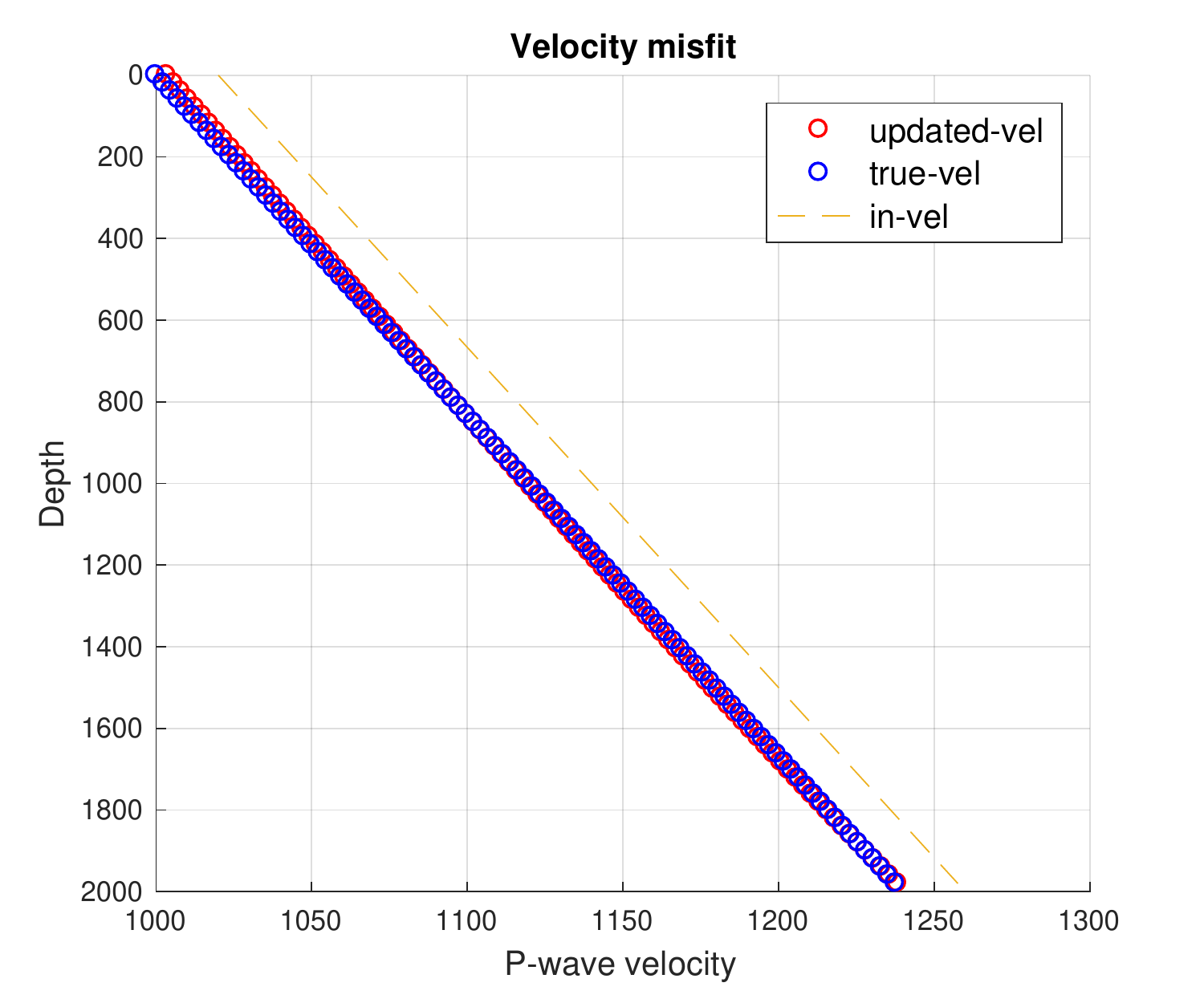}
                \caption{}
                \label{vel_n_10_d_p_101}
        \end{subfigure}
                \begin{subfigure}[b]{.45\textwidth}
                \includegraphics[width=\textwidth]{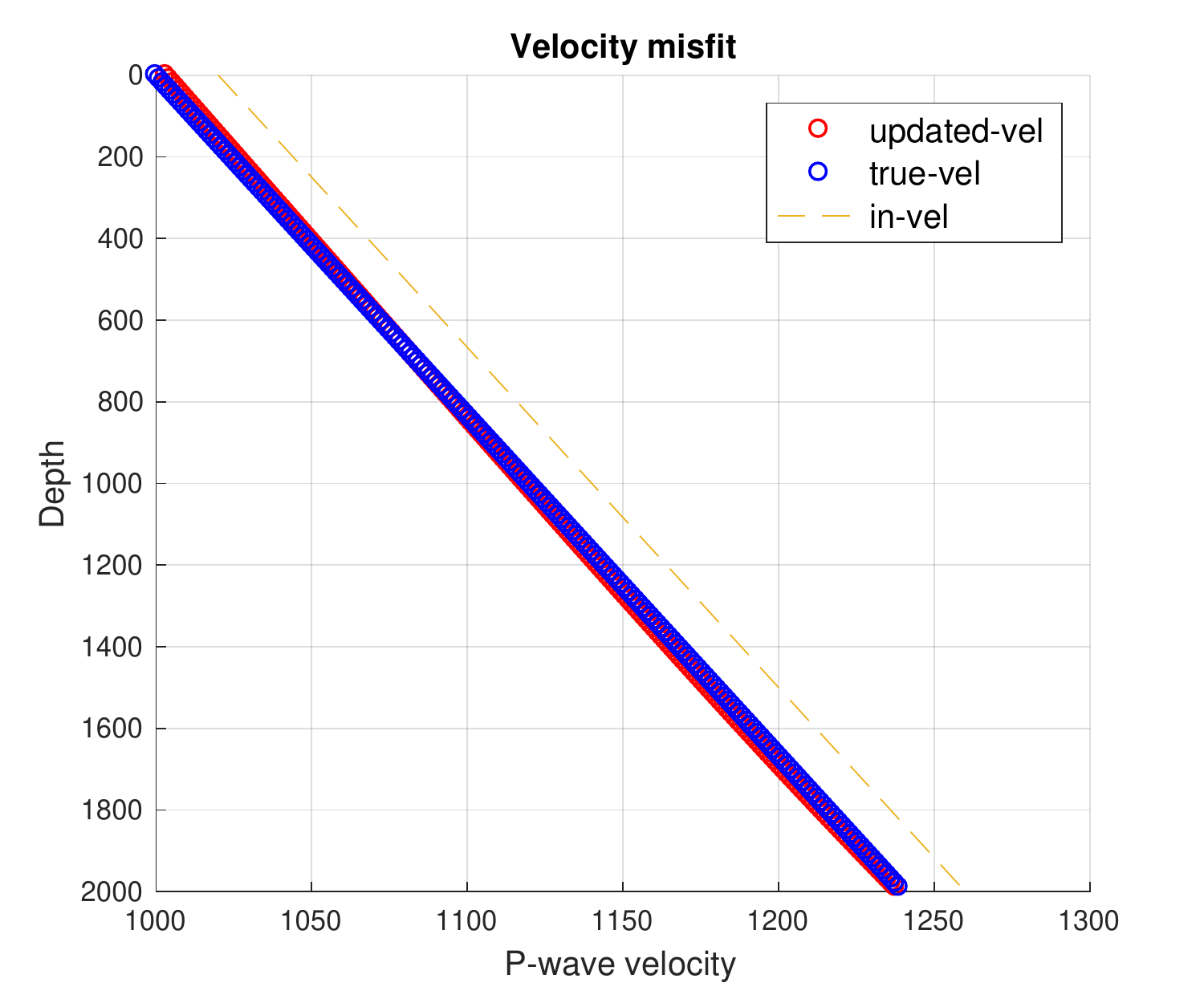}
                \caption{}
                \label{vel_n_10_d_p_202}
        \end{subfigure}       
        \caption{Velocity inversion : variation of noise and number of data points, $v_{in} = v_{true}\pm20$}
        \label{fig:n_d_p_vel_20} 
\end{figure}

\begin{figure}[ht]
        \captionsetup[subfigure]{aboveskip=-.1cm}
        \centering
        \begin{subfigure}[b]{.45\textwidth}
                \includegraphics[width=\textwidth]{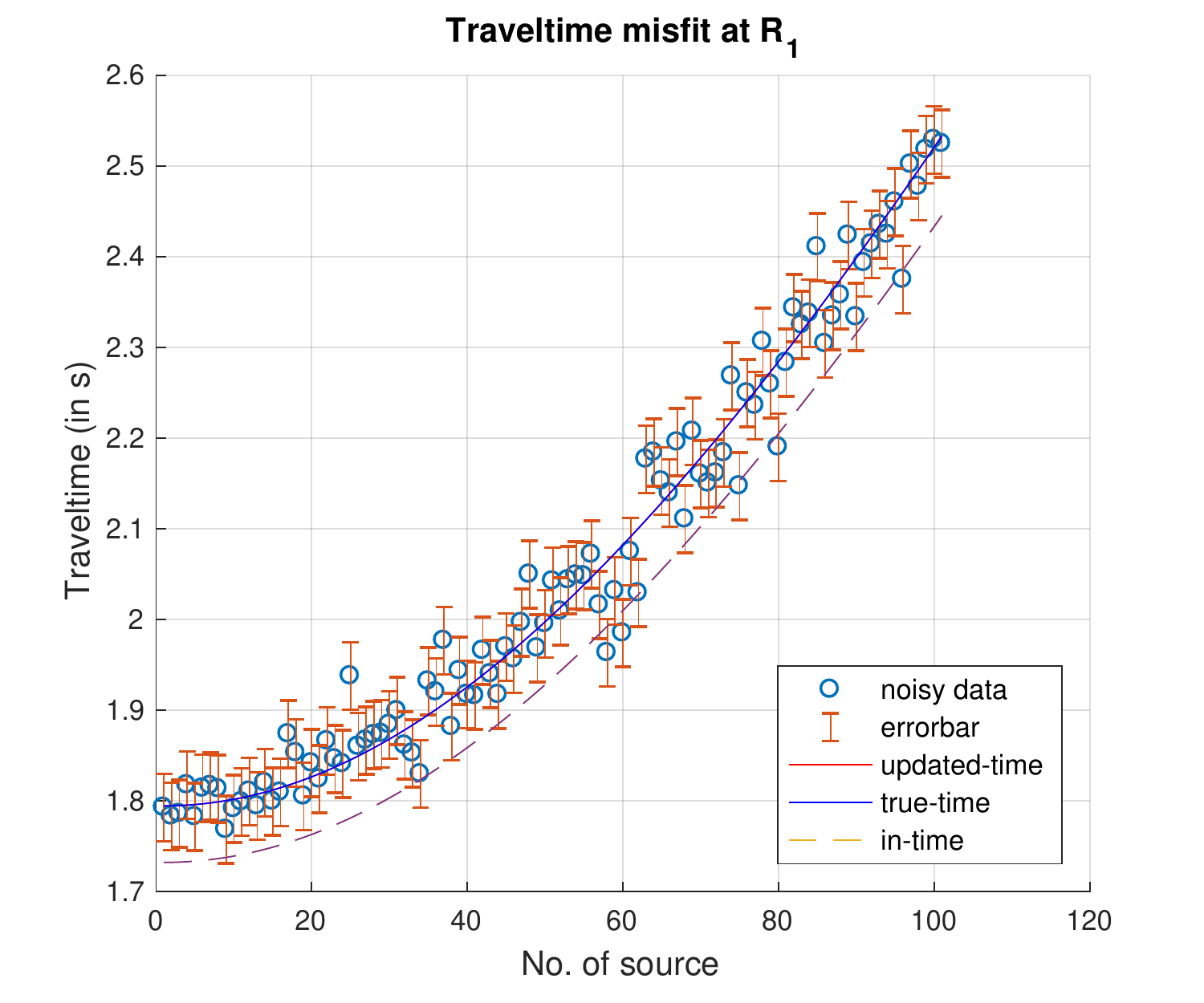}
                \caption{}
                \label{t_t_n2_1_d_p_101}
        \end{subfigure}
        \begin{subfigure}[b]{.45\textwidth}
                \includegraphics[width=\textwidth]{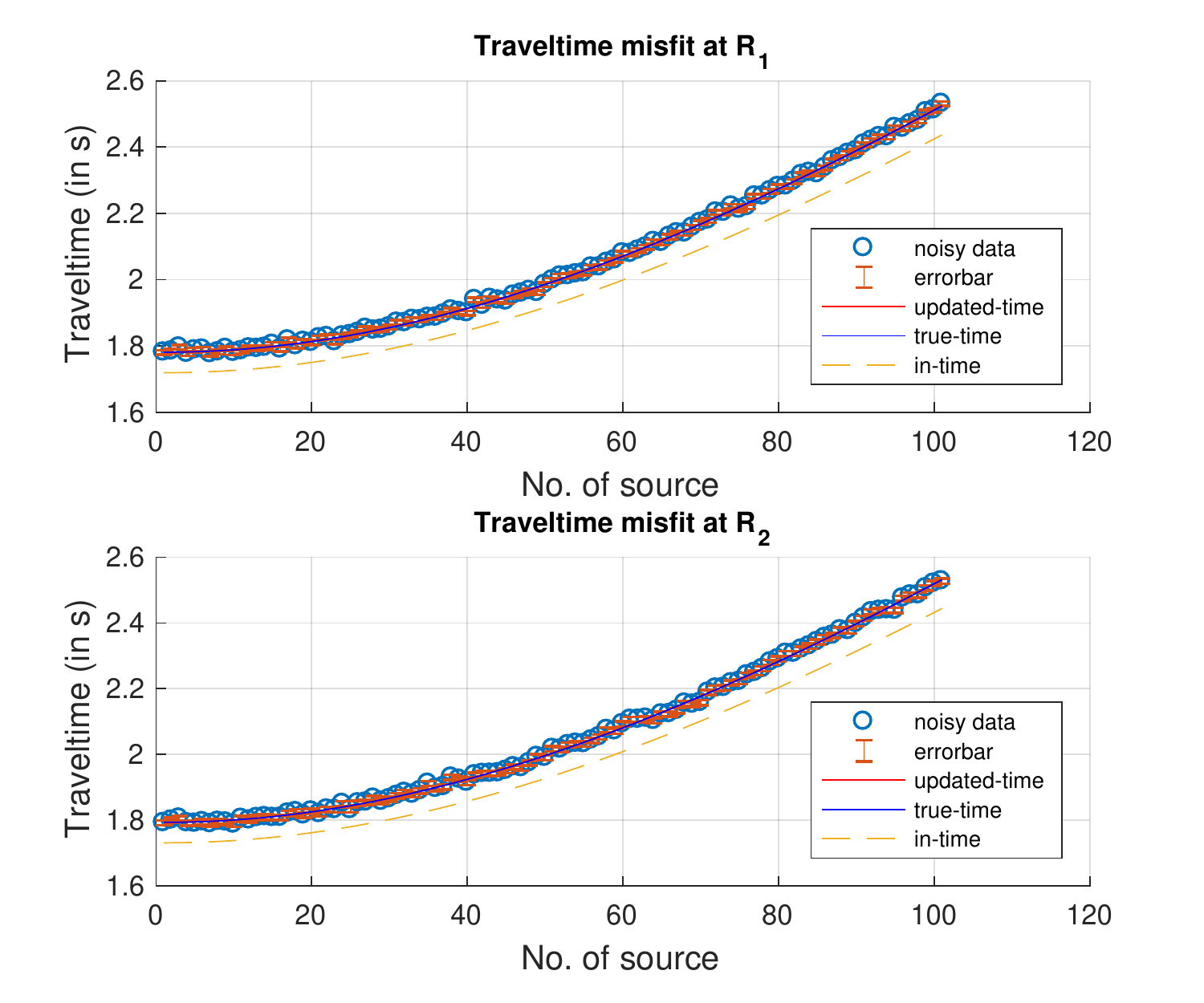}
                \caption{}
                \label{t_t_n2_1_d_p_202}
        \end{subfigure}
        \begin{subfigure}[b]{.45\textwidth}
                \includegraphics[width=\textwidth]{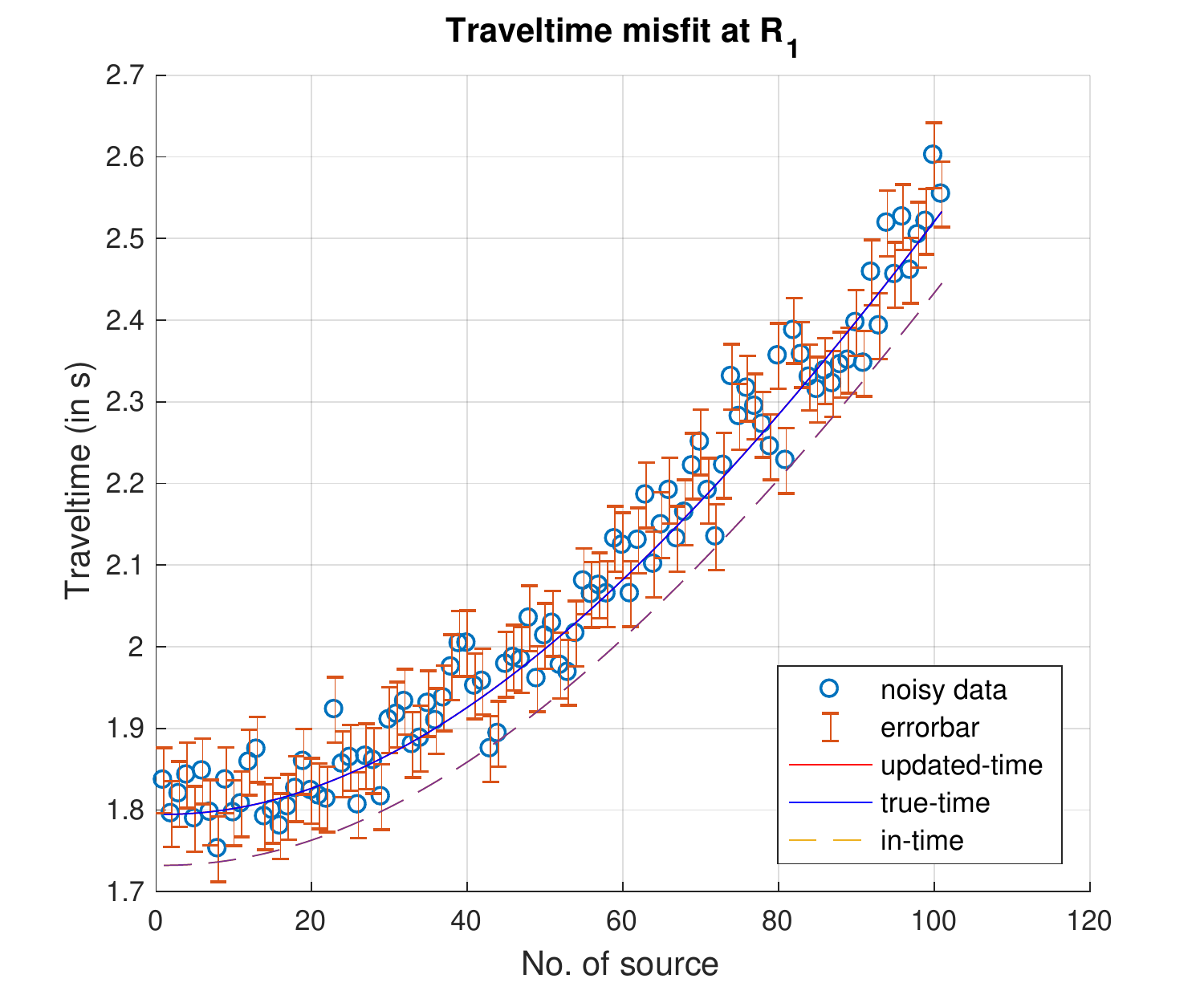}
                \caption{}
                \label{t_t_n2_5_d_p_101}
        \end{subfigure}       
        \begin{subfigure}[b]{.45\textwidth}
                \includegraphics[width=\textwidth]{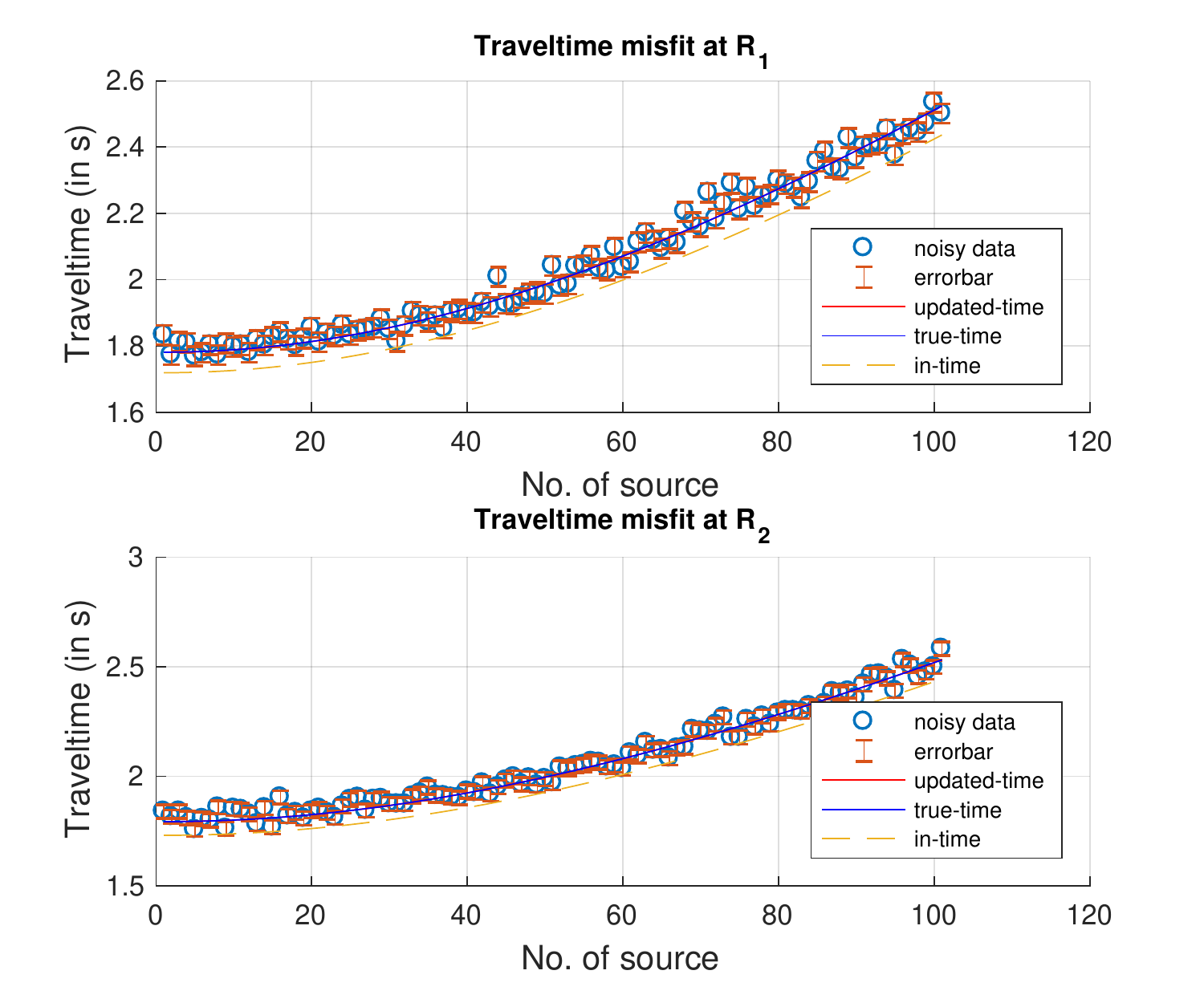}
                \caption{}
                \label{t_t_n2_5_d_p_202}
        \end{subfigure}
        \begin{subfigure}[b]{.45\textwidth}
                \includegraphics[width=\textwidth]{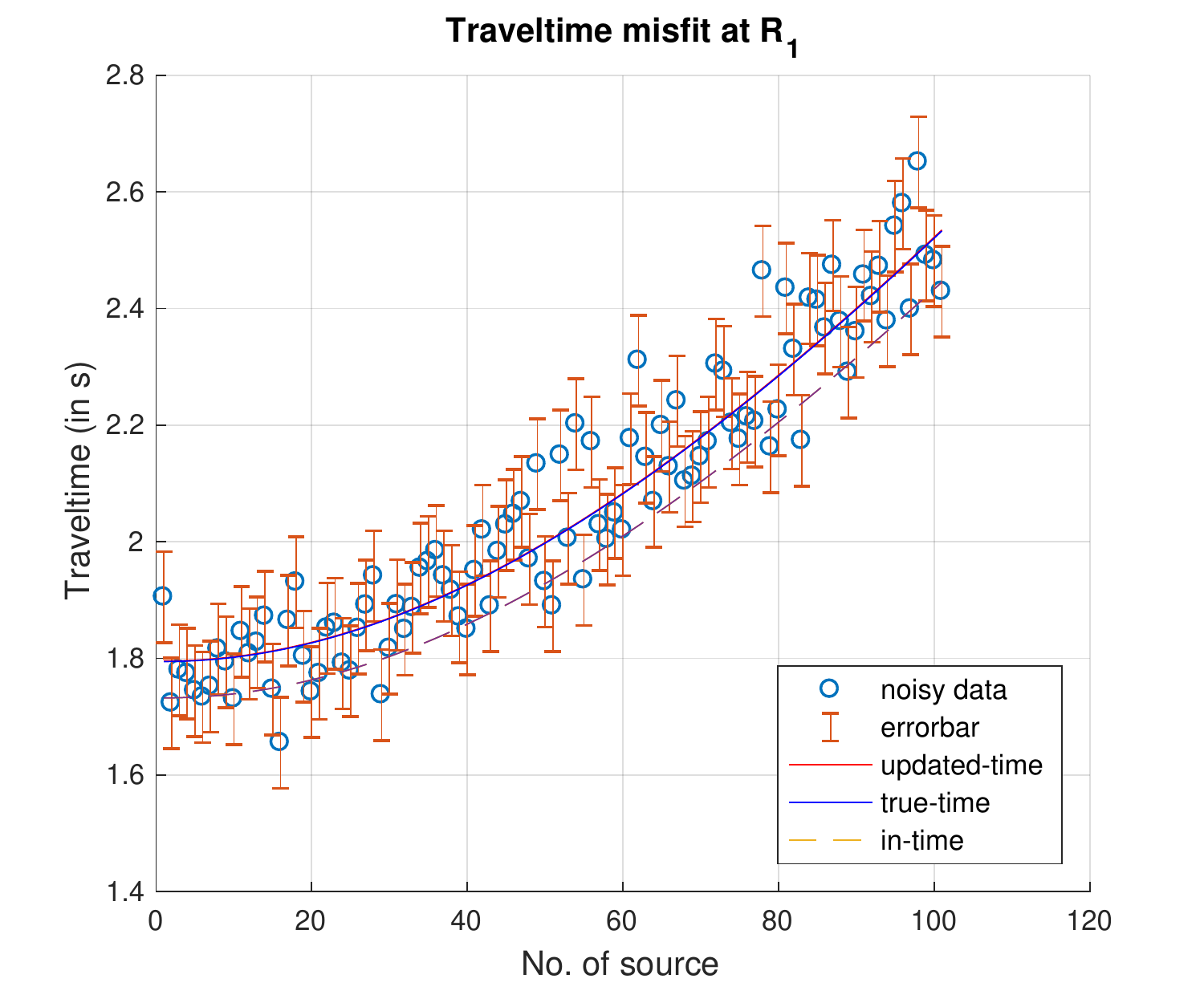}
                \caption{}
                \label{t_t_n2_10_d_p_101}
        \end{subfigure}
                \begin{subfigure}[b]{.45\textwidth}
                \includegraphics[width=\textwidth]{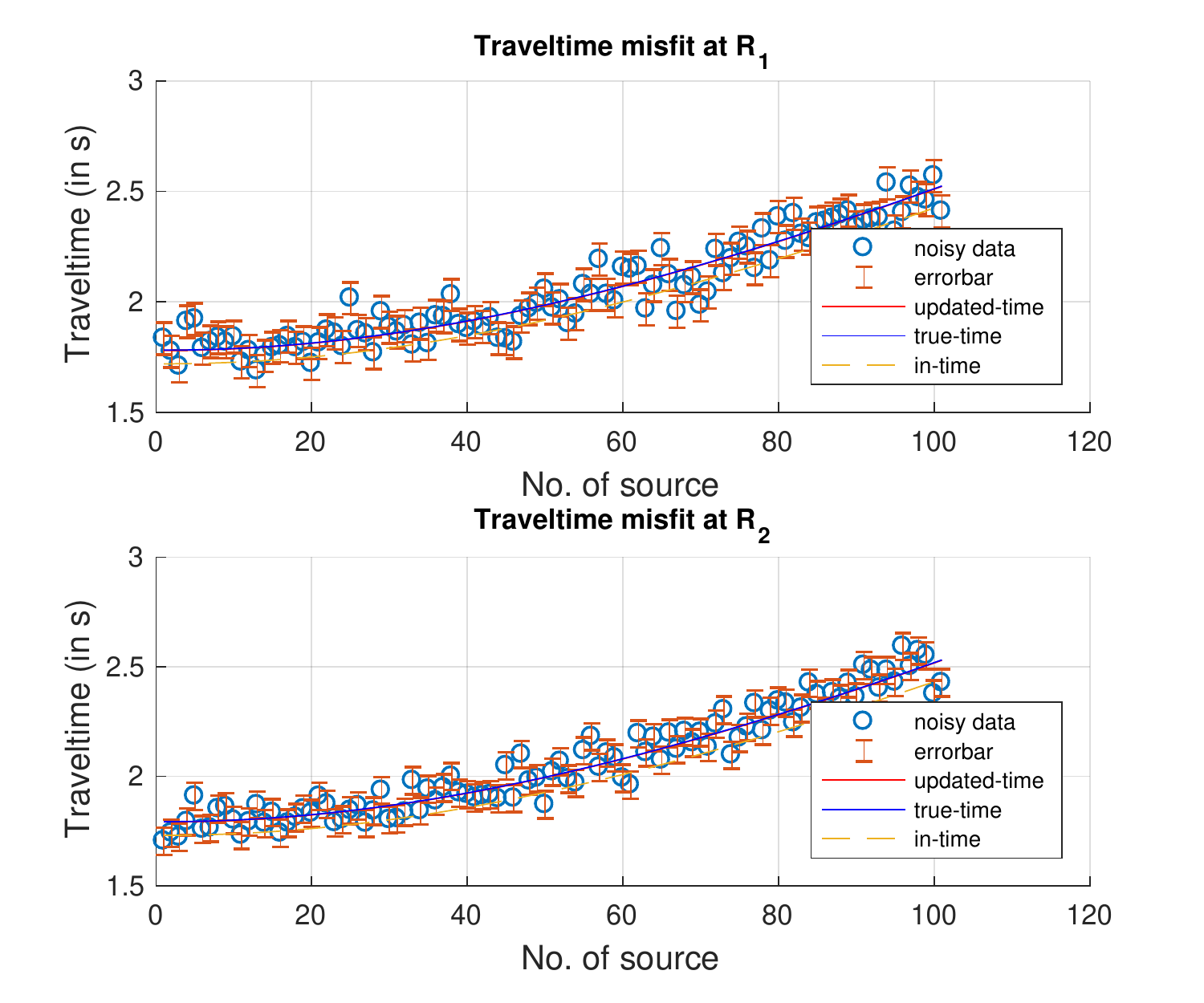}
                \caption{}
                \label{t_t_n2_10_d_p_202}
        \end{subfigure}       
        \caption{Travetime inversion: variation of noise and number of data points, $v_{in} = v_{true}\pm40$}
        \label{fig:n_d_p_tt_40} 
\end{figure}

\begin{figure}[ht]
        \captionsetup[subfigure]{aboveskip=-.1cm,belowskip = 0cm}
        \centering
        \begin{subfigure}[b]{.45\textwidth}
                \includegraphics[width=\textwidth]{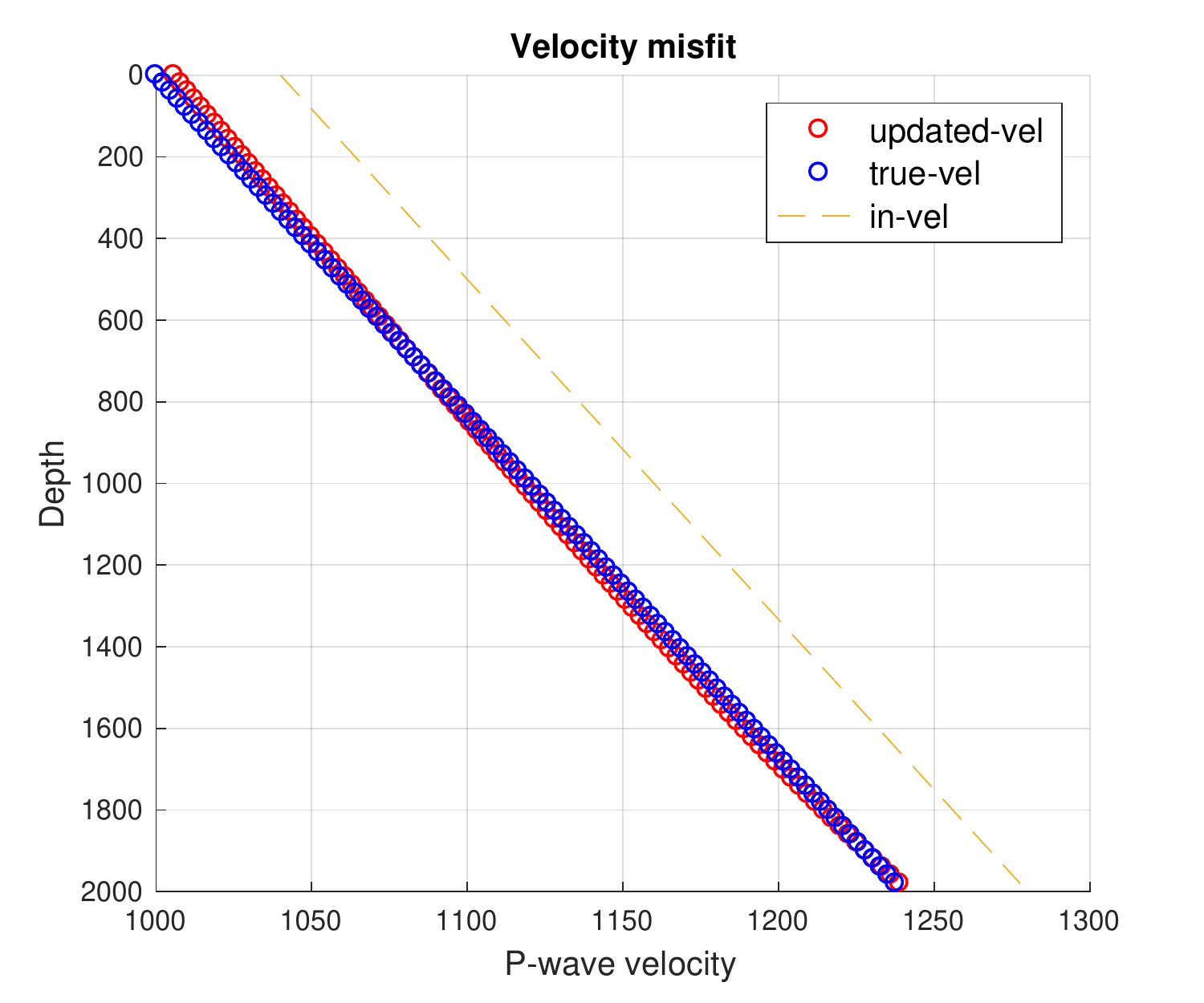}
                \caption{}
                \label{vel_n2_1_d_p_101}
        \end{subfigure}
        \begin{subfigure}[b]{.45\textwidth}
                \includegraphics[width=\textwidth]{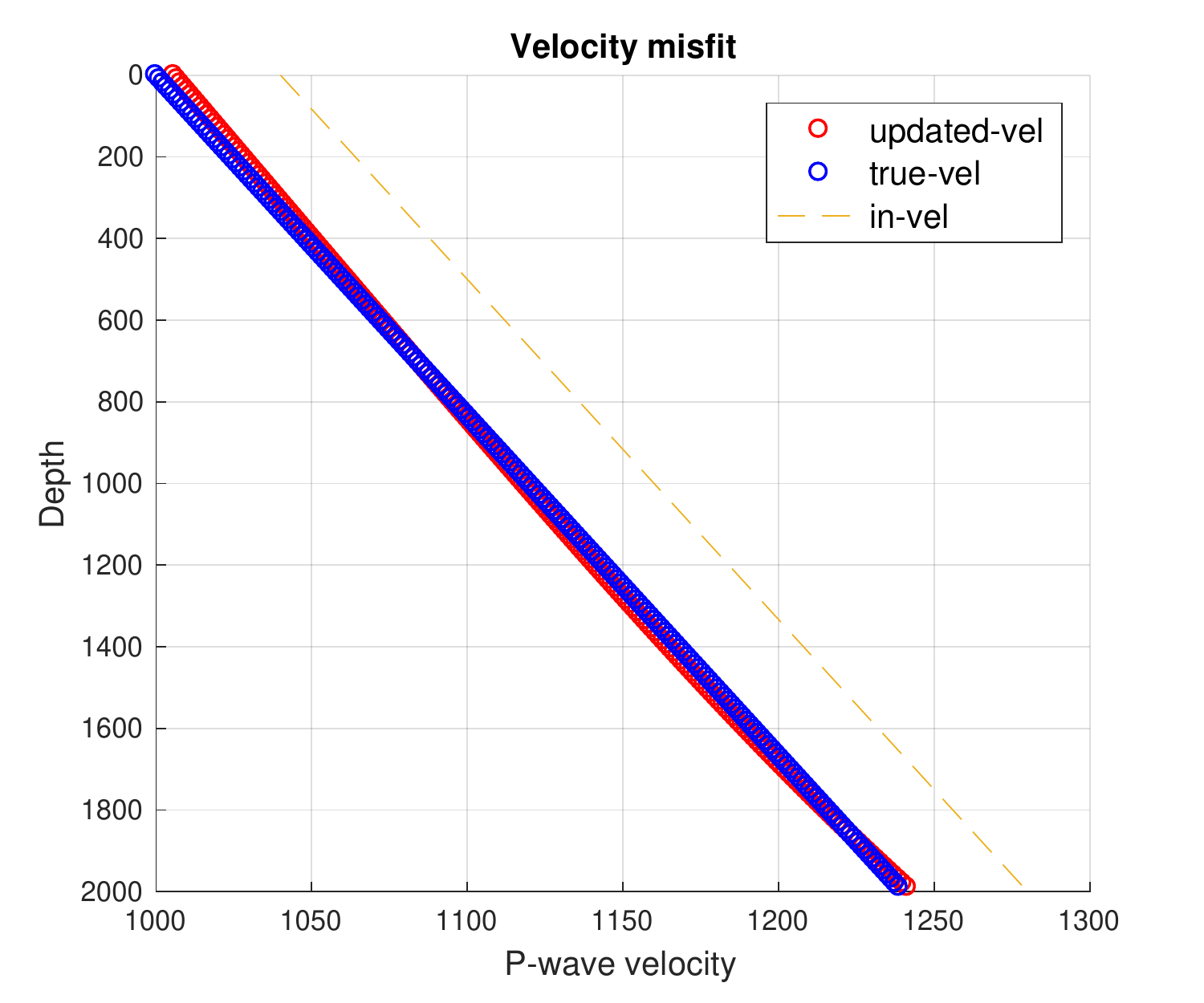}
                \caption{}
                \label{vel_n2_1_d_p_202}
        \end{subfigure}
        \begin{subfigure}[b]{.45\textwidth}
                \includegraphics[width=\textwidth]{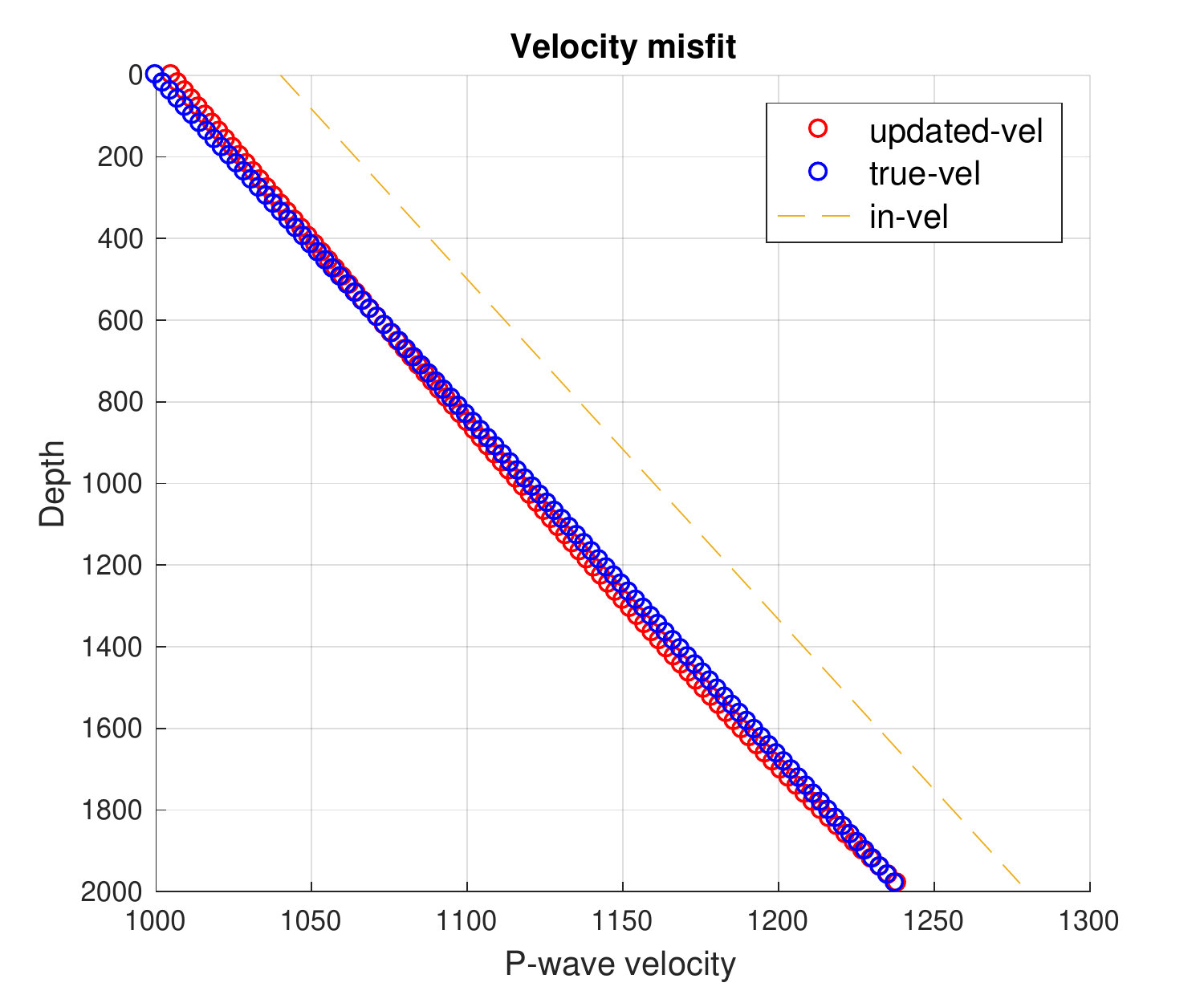}
                \caption{}
                \label{vel_n2_5_d_p_101}
        \end{subfigure}       
        \begin{subfigure}[b]{.45\textwidth}
                \includegraphics[width=\textwidth]{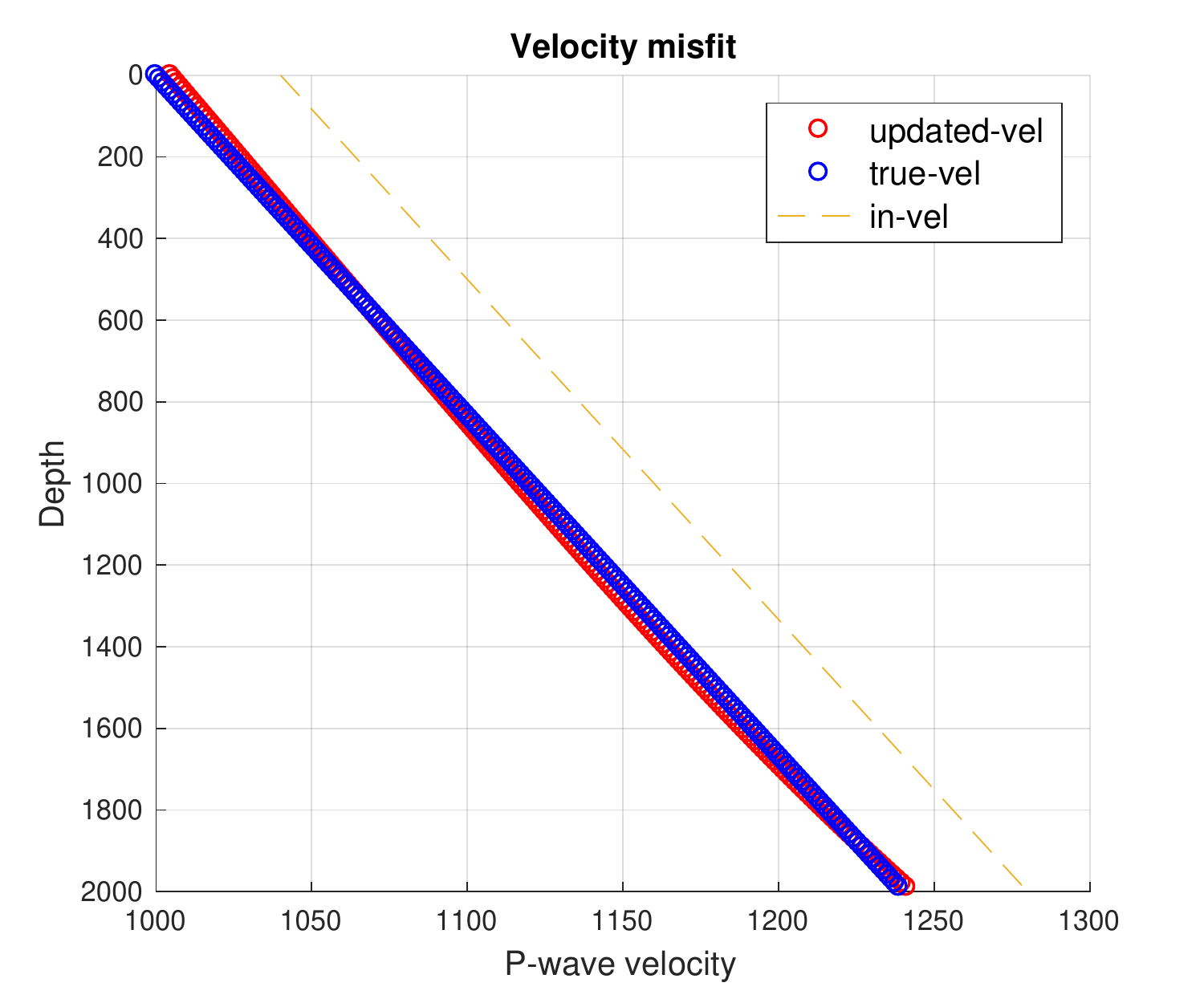}
                \caption{}
                \label{vel_n2_5_d_p_202}
        \end{subfigure}
        \begin{subfigure}[b]{.45\textwidth}
                \includegraphics[width=\textwidth]{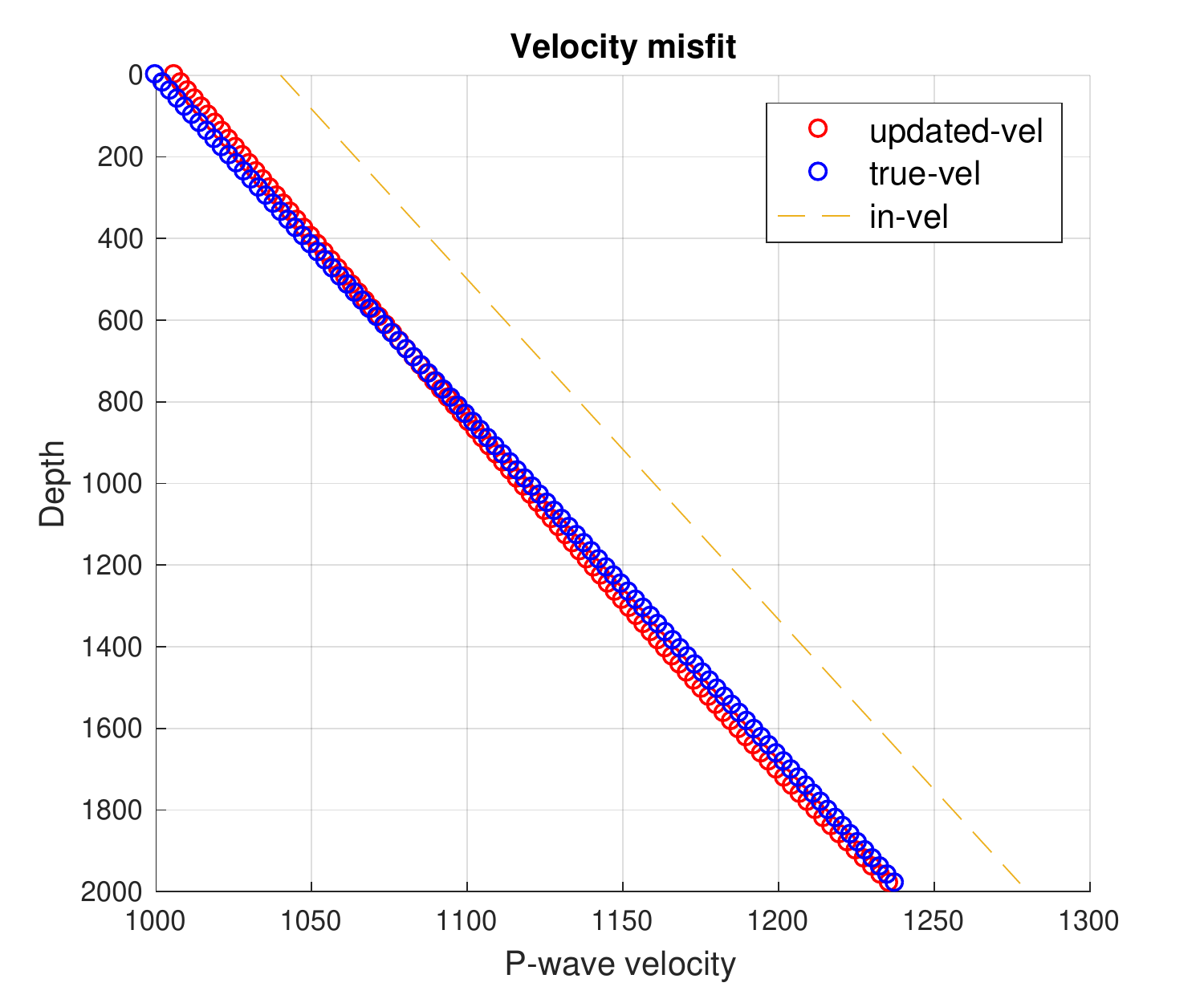}
                \caption{}
                \label{vel_n2_10_d_p_101}
        \end{subfigure}
                \begin{subfigure}[b]{.45\textwidth}
                \includegraphics[width=\textwidth]{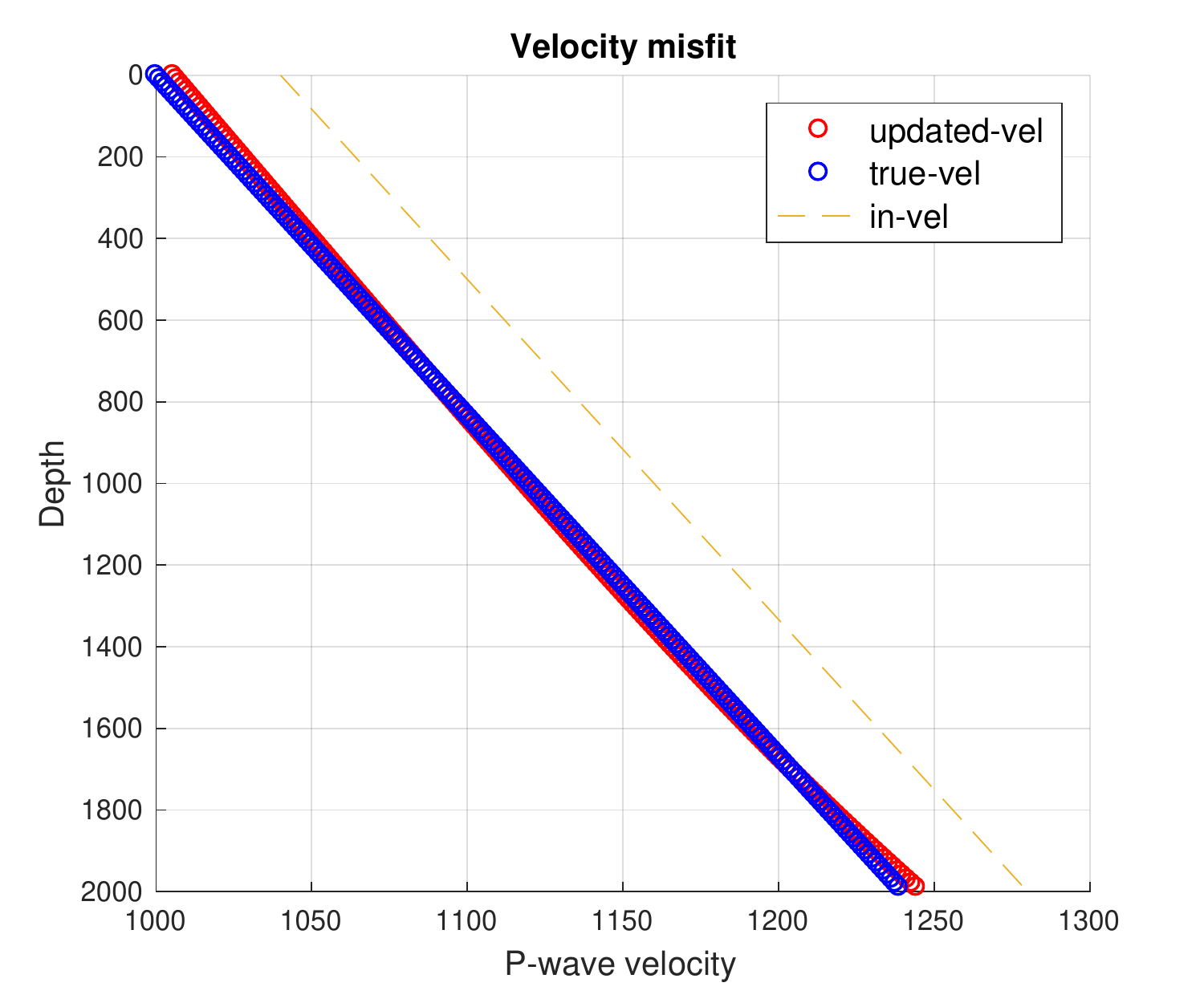}
                \caption{}
                \label{vel_n2_10_d_p_202}
        \end{subfigure}       
        \caption{Velocity inversion : variation of noise and number of data points, $v_{in} = v_{true}\pm40$}
        \label{fig:n_d_p_vel_40} 
\end{figure}

In Table~\ref{tab:syn_n_d_p}, $f({\bf M})$ provides the misfit functional, $a_{inv}$ and $b_{inv}$ present the model parameters after fitting a line to the inverted velocity. 
The traveltime convergence results are shown in Figures~\ref{fig:n_d_p_tt_20} and \ref{fig:n_d_p_tt_40}.
The misfits of the inverted velocity to the reference velocity are shown in Figures~\ref{fig:n_d_p_vel_20} and \ref{fig:n_d_p_vel_40}.
To examine the effect of noise and the number of data points, we add 1$\%$, 5$\%$ and 10$\%$ of random noises and 101 and 202 number of data points. 

\captionsetup[figure]{aboveskip=0cm}

\captionsetup[figure]{aboveskip=0cm}


\captionsetup[figure]{aboveskip=0cm}
%
%
\captionsetup[figure]{aboveskip=0cm,belowskip=0cm}

\FloatBarrier
\subsection{Test of the model parameters $a$ and $b$}
In Table~\ref{tab:syn_b}, we consider the reference velocity model to be a linear function of depth, where parameters $a = 1000 \, {\rm ms^{-1}}$ and $b = 0.12 \, s^{-1}$. 
In contrast to Table~\ref{tab:syn_n_d_p}, here we change the model parameter $b$. 
For the first six tests, the startup velocity for the inverse model is $v_{ref}+\pm 30 \, {\rm ms^{-1}}$, and for the last six tests, the startup velocity is $v_{ref}\pm 60 \, {\rm ms^{-1}}$. 
We set the noise to 1$\%$, the number of data points to 202 and the total number of model parameters to 202.

The purpose of this section to show, for a given noise and data points, the effects of the startup model parameters $a_{in}$ and $b_{in}$ on the inversion. 
For $b_{in}$, we change it from $b_{true} \to b_{true}\pm 0.01$.

\begin{figure}[ht]
       \captionsetup[subfigure]{aboveskip=-.1cm}
        \centering
        \begin{subfigure}[b]{.45\textwidth}
                \includegraphics[width=\textwidth]{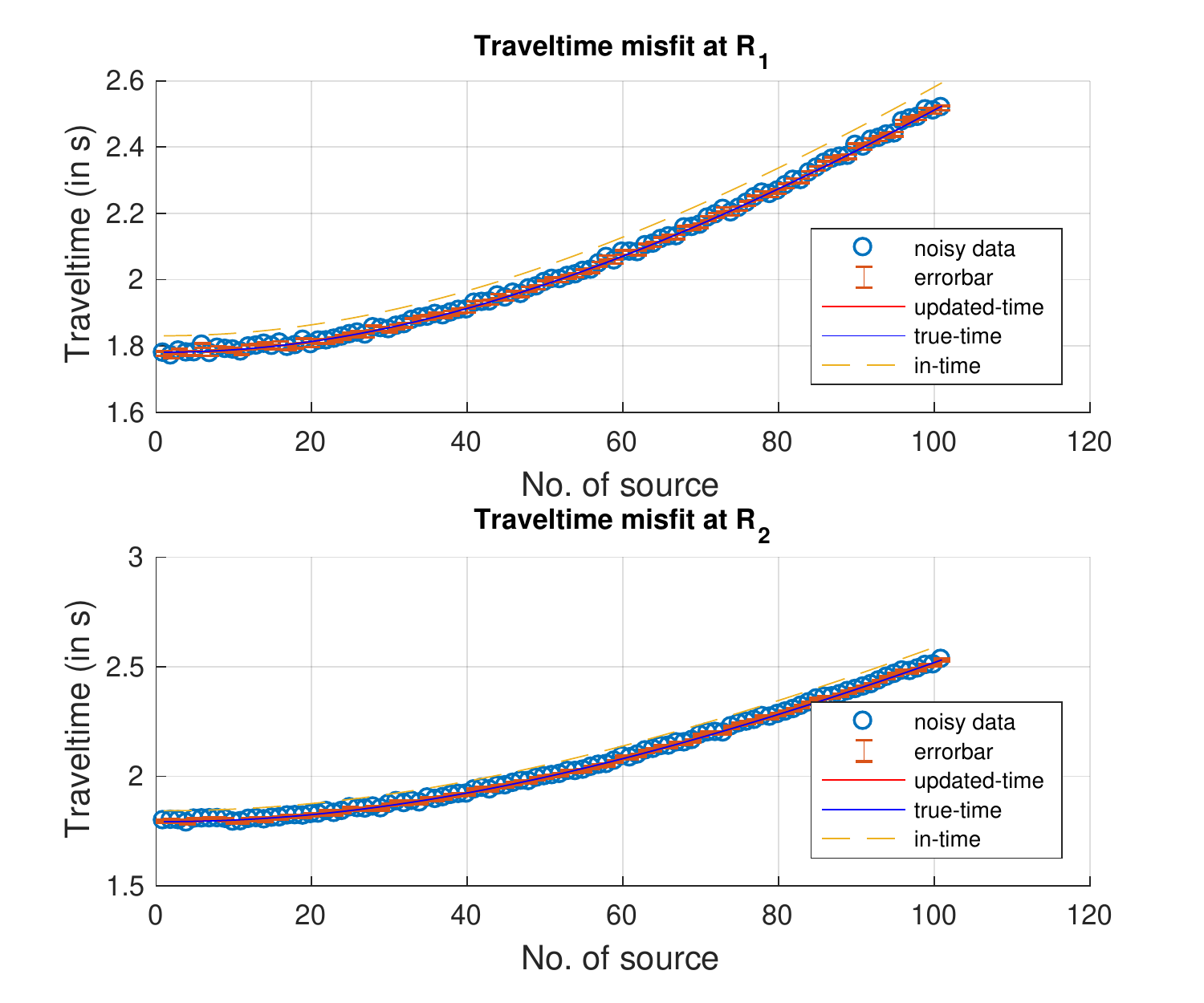}
                \caption{}
                \label{tt_a_m30}
        \end{subfigure}       
        \begin{subfigure}[b]{.45\textwidth}
                \includegraphics[width=\textwidth]{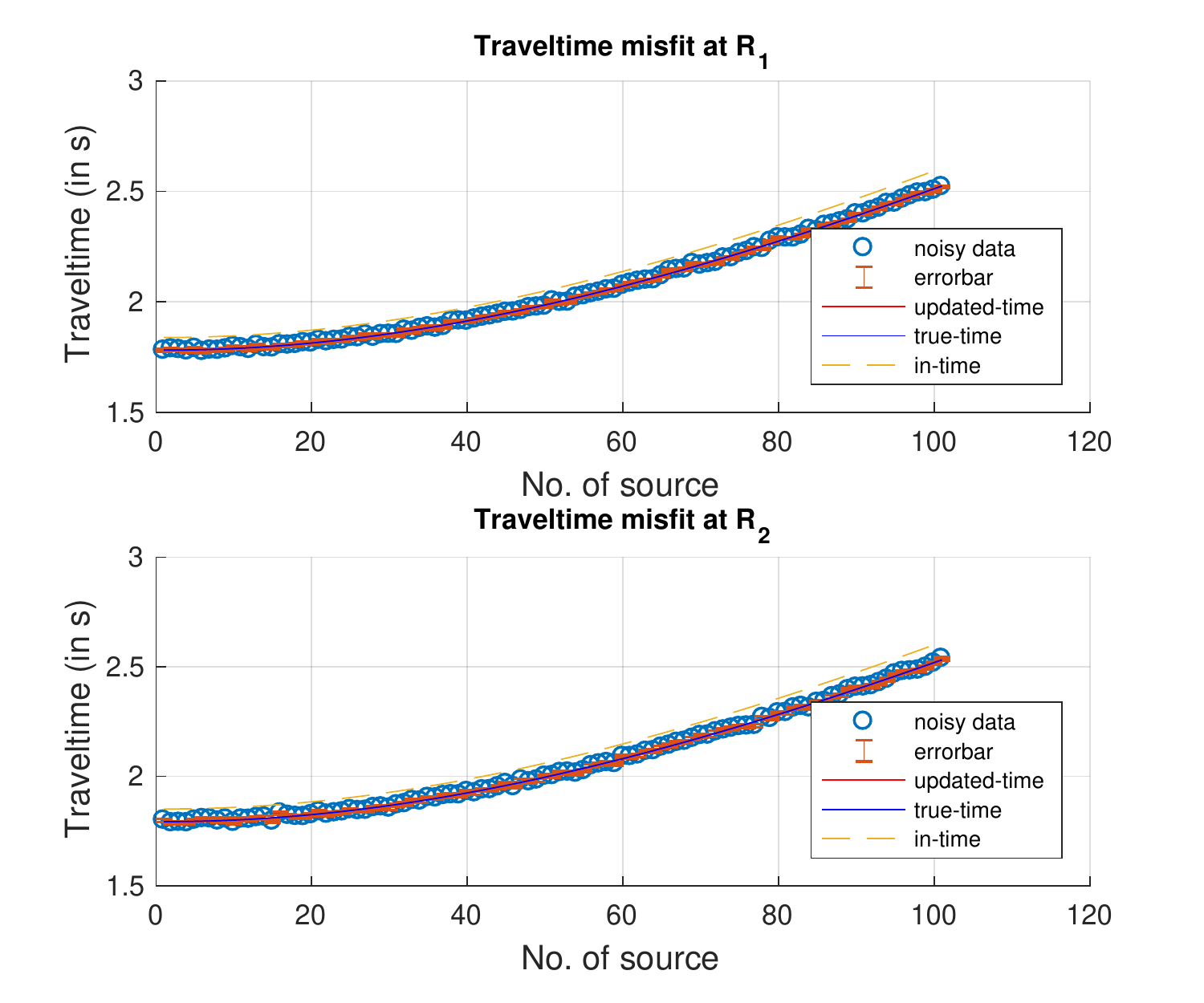}
                \caption{}
                \label{tt_a_m30_b_m005}
        \end{subfigure}
        \begin{subfigure}[b]{.45\textwidth}
                \includegraphics[width=\textwidth]{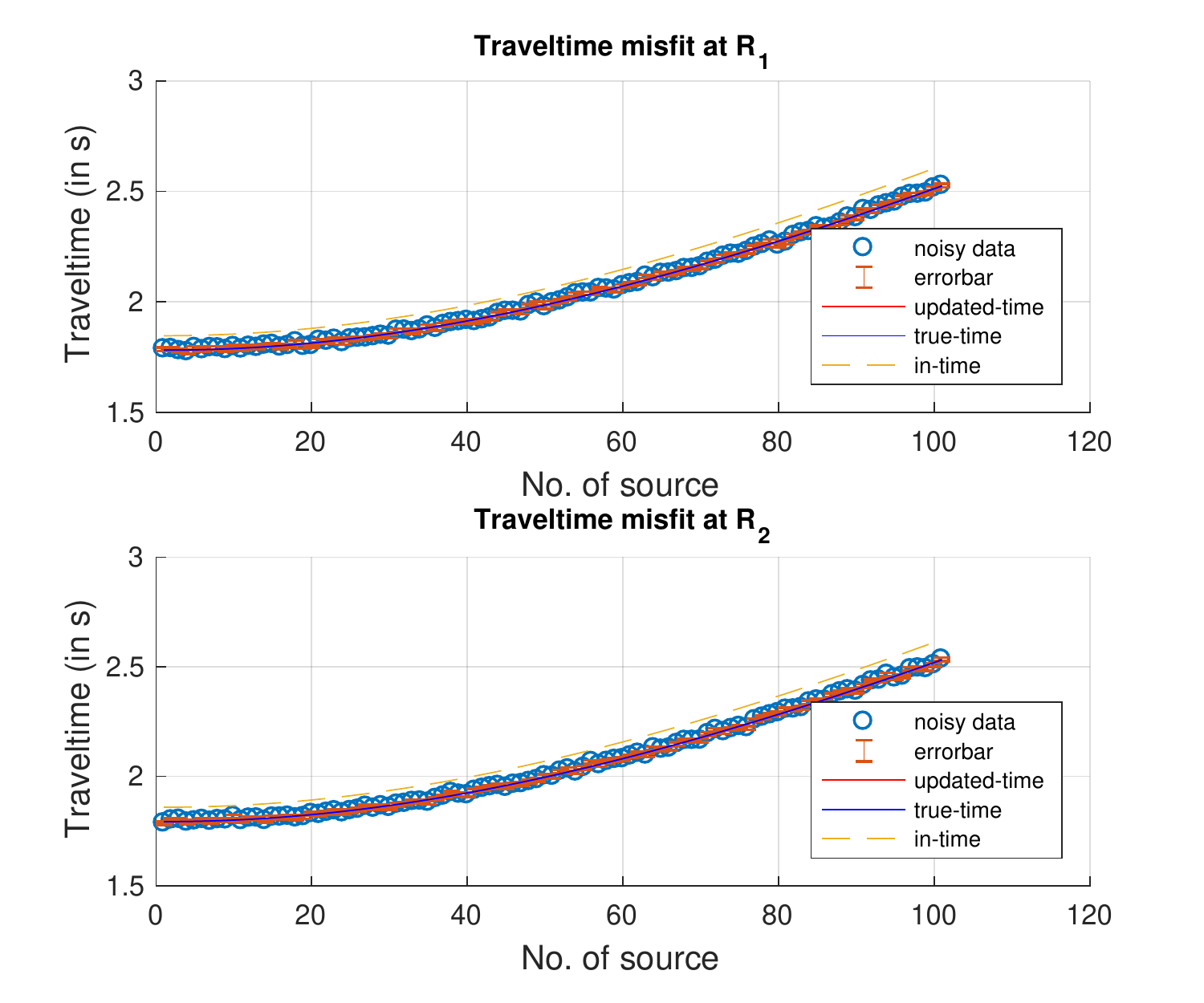}
                \caption{}
                \label{tt_a_m30_b_m01}
        \end{subfigure}
                \begin{subfigure}[b]{.45\textwidth}
                \includegraphics[width=\textwidth]{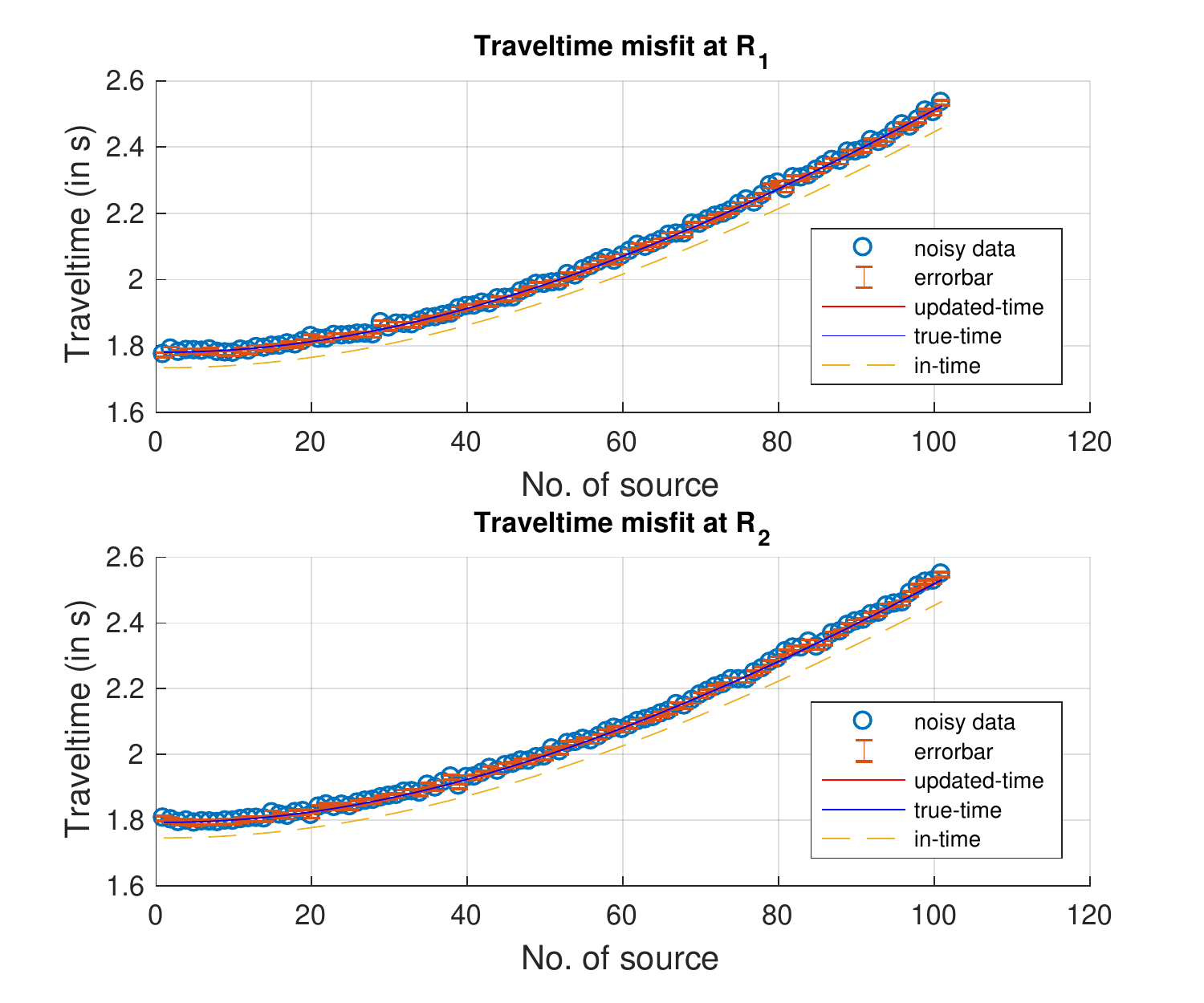}
                \caption{}
                \label{tt_a_30}
        \end{subfigure}       
        \begin{subfigure}[b]{.45\textwidth}
                \includegraphics[width=\textwidth]{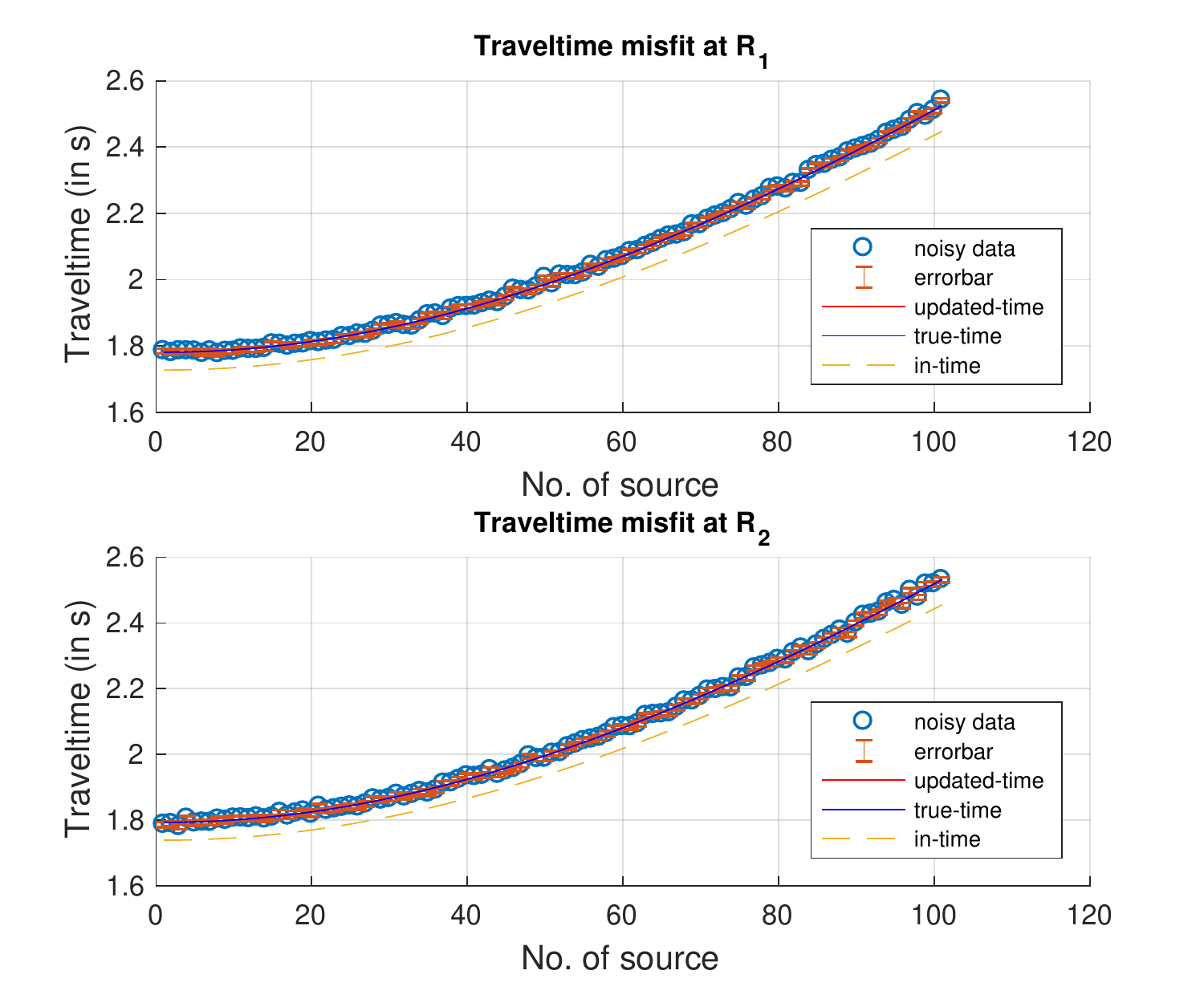}
                \caption{}
                \label{tt_a_30_b_005}
        \end{subfigure}
        \begin{subfigure}[b]{.45\textwidth}
                \includegraphics[width=\textwidth]{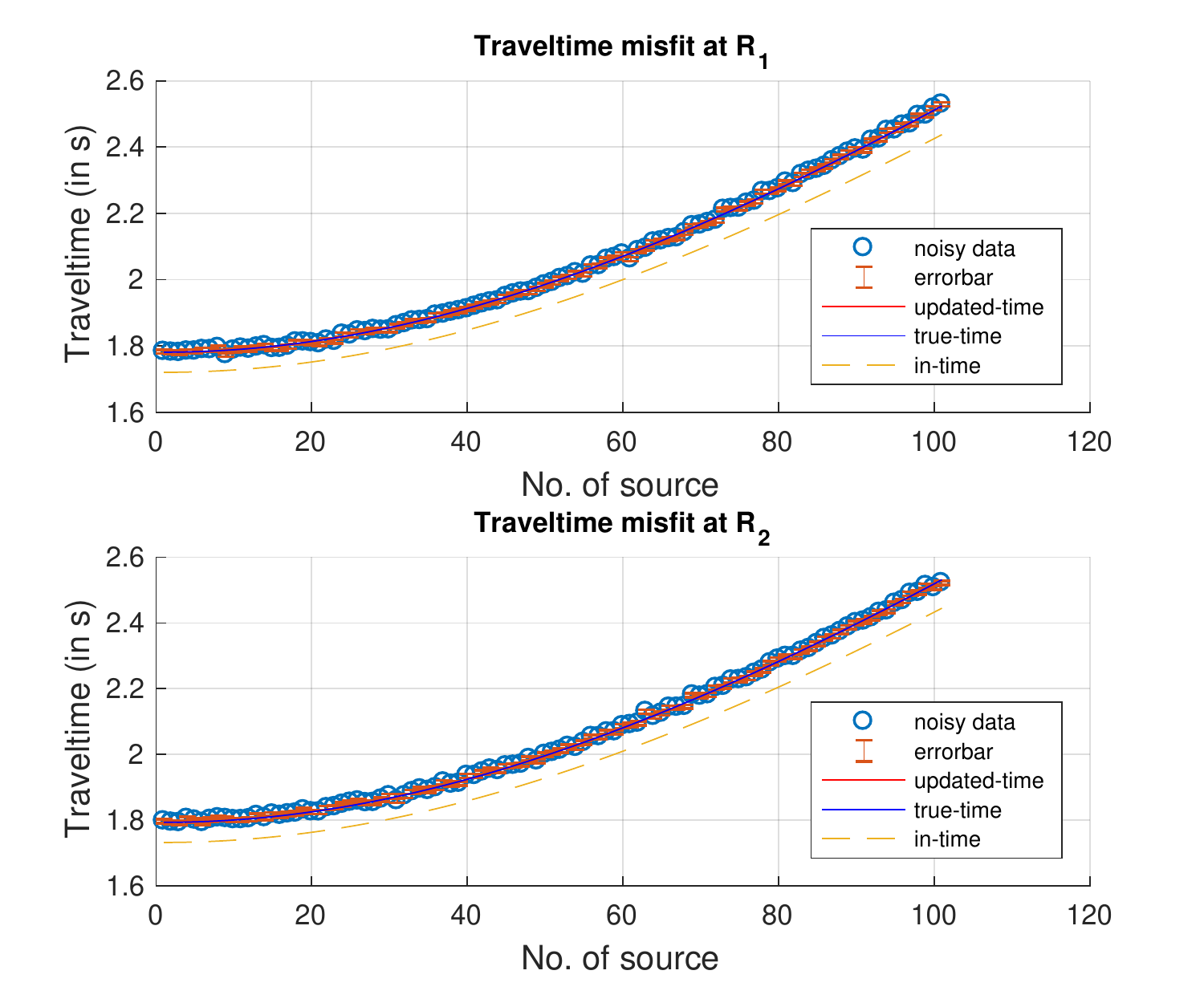}
                \caption{}
                \label{tt_a_30_b_01}
        \end{subfigure}
        \caption{Traveltime inversion for different velocity gradients, $v_{in} = v_{true}\pm30  \, (ms^{-1})$.}
        \label{fig:syn_a_30_b_tt} 
\end{figure}

\begin{figure}[ht]
       \captionsetup[subfigure]{aboveskip=-.1cm}
        \centering
        \begin{subfigure}[b]{.45\textwidth}
                \includegraphics[width=\textwidth]{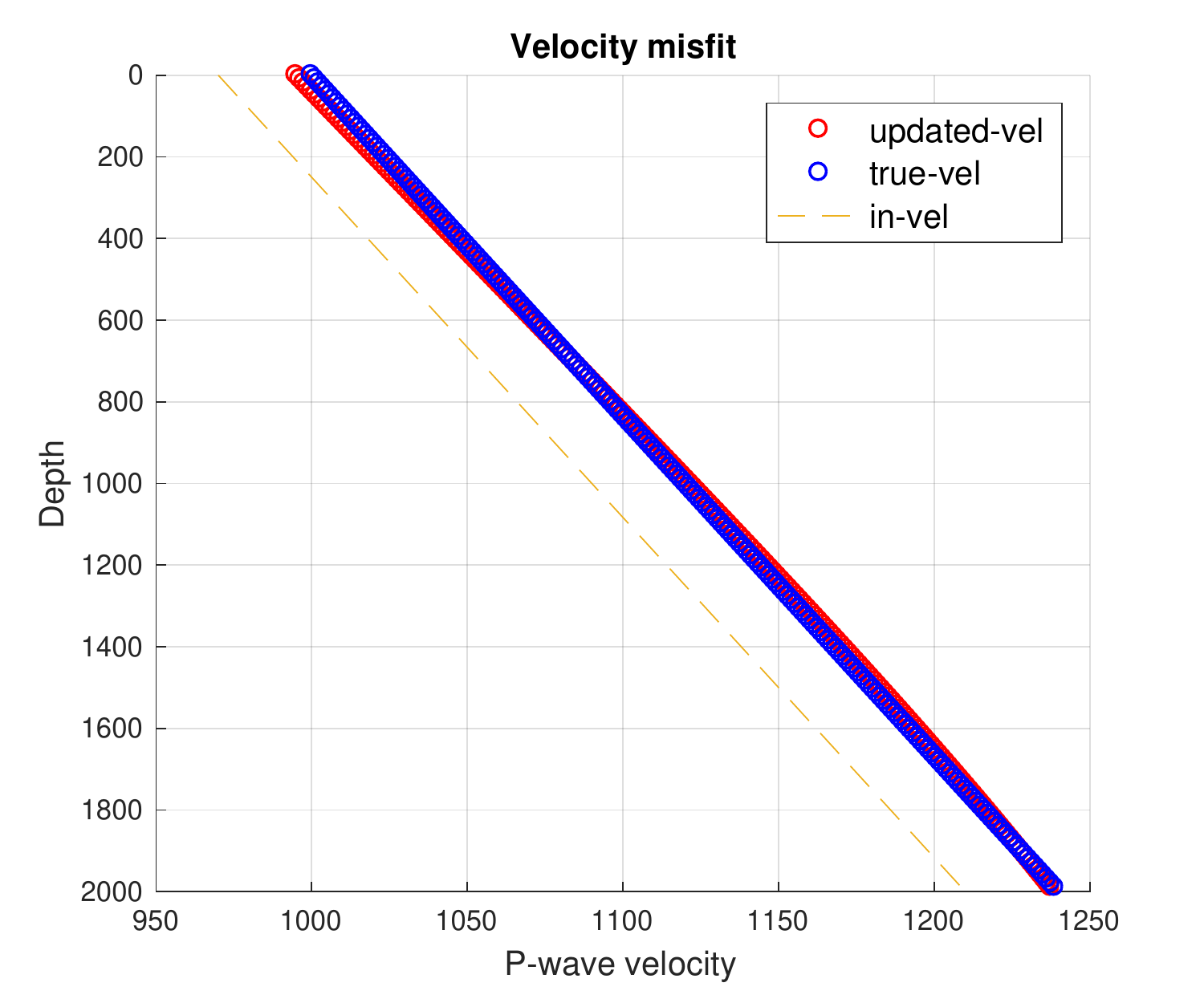}
                \caption{}
                \label{vel_a_m30}
        \end{subfigure}       
        \begin{subfigure}[b]{.45\textwidth}
                \includegraphics[width=\textwidth]{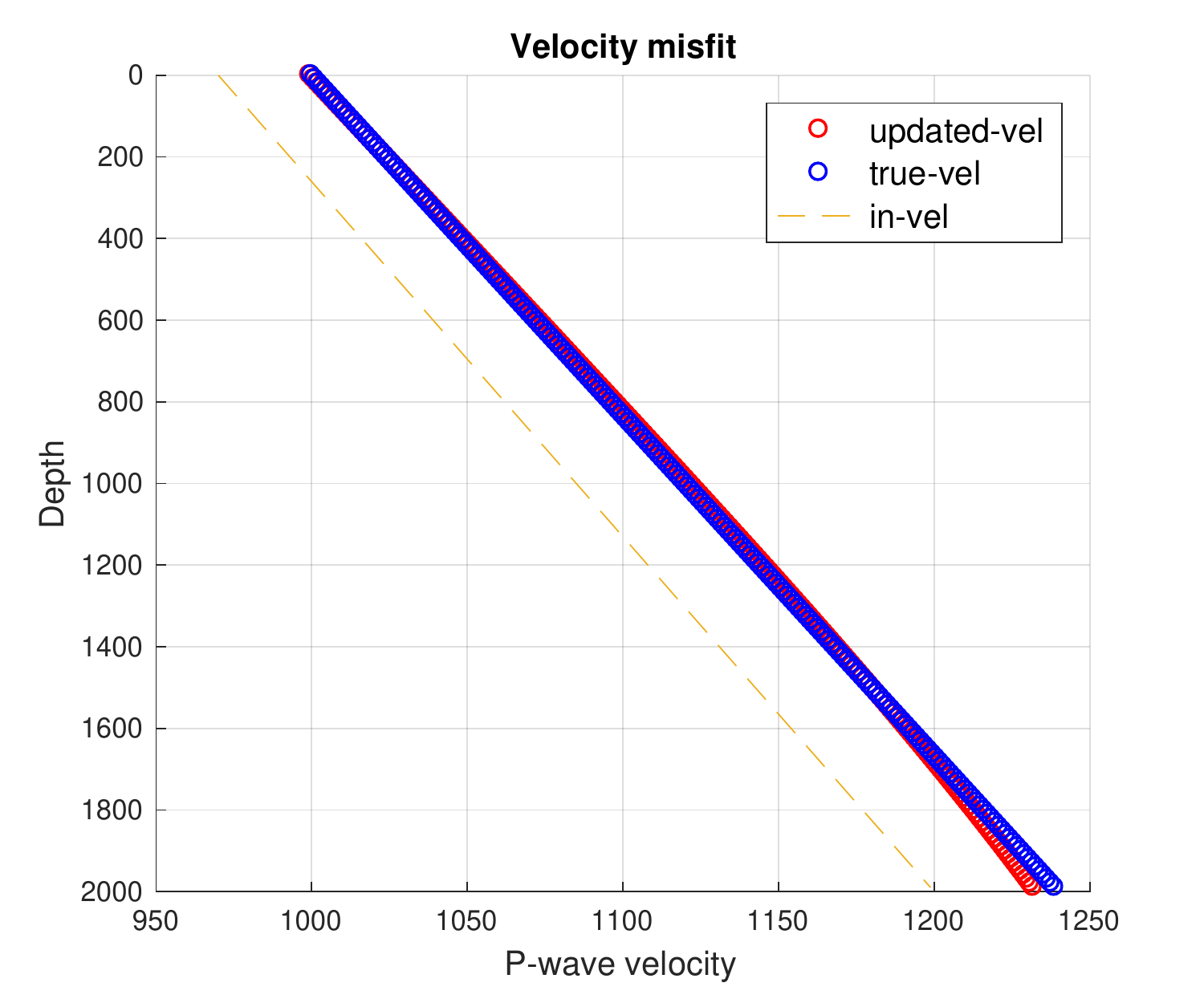}
                \caption{}
                \label{vel_a_m30_b_m005}
        \end{subfigure}
        \begin{subfigure}[b]{.45\textwidth}
                \includegraphics[width=\textwidth]{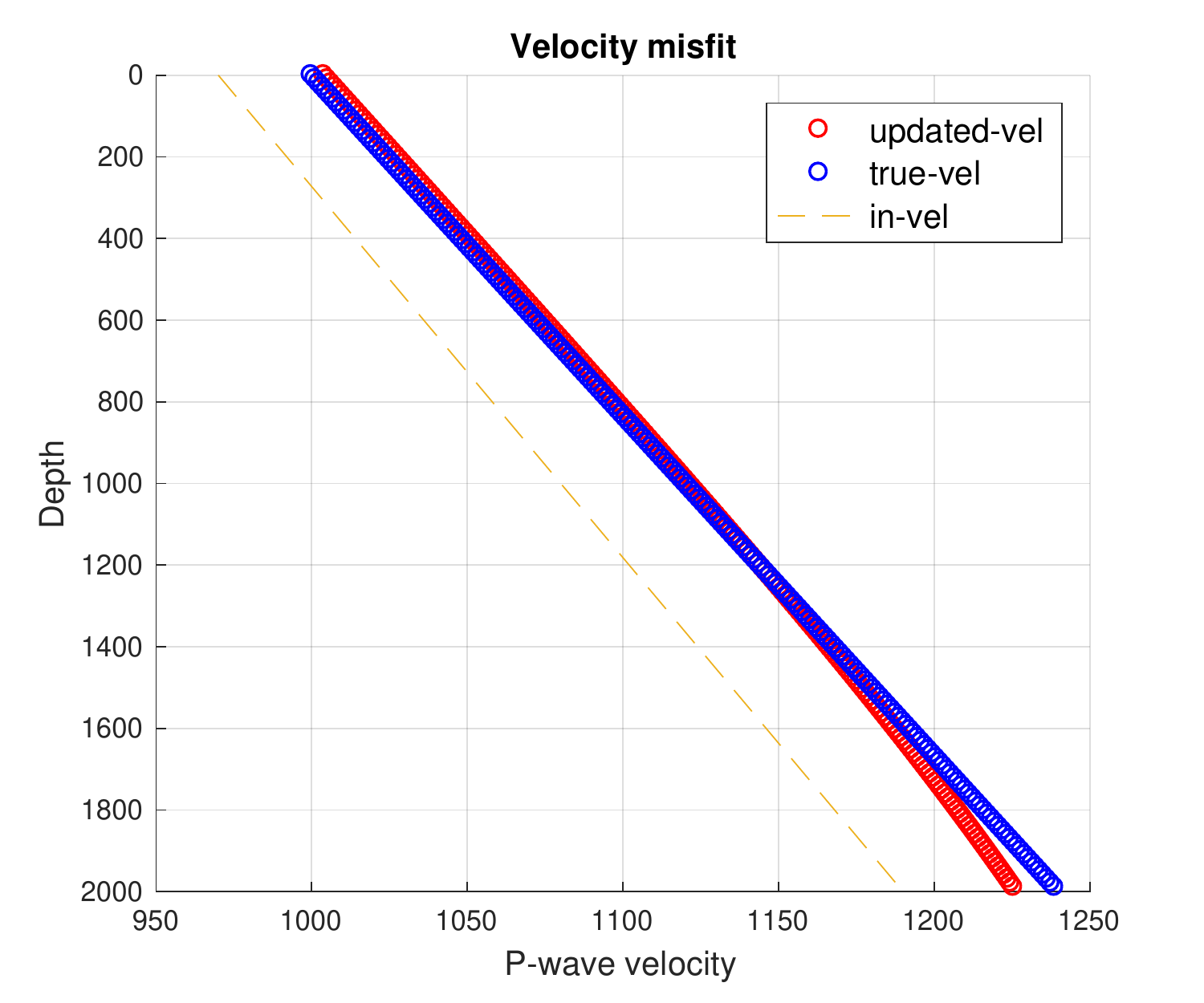}
                \caption{}
                \label{vel_a_m30_b_m01}
        \end{subfigure}
                \begin{subfigure}[b]{.45\textwidth}
                \includegraphics[width=\textwidth]{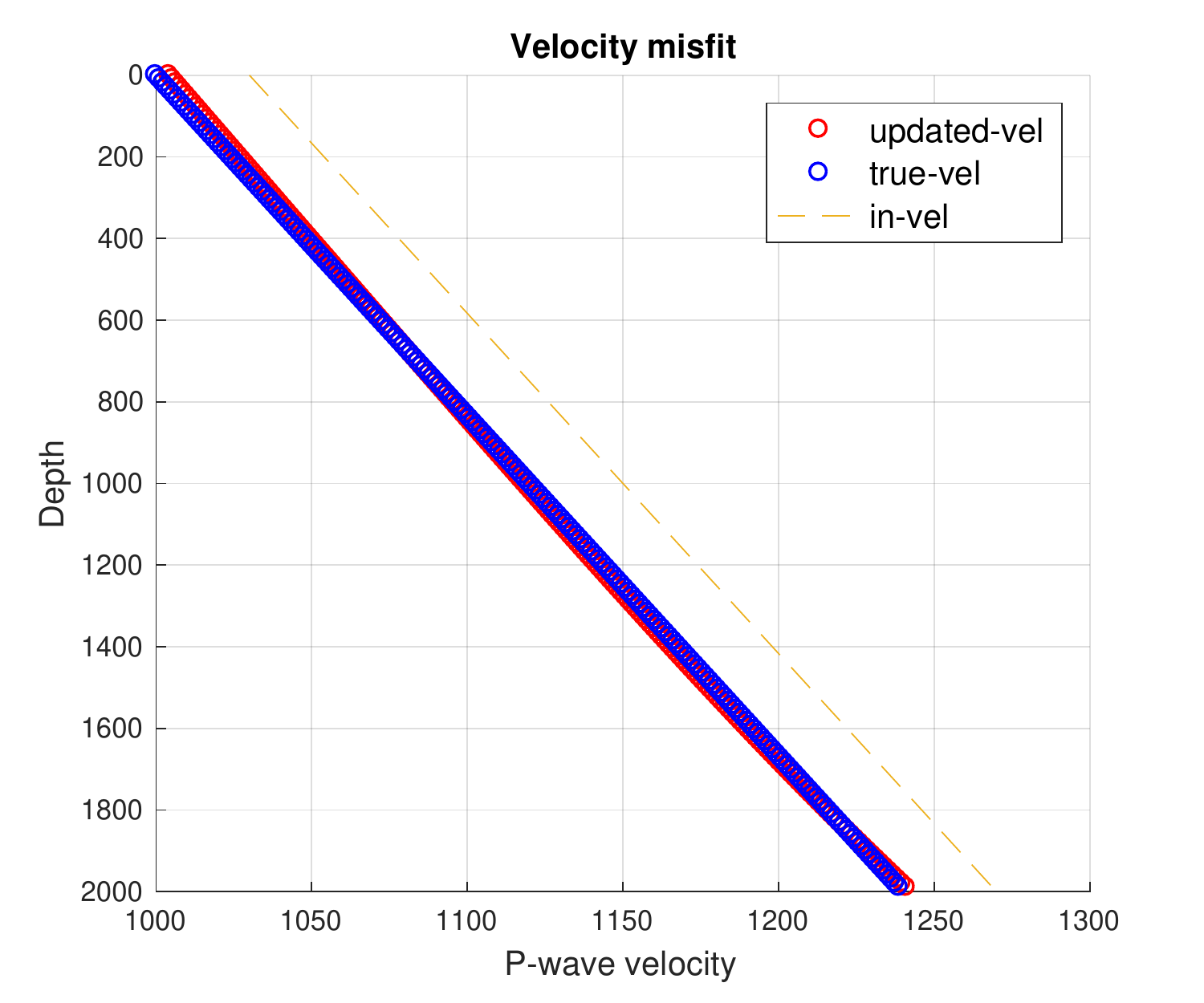}
                \caption{}
                \label{vel_a_30}
        \end{subfigure}       
        \begin{subfigure}[b]{.45\textwidth}
                \includegraphics[width=\textwidth]{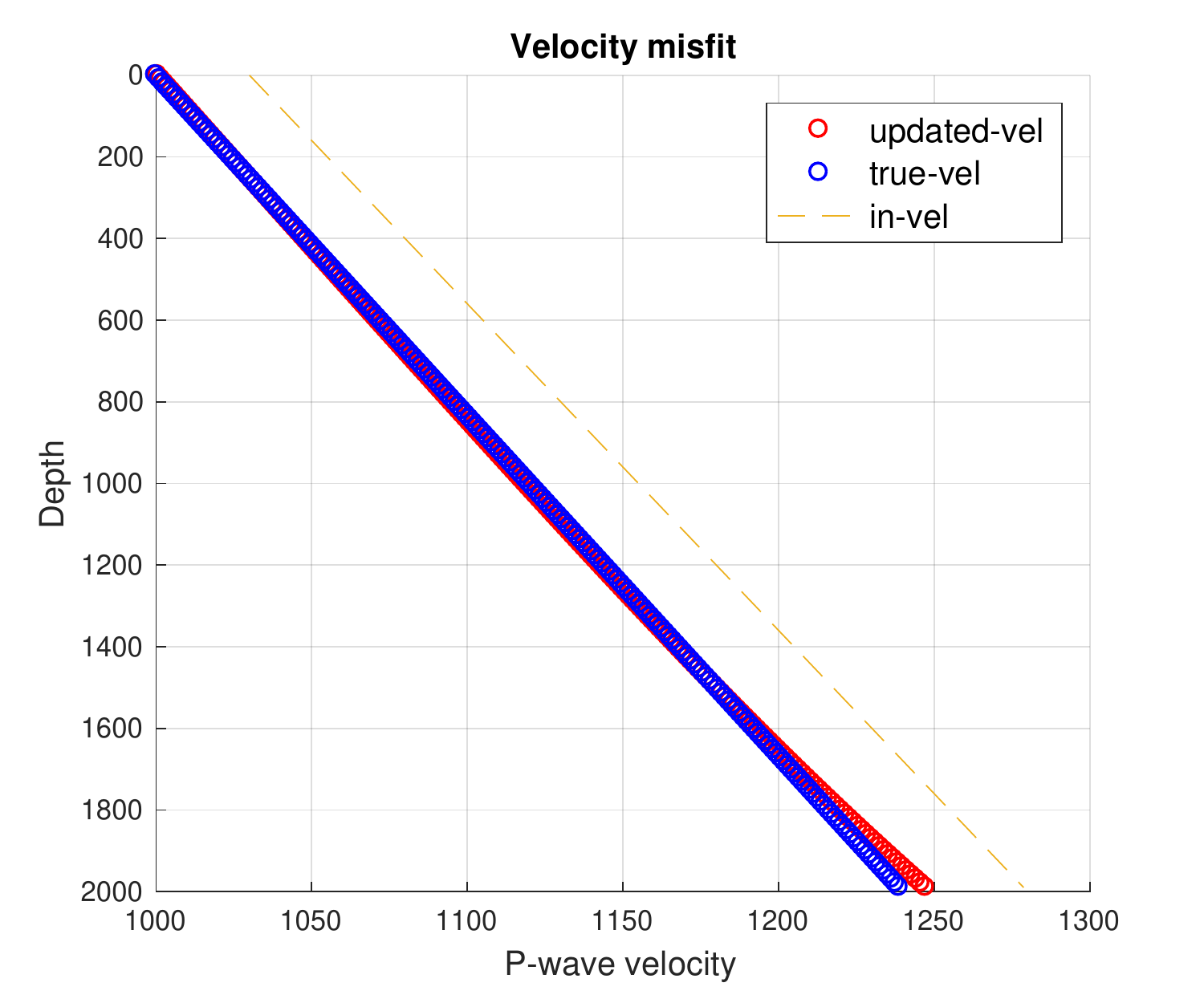}
                \caption{}
                \label{vel_a_30_b_005}
        \end{subfigure}
        \begin{subfigure}[b]{.45\textwidth}
                \includegraphics[width=\textwidth]{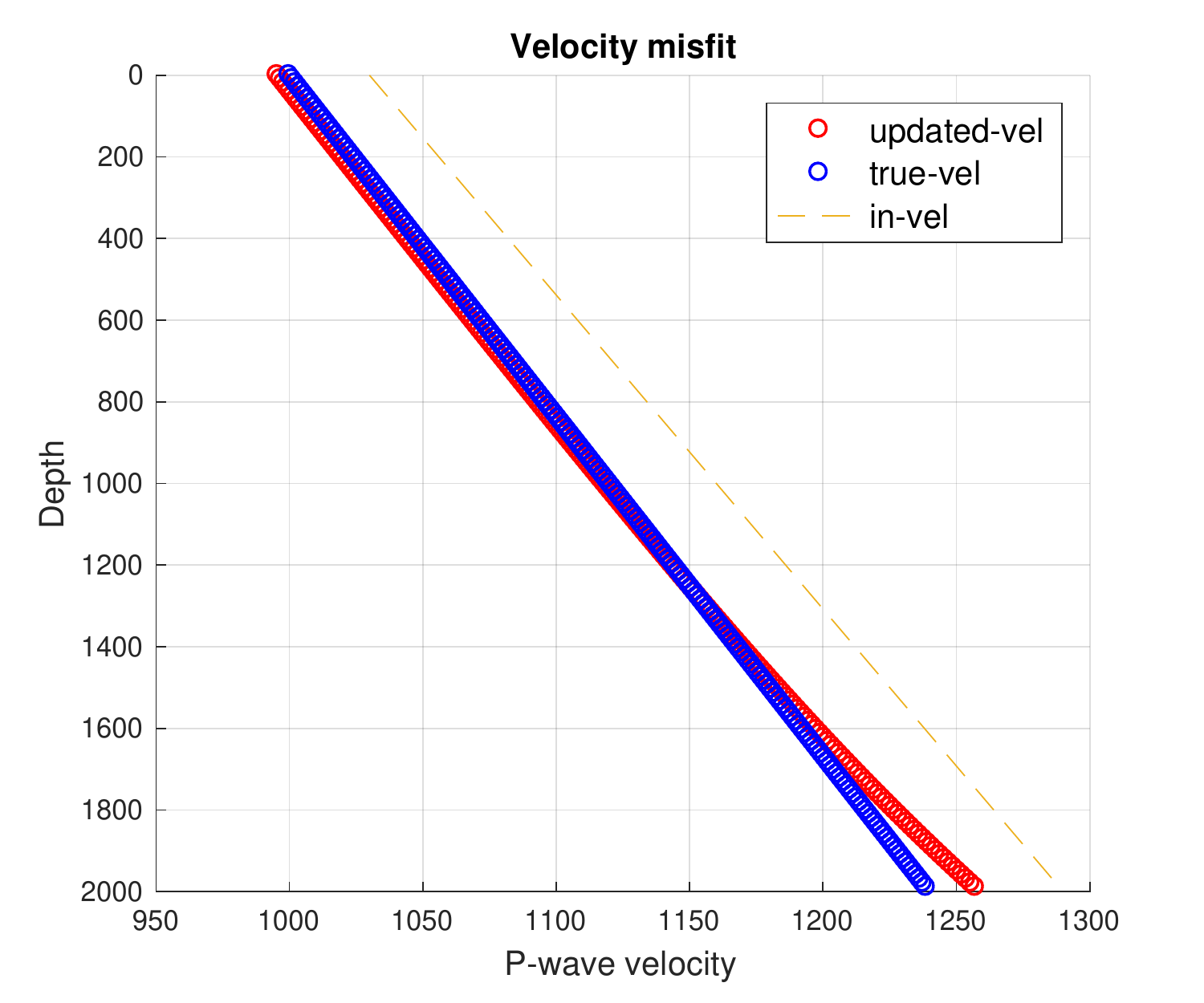}
                \caption{}
                \label{vel_a_30_b_01}
        \end{subfigure}
        \caption{Velocity model for different velocity gradients, $v_{in} = v_{true}\pm30  \, (ms^{-1})$.}
        \label{fig:syn_a_30_b_vel} 
\end{figure}

\begin{figure}[ht]
        \captionsetup[subfigure]{aboveskip=-.1cm}
        \centering
        \begin{subfigure}[b]{.45\textwidth}
                \includegraphics[width=\textwidth]{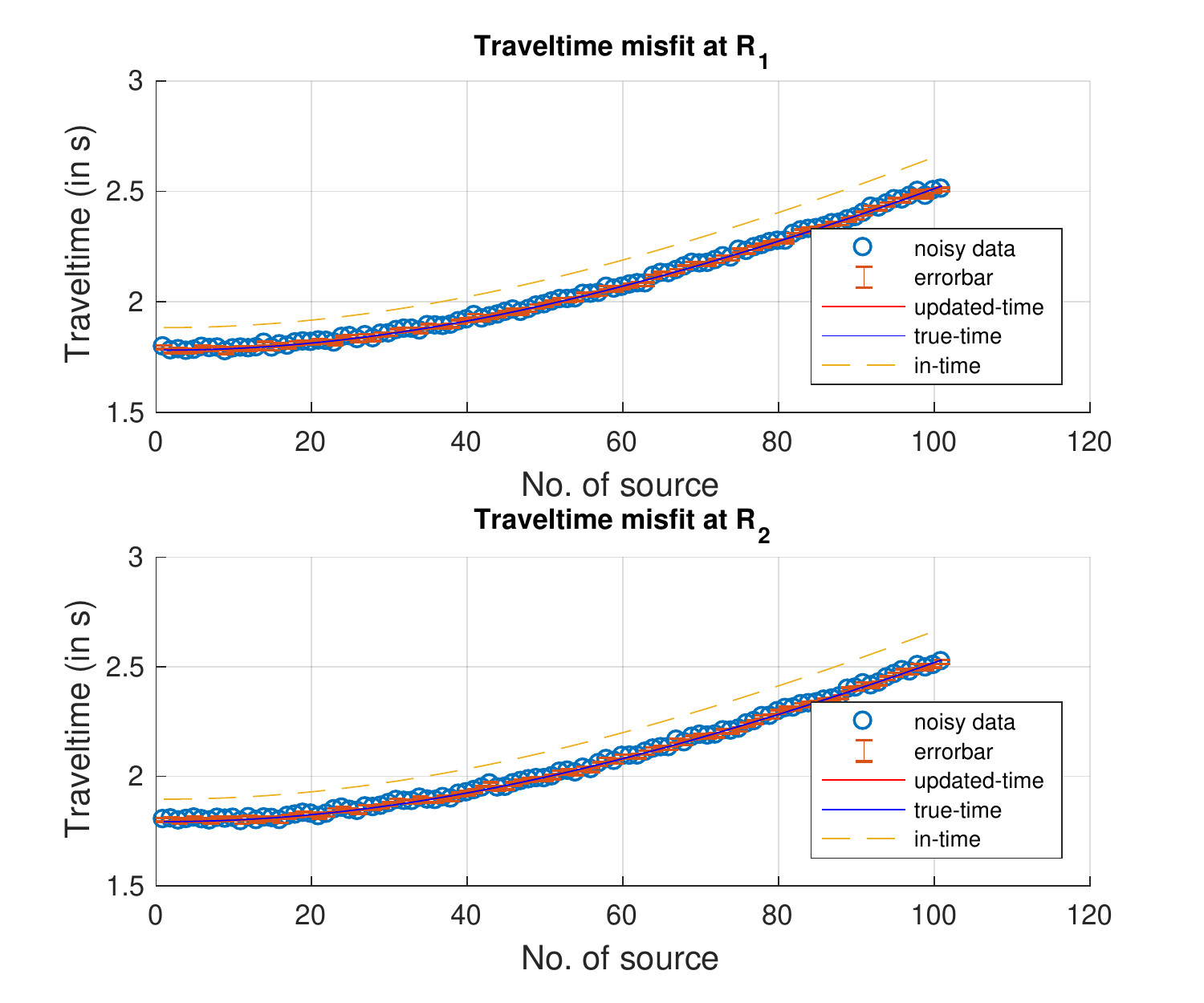}
                \caption{}
                \label{tt_a_m60}
        \end{subfigure}       
        \begin{subfigure}[b]{.45\textwidth}
                \includegraphics[width=\textwidth]{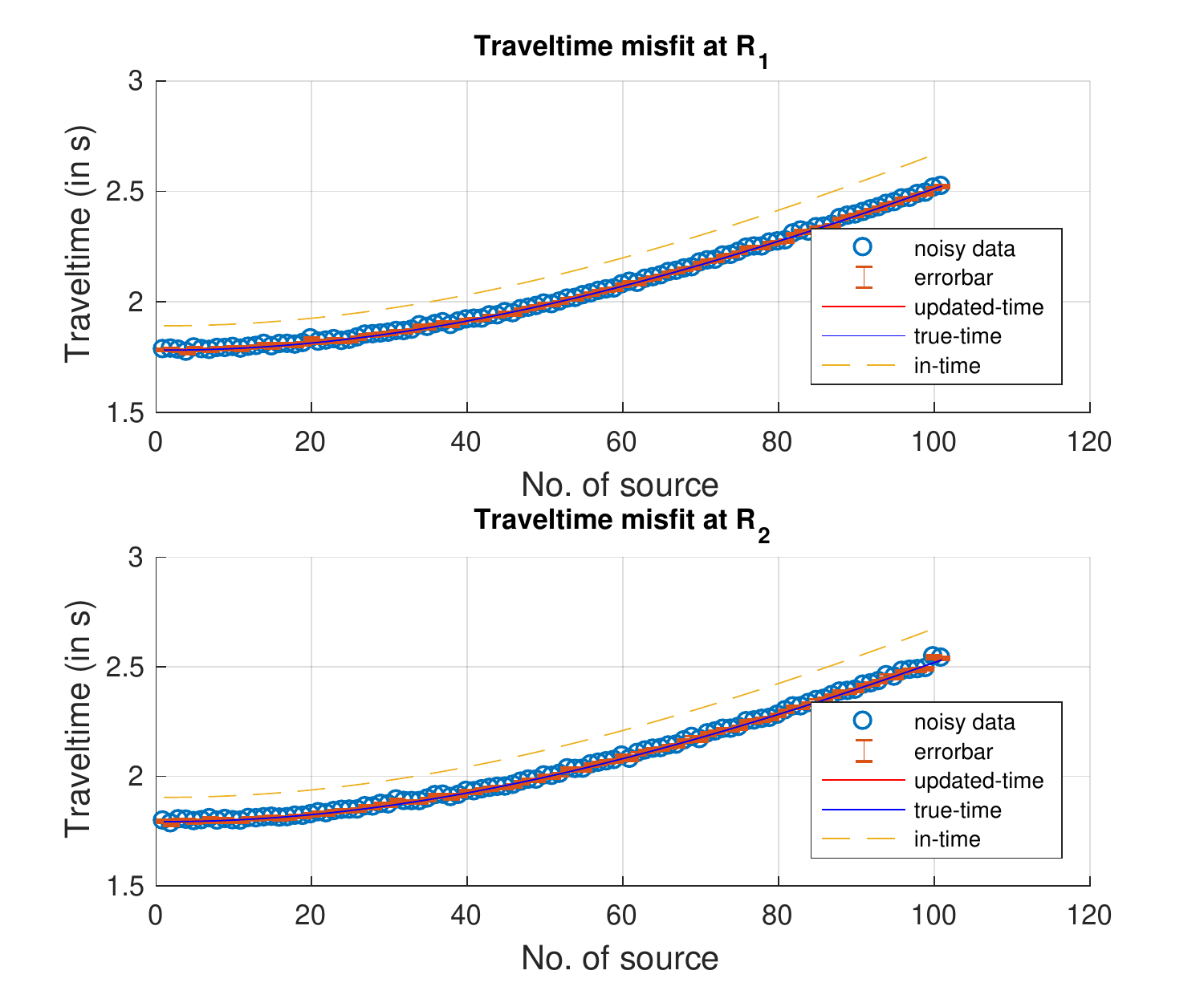}
                \caption{}
                \label{tt_a_m60_b_m005}
        \end{subfigure}
        \begin{subfigure}[b]{.45\textwidth}
                \includegraphics[width=\textwidth]{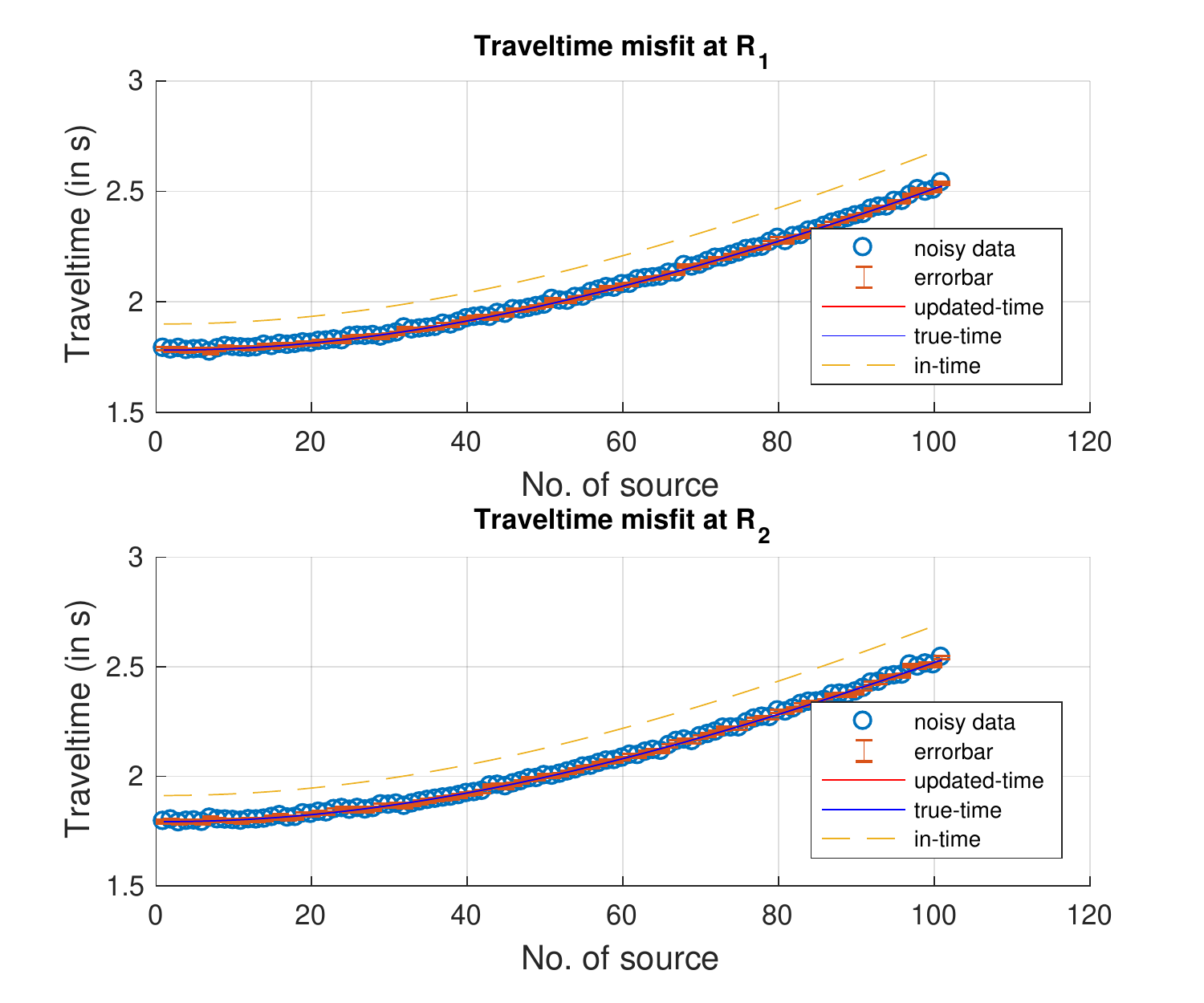}
                \caption{}
                \label{tt_a_m60_b_m01}
        \end{subfigure}
                \begin{subfigure}[b]{.45\textwidth}
                \includegraphics[width=\textwidth]{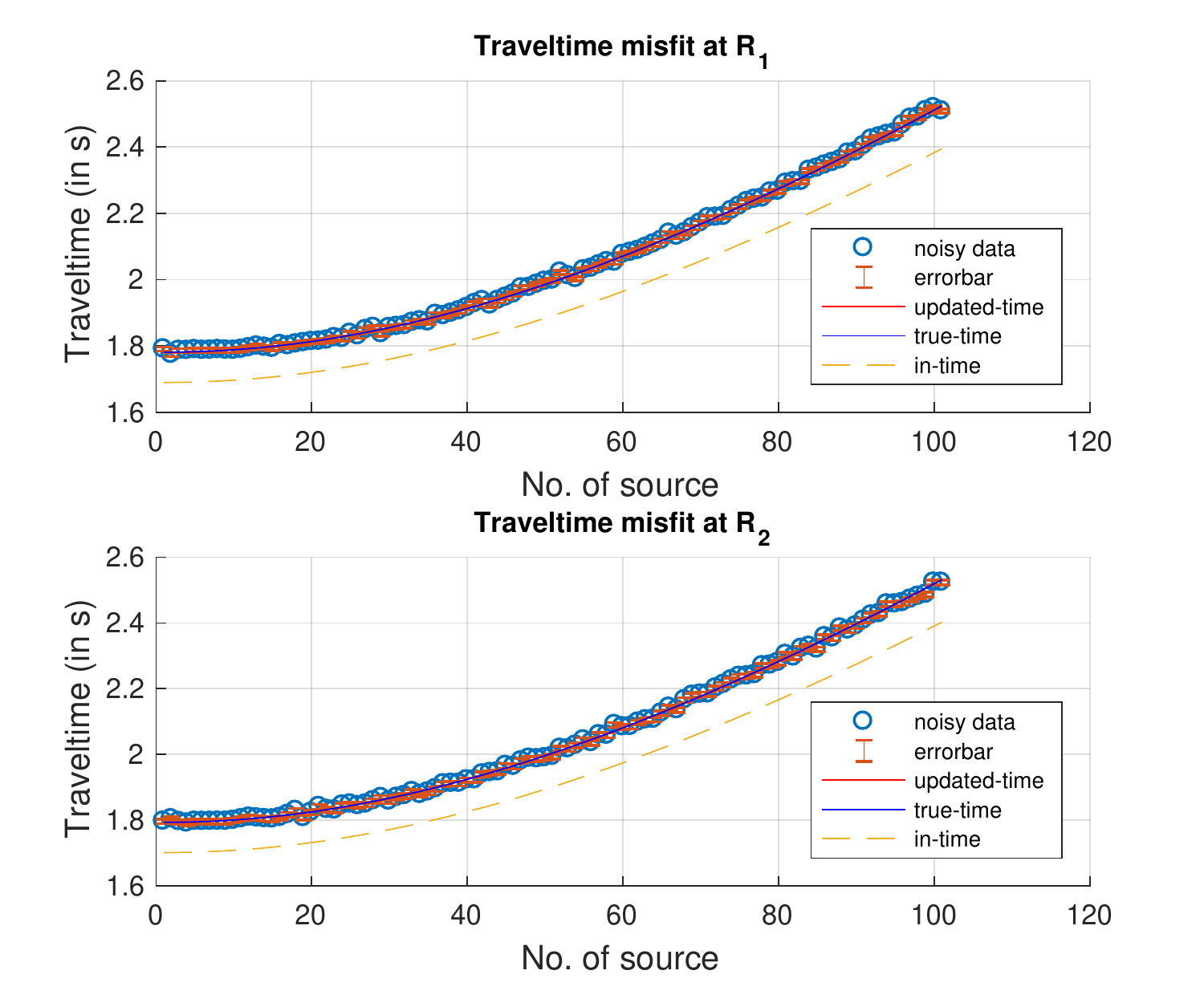}
                \caption{}
                \label{tt_a_60}
        \end{subfigure}       
        \begin{subfigure}[b]{.45\textwidth}
                \includegraphics[width=\textwidth]{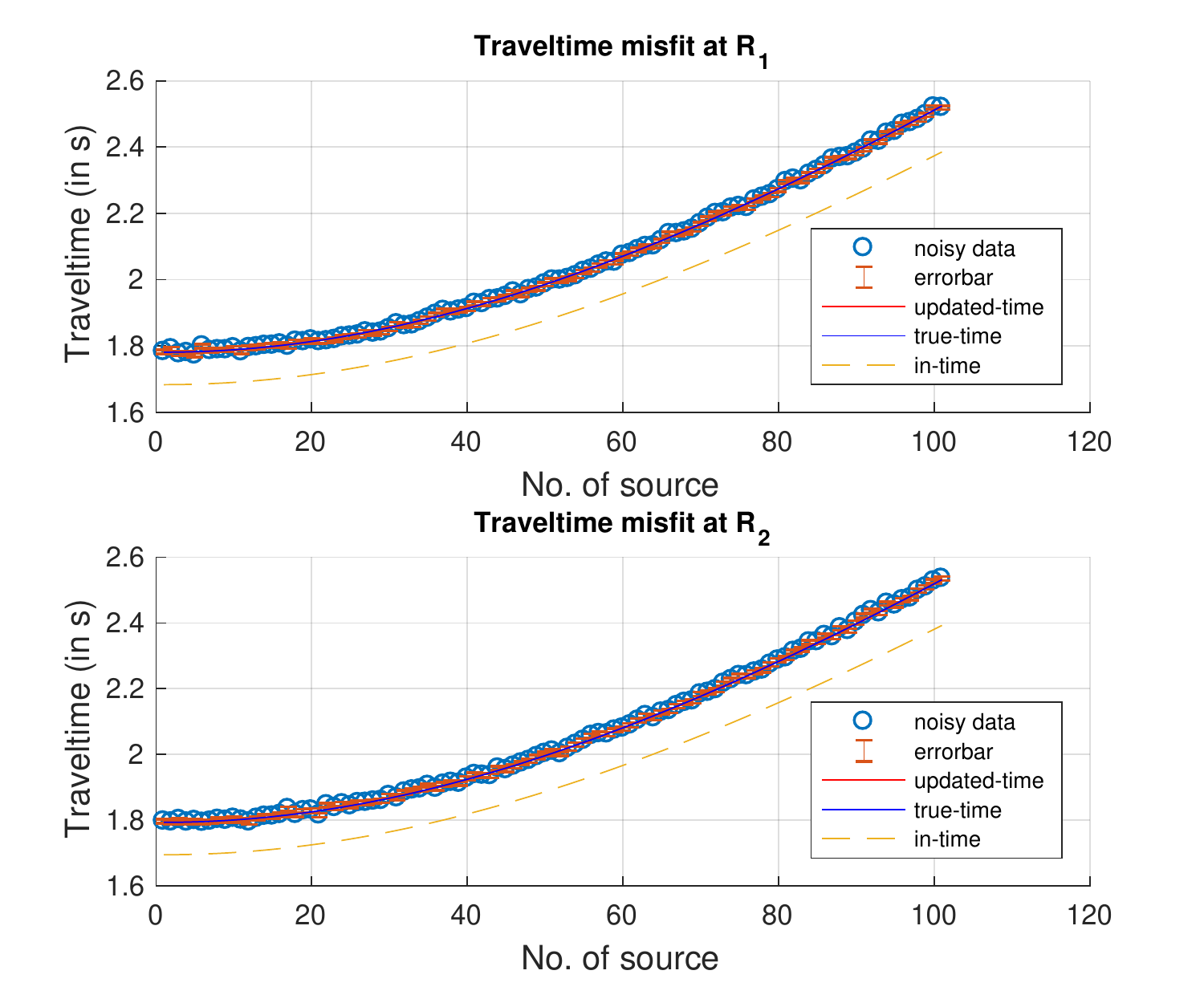}
                \caption{}
                \label{tt_a_60_b_005}
        \end{subfigure}
        \begin{subfigure}[b]{.45\textwidth}
                \includegraphics[width=\textwidth]{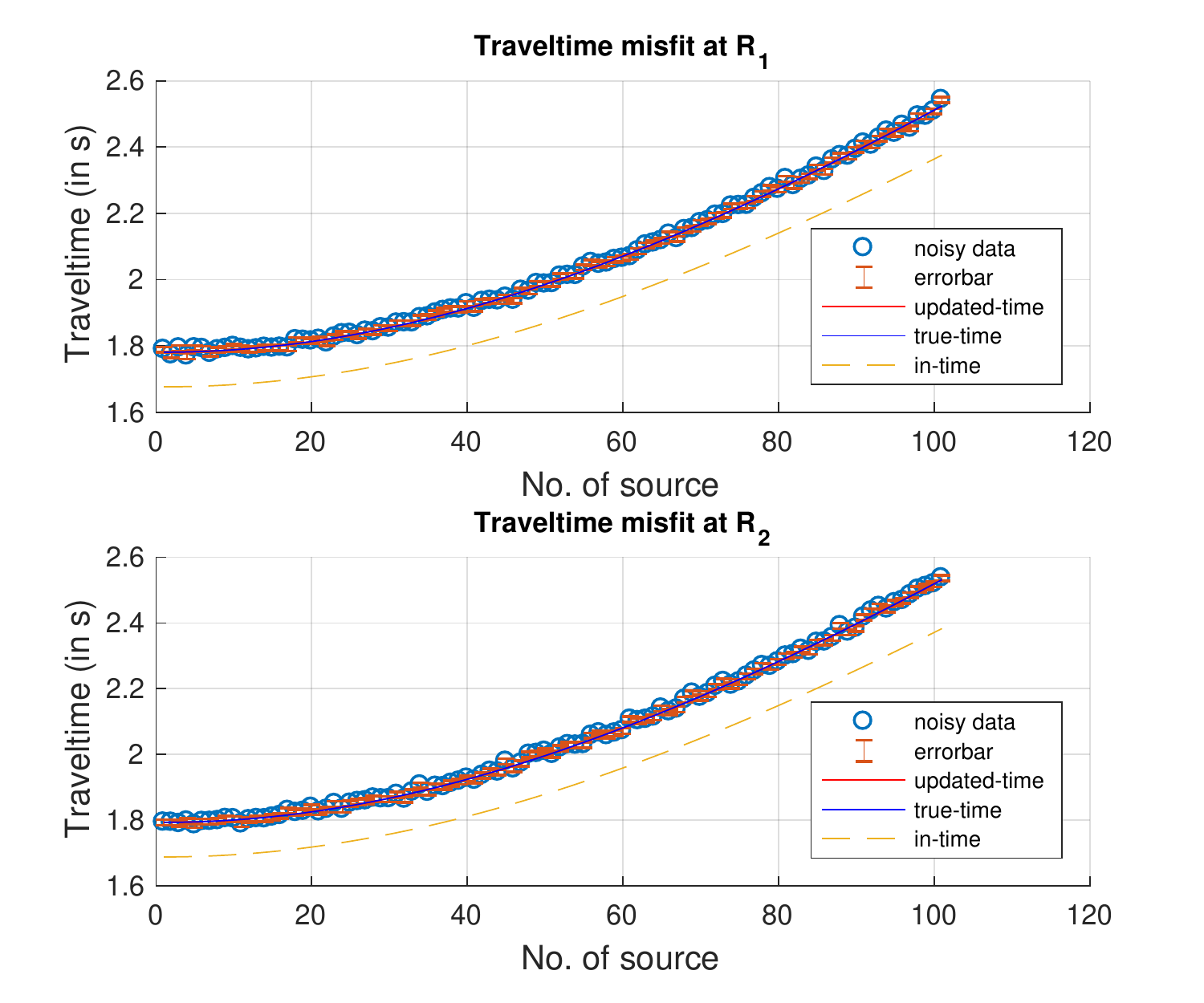}
                \caption{}
                \label{tt_a_60_b_01}
        \end{subfigure}
        \caption{Traveltime inversion for different velocity gradients, $v_{in} = v_{true}\pm60  \, (ms^{-1})$.}
        \label{fig:syn_a_60_b_tt} 
\end{figure}

\begin{figure}[ht]
        \captionsetup[subfigure]{aboveskip=-.1cm}
        \centering
        \begin{subfigure}[b]{.45\textwidth}
                \includegraphics[width=\textwidth]{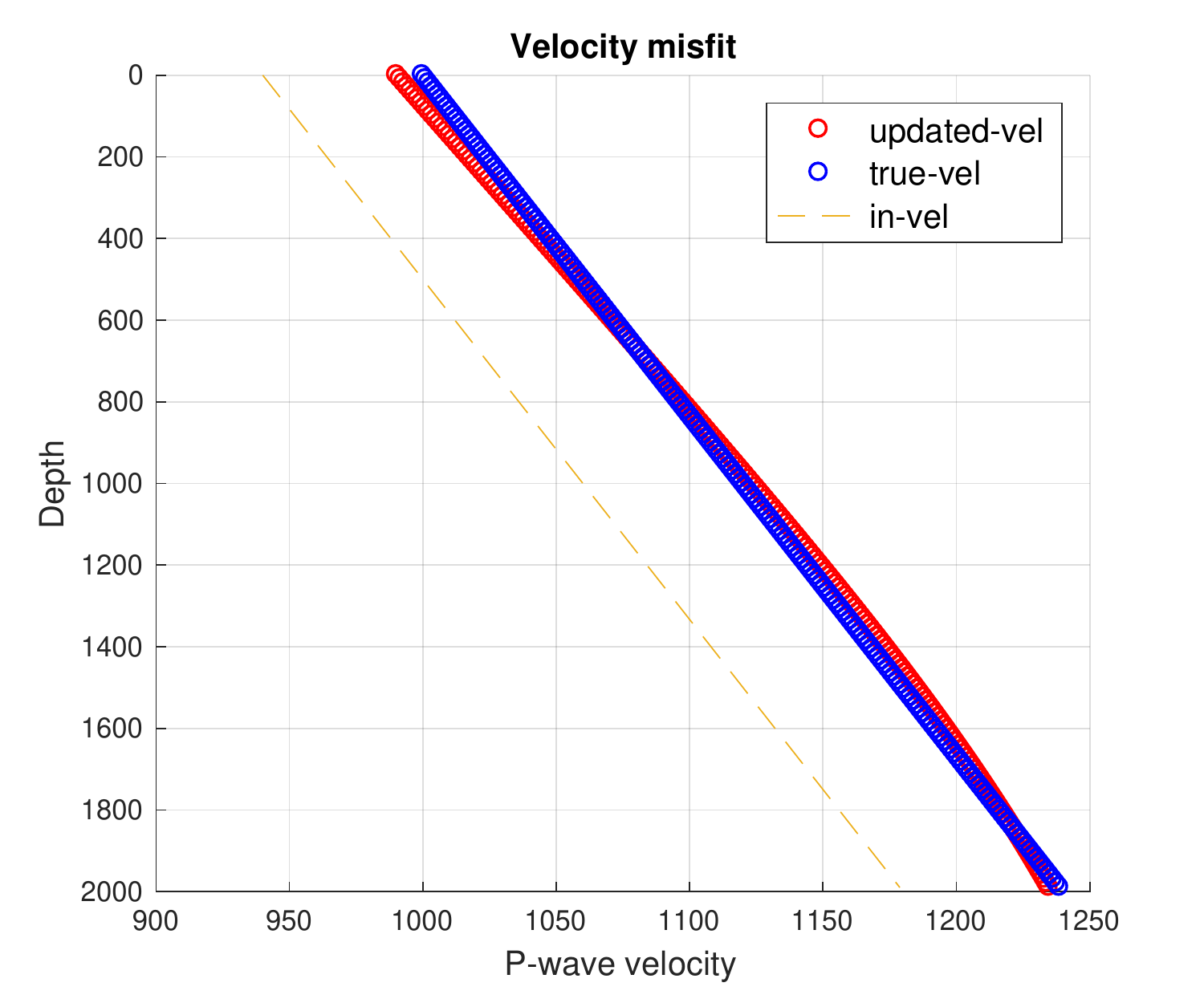}
                \caption{}
                \label{vel_a_m60}
        \end{subfigure}       
        \begin{subfigure}[b]{.45\textwidth}
                \includegraphics[width=\textwidth]{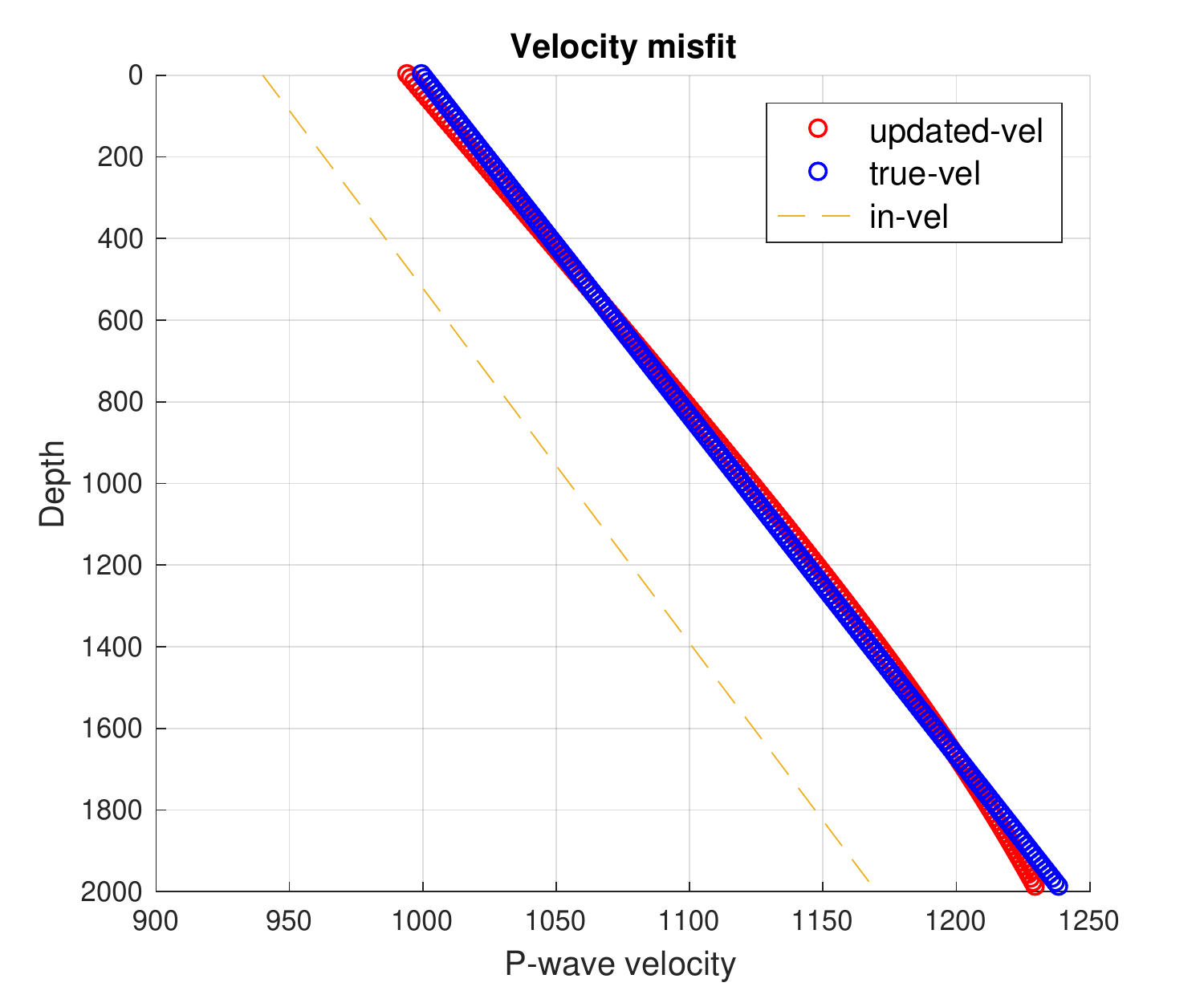}
                \caption{}
                \label{vel_a_m60_b_m005}
        \end{subfigure}
        \begin{subfigure}[b]{.45\textwidth}
                \includegraphics[width=\textwidth]{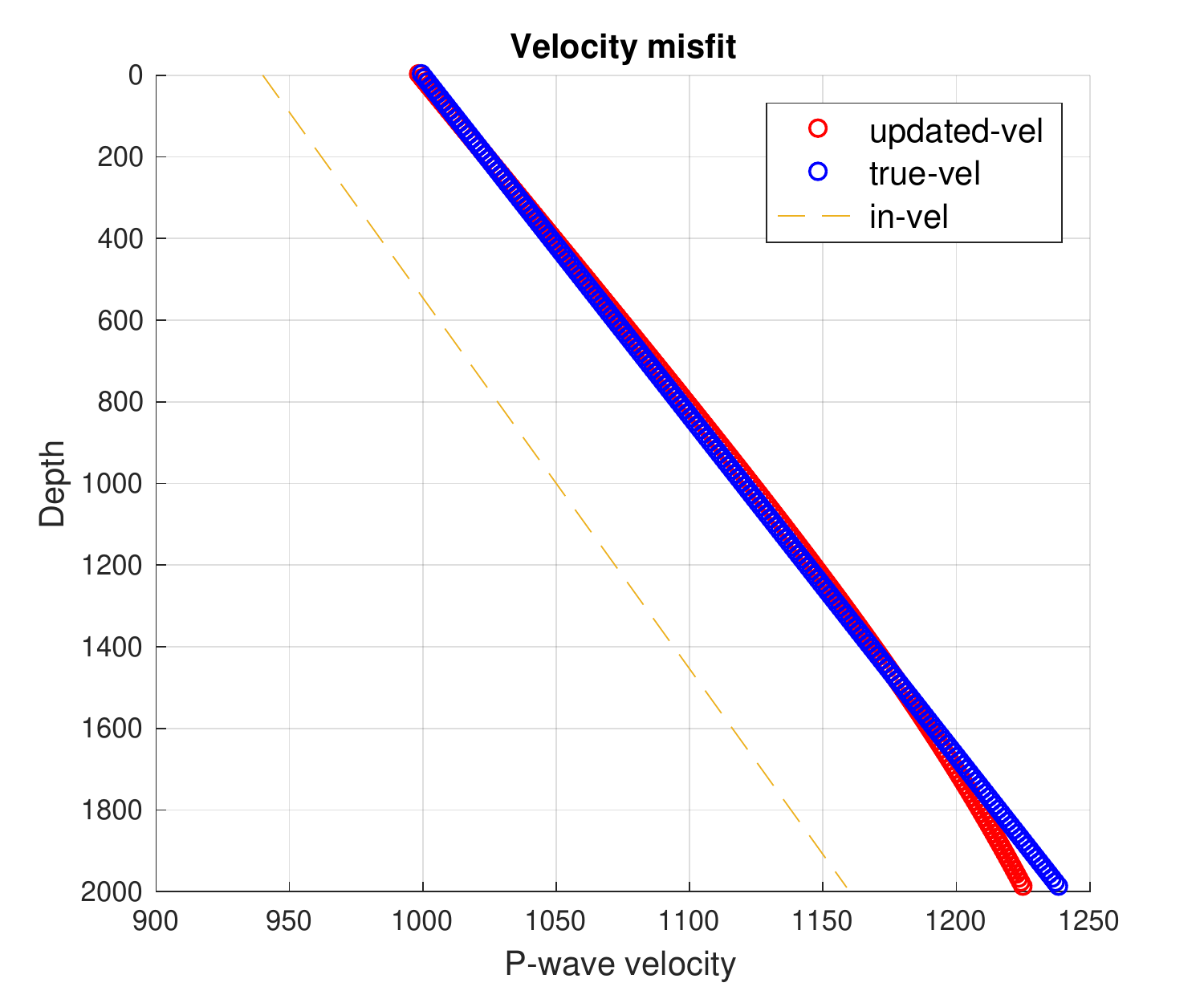}
                \caption{}
                \label{vel_a_m60_b_m01}
        \end{subfigure}
                \begin{subfigure}[b]{.45\textwidth}
                \includegraphics[width=\textwidth]{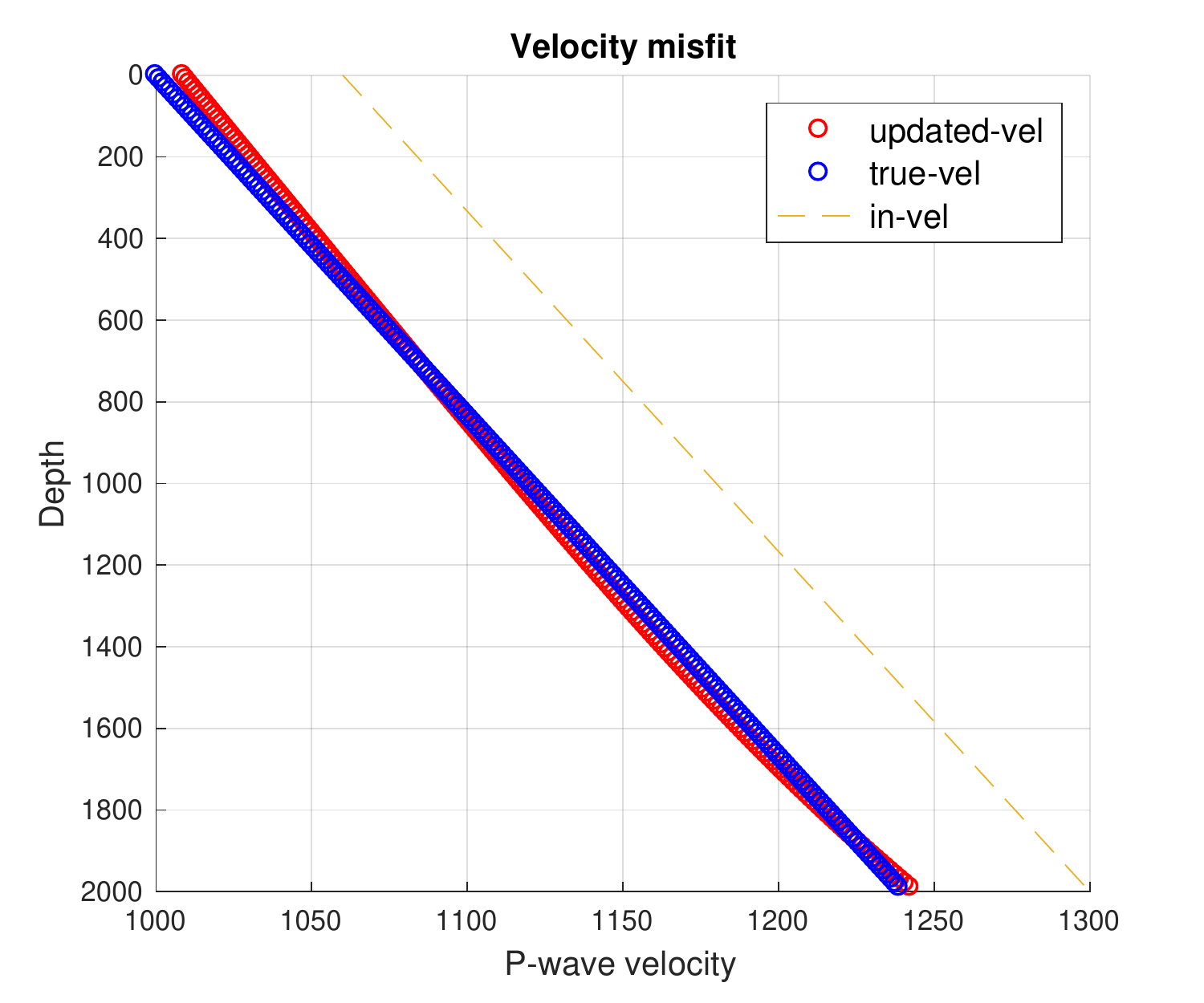}
                \caption{}
                \label{vel_a_60}
        \end{subfigure}       
        \begin{subfigure}[b]{.45\textwidth}
                \includegraphics[width=\textwidth]{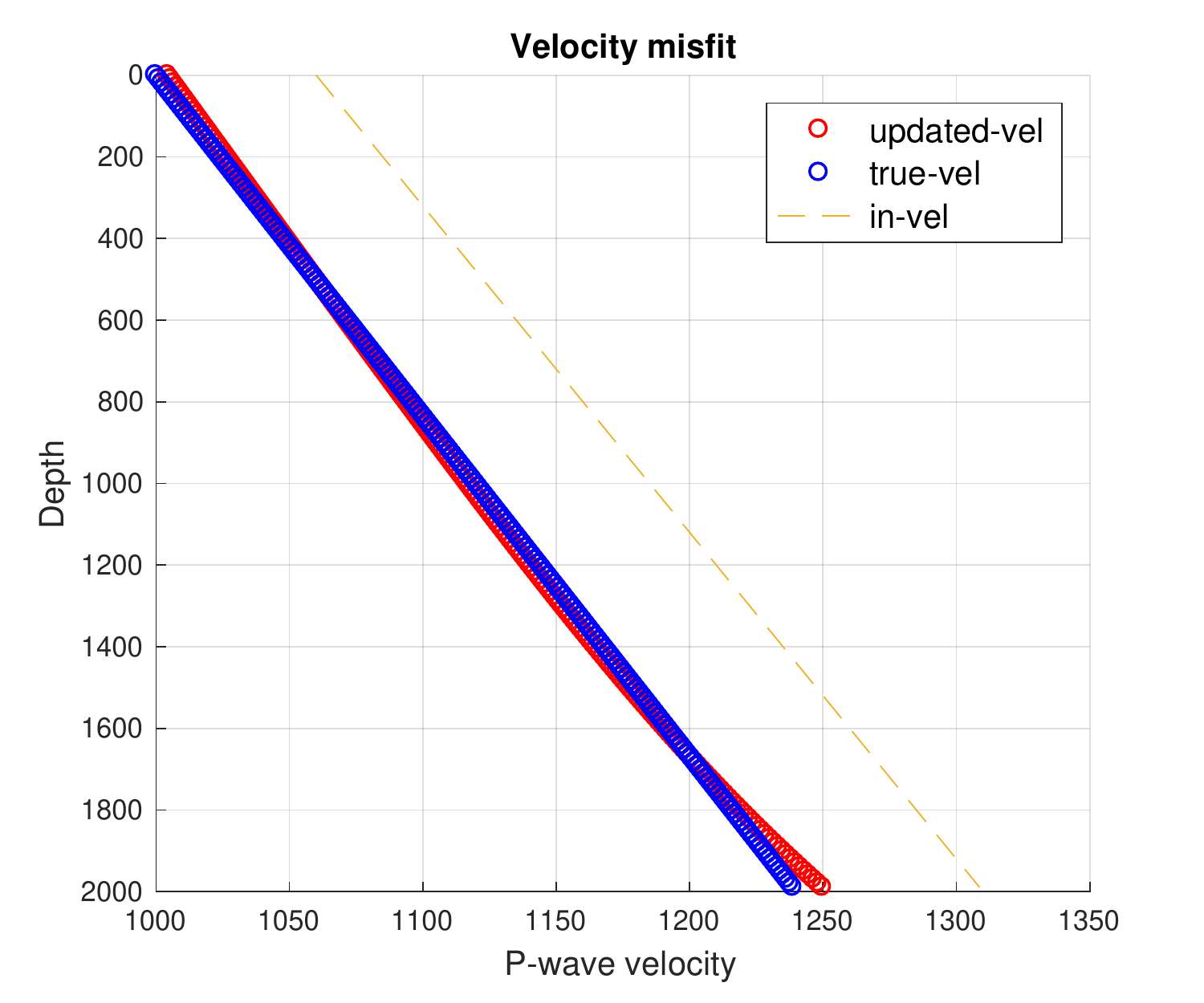}
                \caption{}
                \label{vel_a_60_b_005}
        \end{subfigure}
        \begin{subfigure}[b]{.45\textwidth}
                \includegraphics[width=\textwidth]{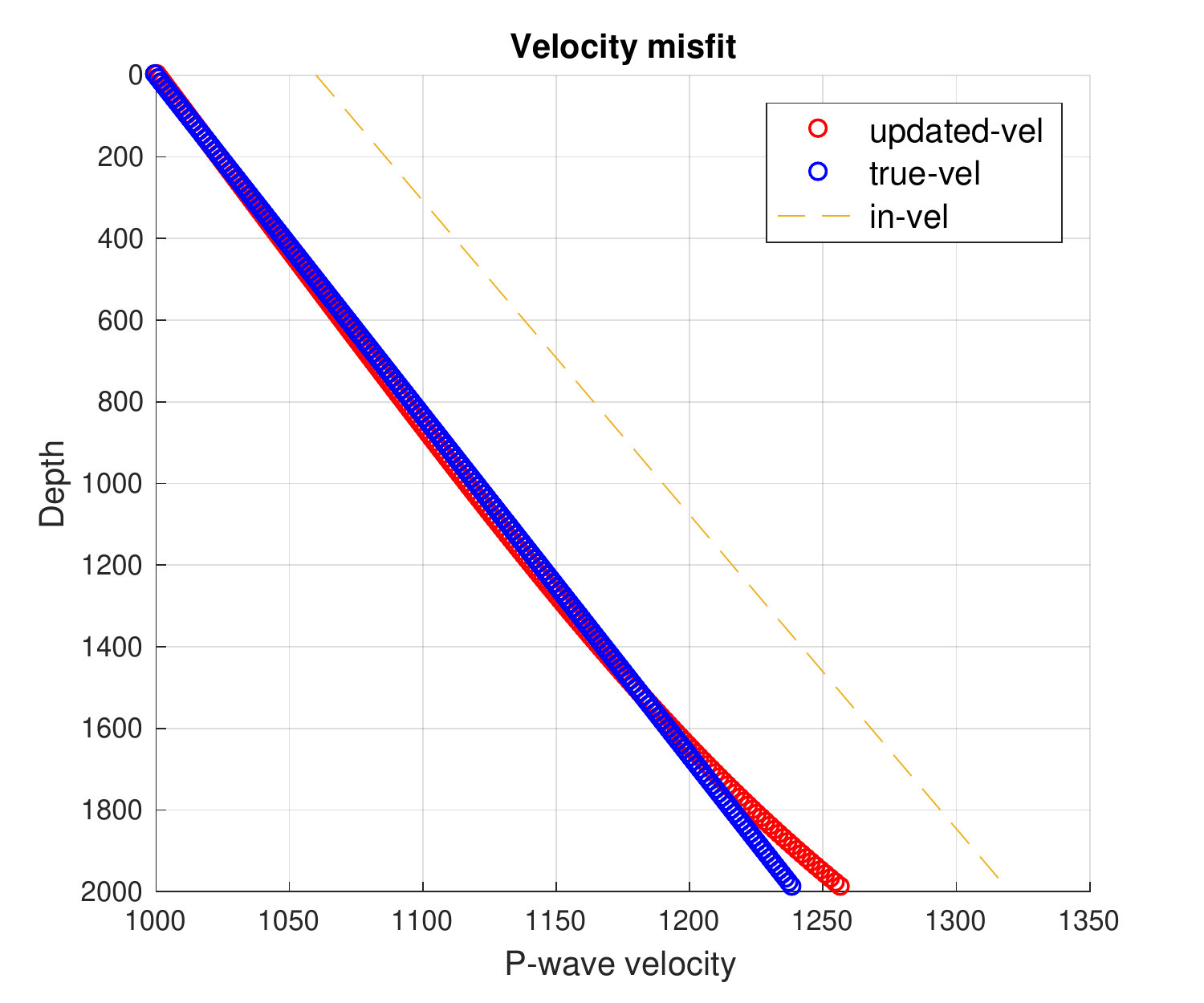}
                \caption{}
                \label{vel_a_60_b_01}
        \end{subfigure}
        \caption{Velocity model for different velocity gradients, $v_{in} = v_{true}\pm60  \, (ms^{-1})$.}
        \label{fig:syn_a_60_b_vel} 
\end{figure}

The traveltime convergence results are shown in Figures~\ref{fig:syn_a_30_b_tt} and \ref{fig:syn_a_60_b_tt}.
The velocity misfits are shown in Figures~\ref{fig:syn_a_30_b_vel} and \ref{fig:syn_a_60_b_vel}.

\paragraph{}

\begin{table}[h]
\centering
\begin{tabular}{c*{6}{c}}
\toprule
Test &
$a_{in}$ &
$b_{in}$ &
$a_{inv}$ &
$b_{inv}$ &
$f({\bf M})$ &
Figure
\\[2pt]
\toprule
1 & 970 & 0.1200 & 997.77 & 0.1225 & 201.07 & \ref{tt_a_m30},\ref{vel_a_m30}\\
2 & 970 & 0.1150 & 1002.05 & 0.1178 & 200.53 & \ref{tt_a_m30_b_m005},\ref{vel_a_m30_b_m005} \\
3 & 970 & 0.1100 & 1006.44 & 0.1130 & 201.56 & \ref{tt_a_m30_b_m01},\ref{vel_a_m30_b_m01}\\
4 & 1030 & 0.1200 & 1002.11 & 0.1178 & 199.09 & \ref{tt_a_30},\ref{vel_a_30}\\
5 & 1030 & 0.1250 & 998.19 & 0.1220 & 199.71 & \ref{tt_a_30_b_005},\ref{vel_a_30_b_005}\\
6 & 1030 & 0.1300 & 993.00 & 0.1275 & 198.04 & \ref{tt_a_30_b_01},\ref{vel_a_30_b_01}\\
\cmidrule{1-7}
7 & 940 & 0.1200 & 994.58 & 0.1257 & 201.30 & \ref{tt_a_m60},\ref{vel_a_m60}\\
8 & 940 & 0.1150 & 999.20 & 0.1209 & 196.79 & \ref{tt_a_m60_b_m005},\ref{vel_a_m60_b_m005}\\
9 & 940 & 0.1100 & 1003.75 & 0.1160 & 201.69 & \ref{tt_a_m60_b_m01},\ref{vel_a_m60_b_m01}\\
10 & 1060 & 0.1200 & 1005.02 & 0.1145 & 201.03 & \ref{tt_a_60},\ref{vel_a_60}\\
11 & 1060 & 0.1250 & 1000.29 & 0.1199 & 199.32 & \ref{tt_a_60_b_005},\ref{vel_a_60_b_005}\\
12 & 1060 & 0.1300 & 994.62 & 0.1259 & 198.07 & \ref{tt_a_60_b_01},\ref{vel_a_60_b_01}\\
\bottomrule
\end{tabular}
\caption[Varying model parameters]{Model set-up: number of data points $=$ 202, added noise up to 1$\%$. 
test of the first six, $a_{in} = a_{true} \pm 30$ and test of the last six, $a_{in} = a_{true} \pm 60$ (units of $a$ and $b$ are ${\rm ms^{-1}}$ and ${\rm s^{-1}}$)}
\label{tab:syn_b}
\end{table}

Table~\ref{tab:syn_b} shows the inversion results to be more sensitive to the parameter $b$ compared to the parameter $a$.
However, the synthetic experiments show that the inversion method produces the reference velocity consistently within a small range of error. 
If we apply a good startup model and sufficient data points, the synthetic results show that the inversion method can produce a reasonable velocity model of a medium.

\captionsetup[figure]{aboveskip=0cm}
%
%
\captionsetup[figure]{aboveskip=0cm}

\captionsetup[figure]{aboveskip=0cm}
%
%
\captionsetup[figure]{aboveskip=0cm}
\FloatBarrier
\section{1-D tomography : Application in real data}
In this section, we apply the 1-D tomography and two-parameter inversion methods to a field data~(\ref{sec:data_des}).
In the two-parameter inversion, the traveltime expression is used from~\cite{SlawinskiSlawinski1999}. 
We develop the codes for both methods in Matlab and provide the source codes in the appendices~\ref{sec:code_syn}, \ref{sec:code_real}, and \ref{sec:code_ab}.

In Table~\ref{tab:real}, we use the traveltime data from Appendix~\ref{app:ttdata}.
The total number of data points is 54, and the receivers are located up to the depth of 2650.20 ${\rm m}$. 
In a real case study, the velocity results from 1-D traveltime tomography can be in any order with depth. 
To get the linear inhomogeneity parameters, we use linear regression on the inverted velocity. 

We also apply the real data on the $ab$ model to calculate a global $a$ and $b$. 
In Table~\ref{tab:real}, for the range of startup values, the two-parameter velocity inversion results do not change.
The values of $a$ and $b$ are $1247.07\, {\rm ms^{-1}}$ and $0.4384 \, s^{-1}$.
However, the inversion results of the tomography are sensitive to the startup values. 
The low number of data points makes the inversion problem more sensitive to startup values.  

\begin{figure}[ht]
        \captionsetup[subfigure]{aboveskip=-.1cm}
        \centering
        \begin{subfigure}[b]{.45\textwidth}
                \includegraphics[width=\textwidth]{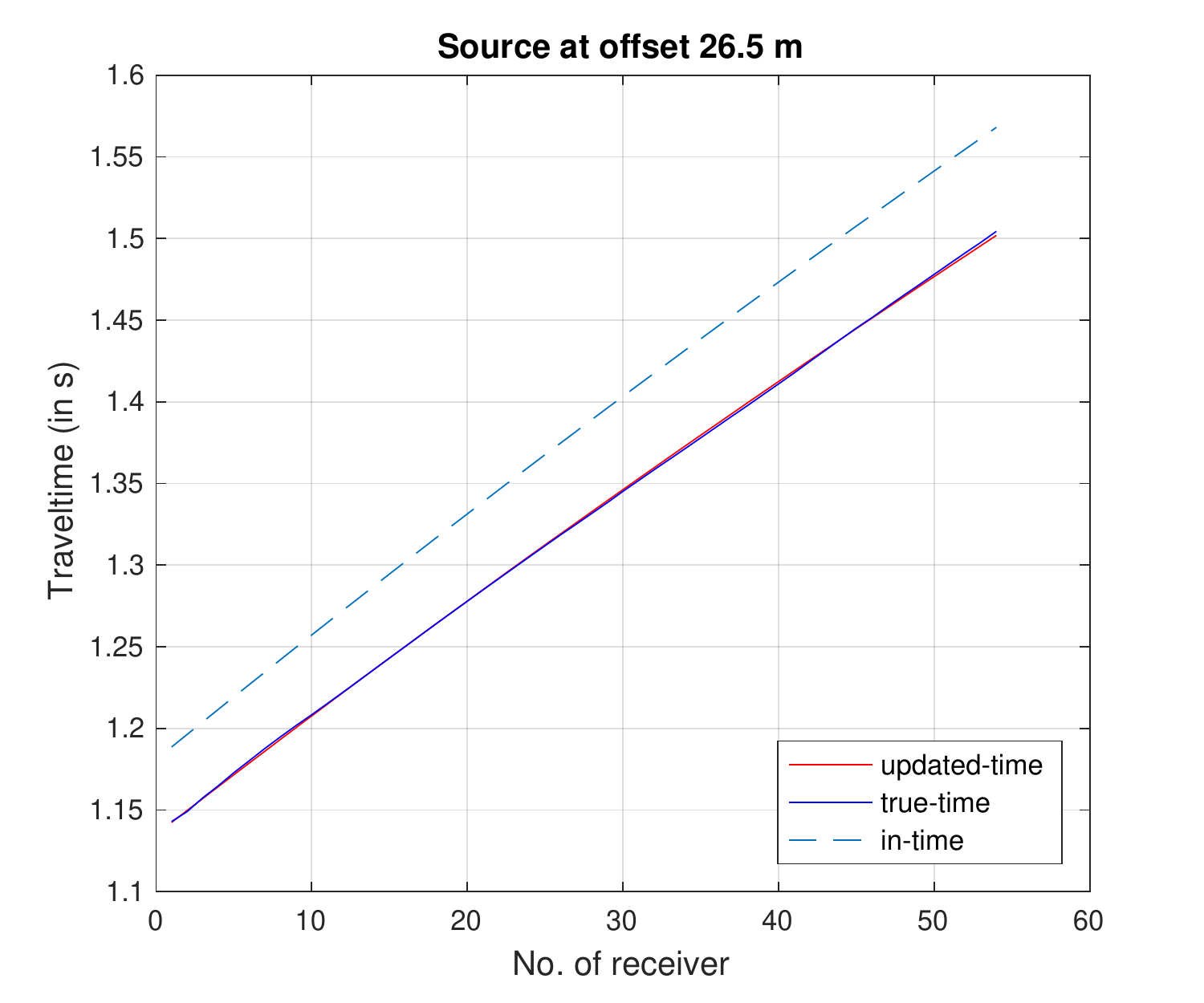}
                \caption{}
                \label{t_t_a_1225_b_40}
        \end{subfigure}  
        \begin{subfigure}[b]{.45\textwidth}
                \includegraphics[width=\textwidth]{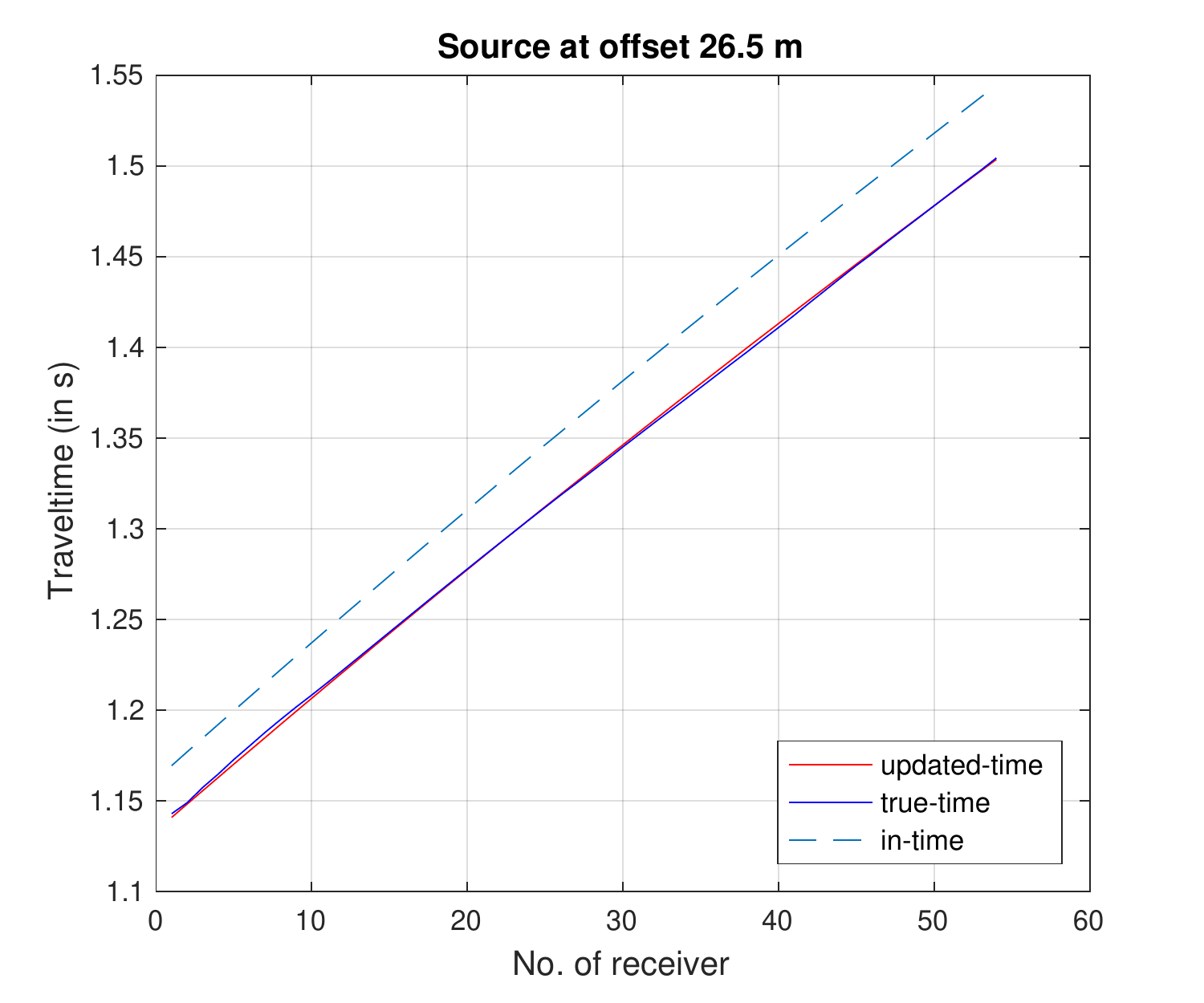}
                \caption{}
                \label{t_t_a_1250_b_40}
        \end{subfigure}  
        \begin{subfigure}[b]{.45\textwidth}
                \includegraphics[width=\textwidth]{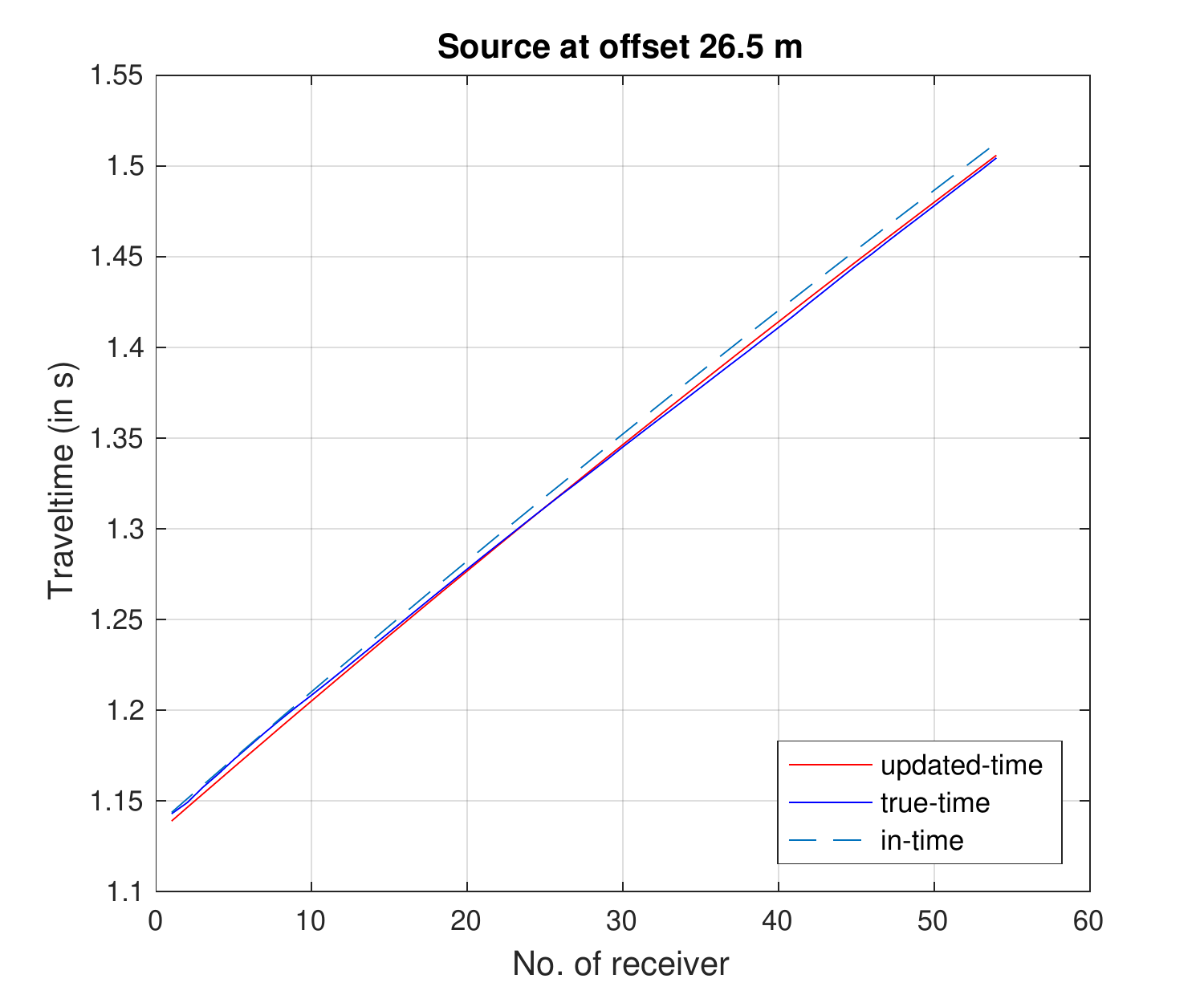}
                \caption{}
                \label{t_t_a_1285_b_40}
        \end{subfigure}   
        \begin{subfigure}[b]{.45\textwidth}
                \includegraphics[width=\textwidth]{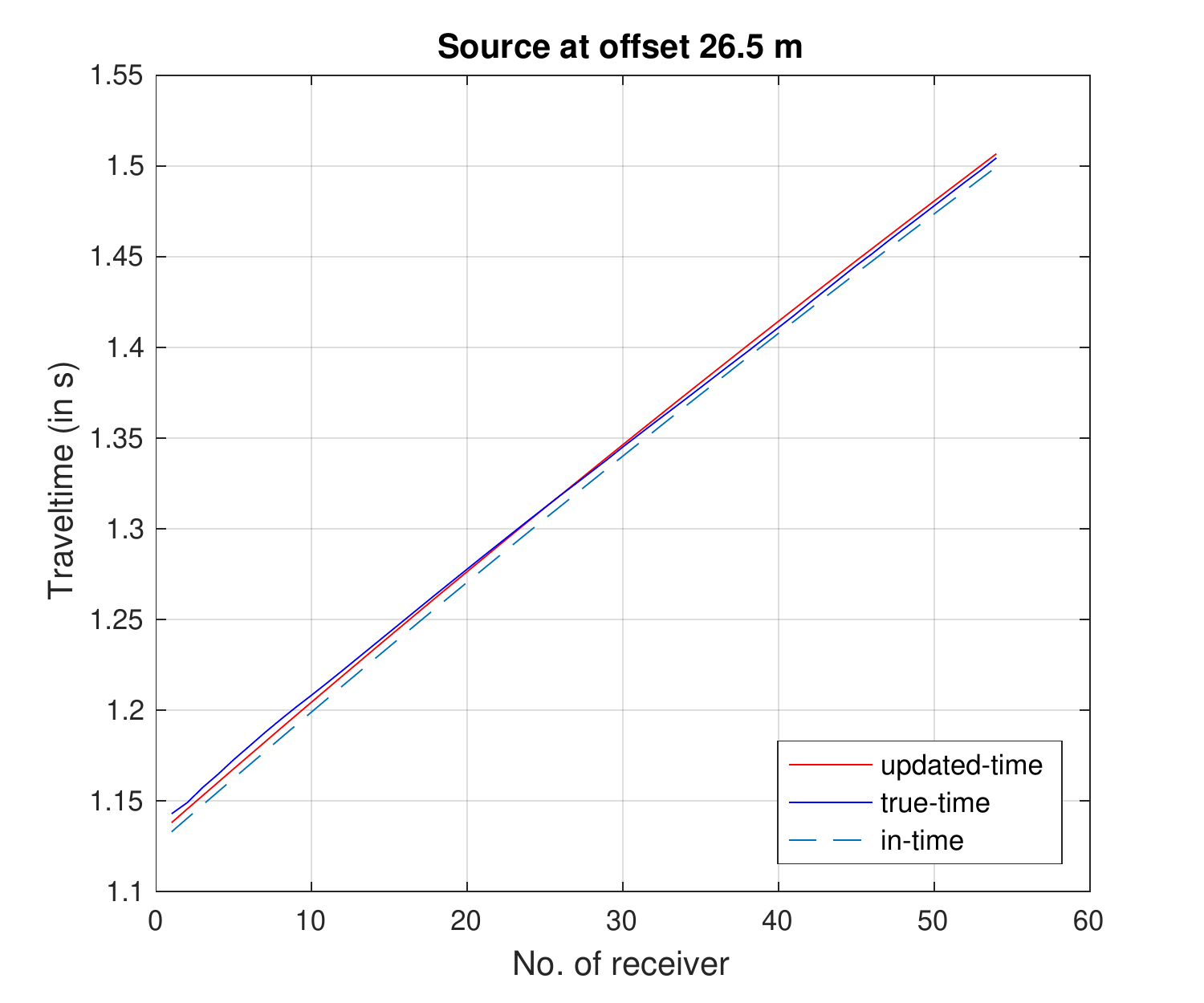}
                \caption{}
                \label{t_t_a_1300_b_40}
        \end{subfigure}  
        \begin{subfigure}[b]{.45\textwidth}
                \includegraphics[width=\textwidth]{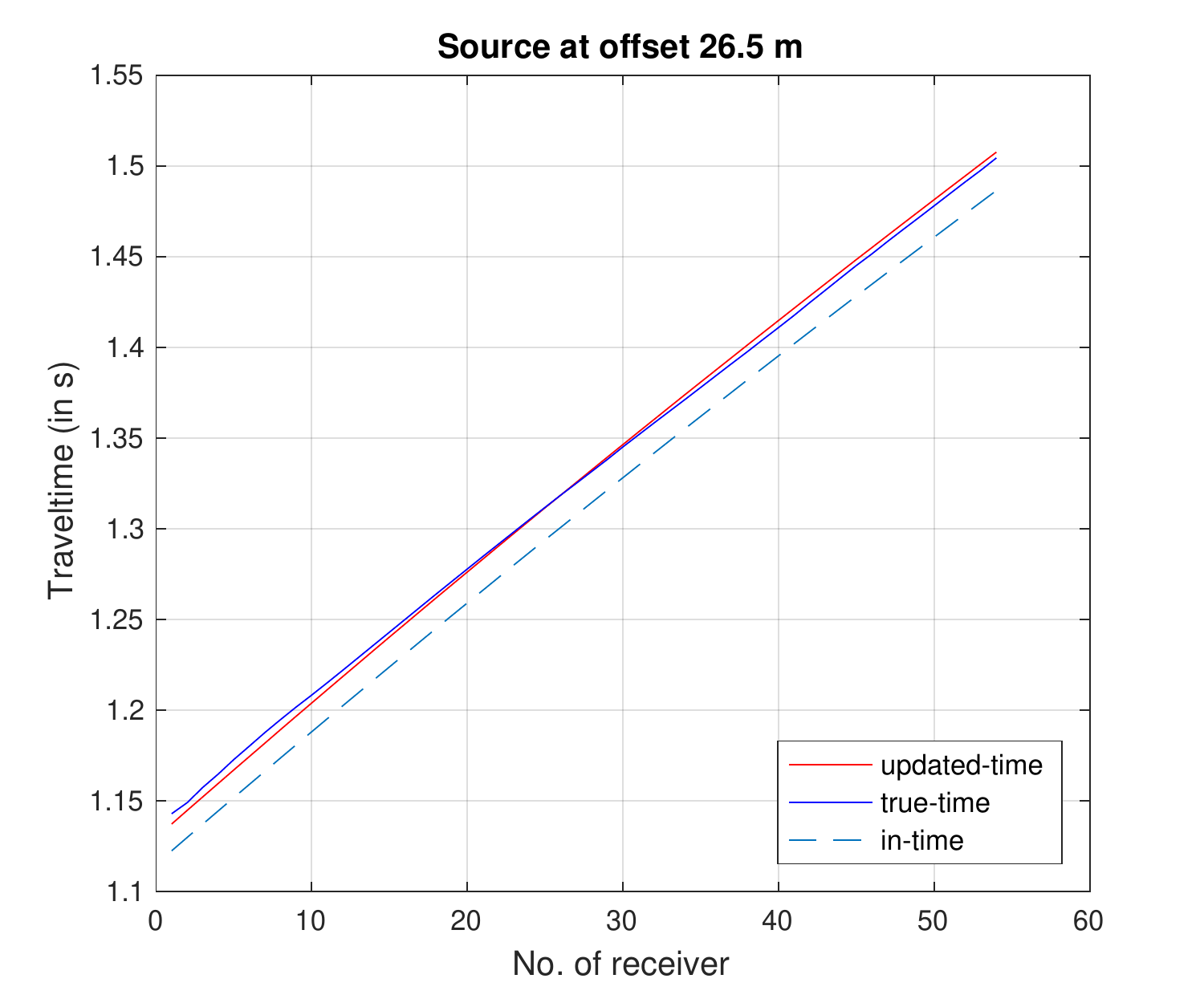}
                \caption{}
                \label{t_t_a_1315_b_40}
        \end{subfigure}      
        \begin{subfigure}[b]{.45\textwidth}
                \includegraphics[width=\textwidth]{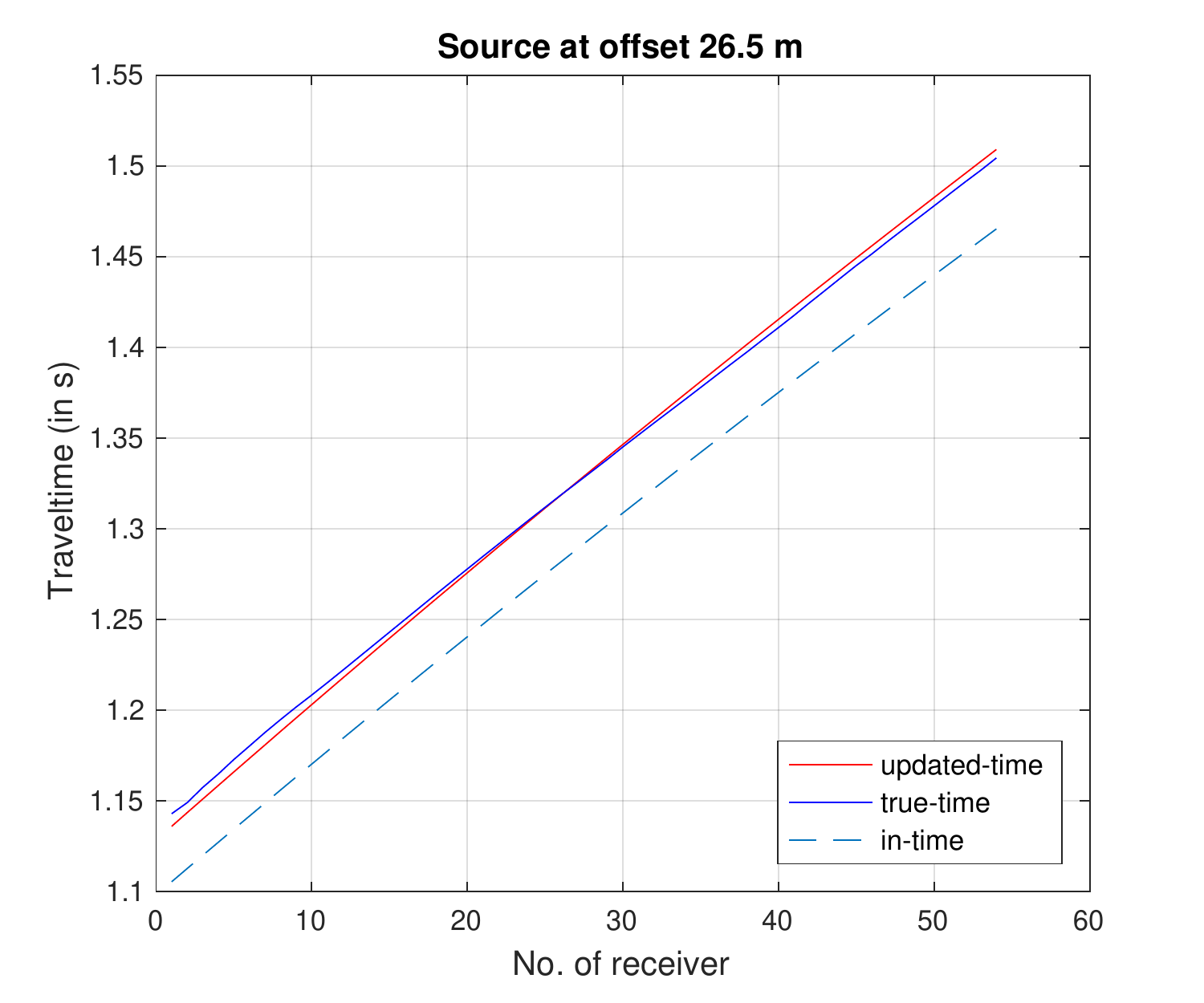}
                \caption{}
                \label{t_t_a_1340_b_40}
        \end{subfigure}   
        \caption{Traveltime inversion for different velocity gradients, $v_{in} = v_{true}\pm60 \, (ms^{-1})$.}
        \label{fig:real_tt} 
\end{figure}

\begin{figure}[ht]
       \captionsetup[subfigure]{aboveskip=-.1cm}
        \centering
        \begin{subfigure}[b]{.45\textwidth}
                \includegraphics[width=\textwidth]{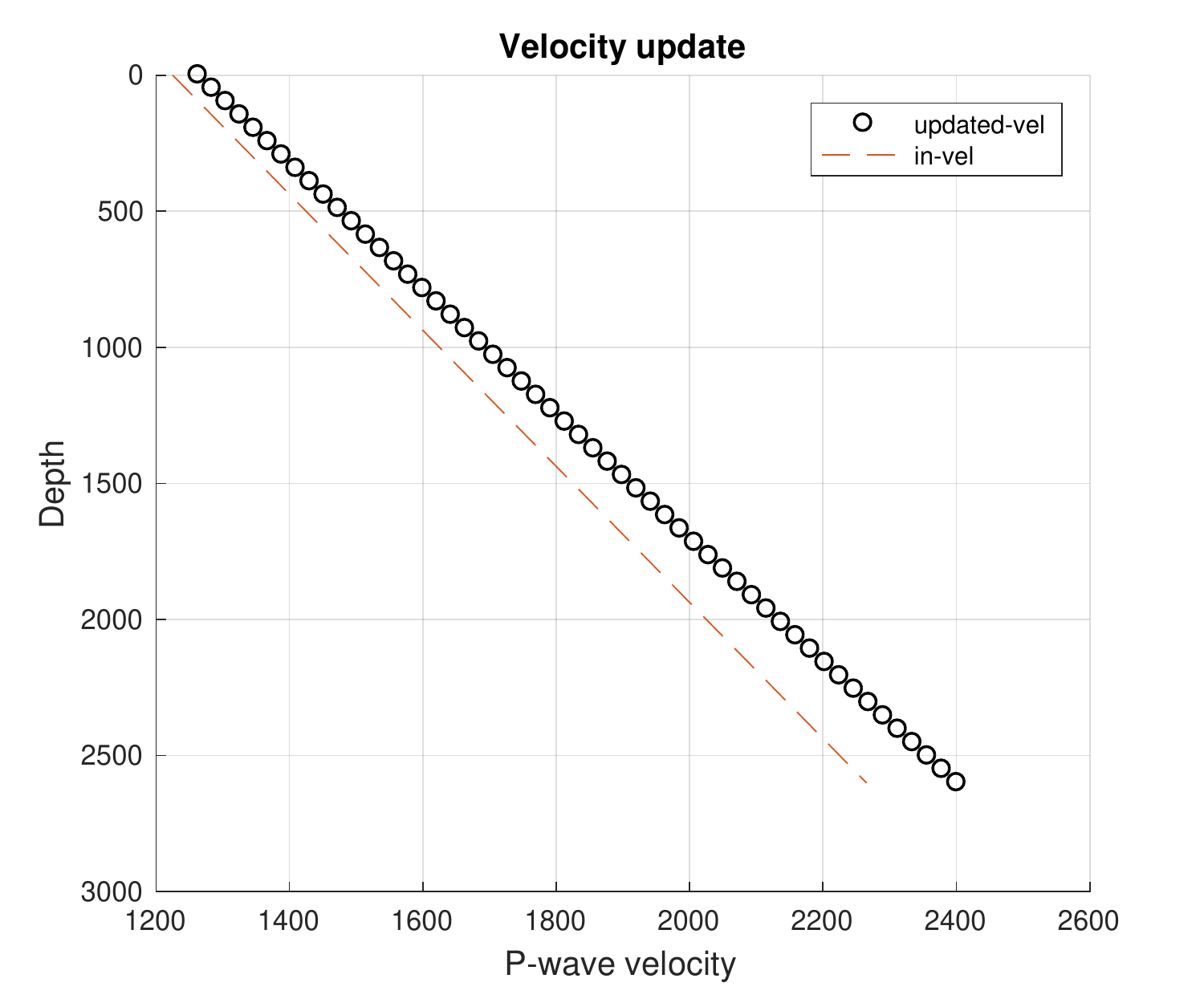}
                \caption{}
                \label{vel_a_1225_b_40}
        \end{subfigure} 
        \begin{subfigure}[b]{.45\textwidth}
                \includegraphics[width=\textwidth]{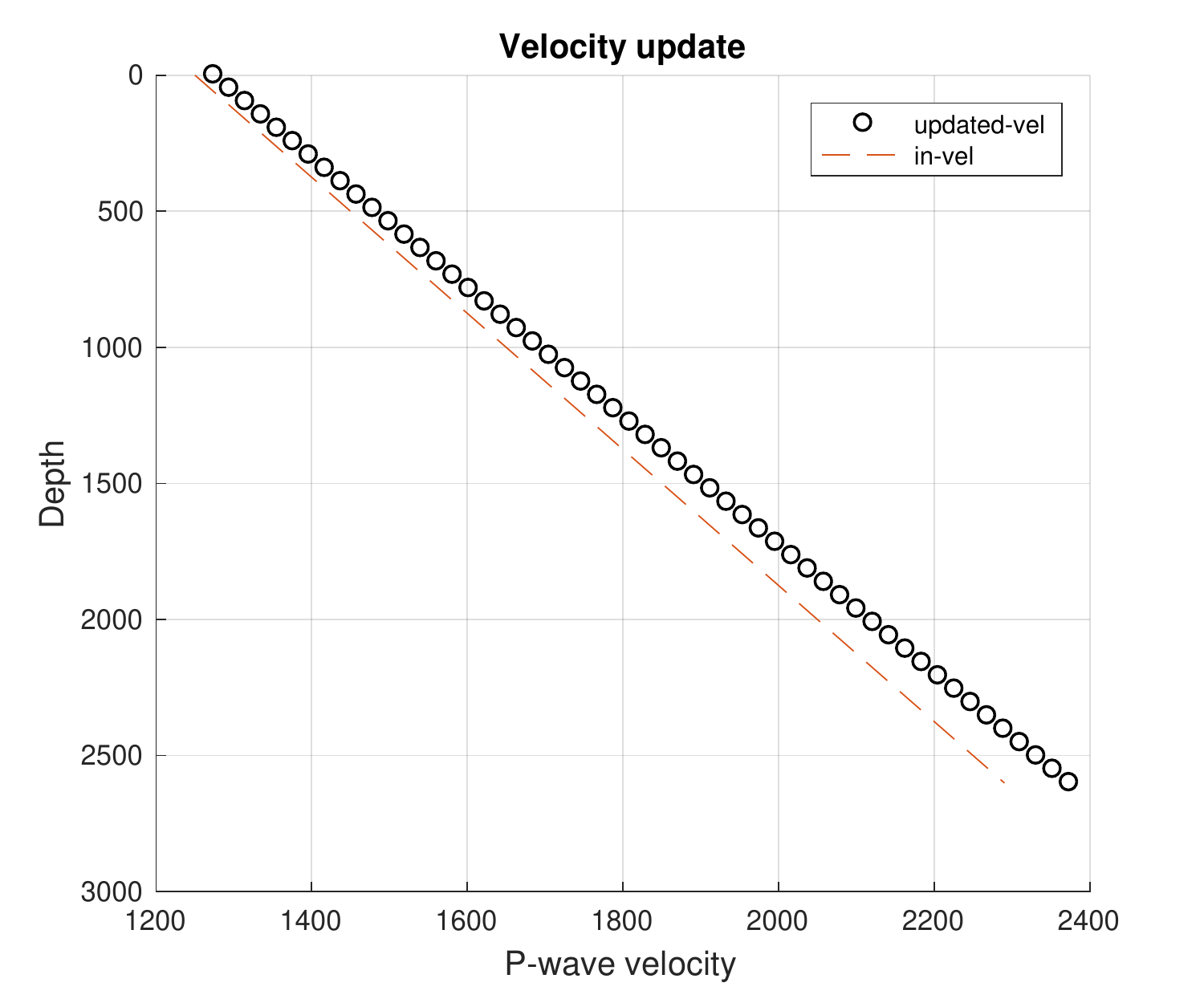}
                \caption{}
                \label{vel_a_1250_b_40}
        \end{subfigure}
         \begin{subfigure}[b]{.45\textwidth}
                \includegraphics[width=\textwidth]{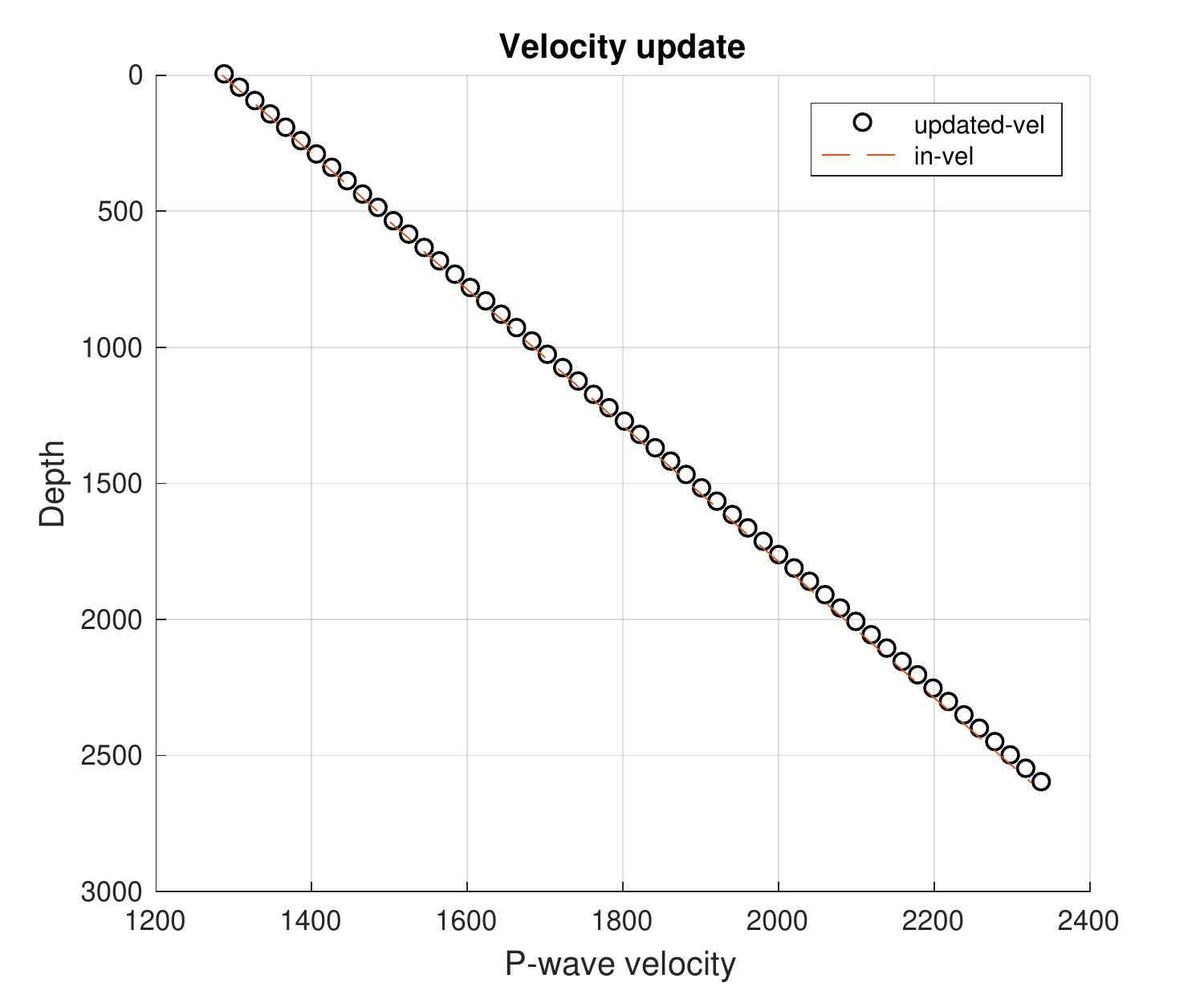}
                \caption{}
                \label{vel_a_1285_b_40}
        \end{subfigure} 
        \begin{subfigure}[b]{.45\textwidth}
                \includegraphics[width=\textwidth]{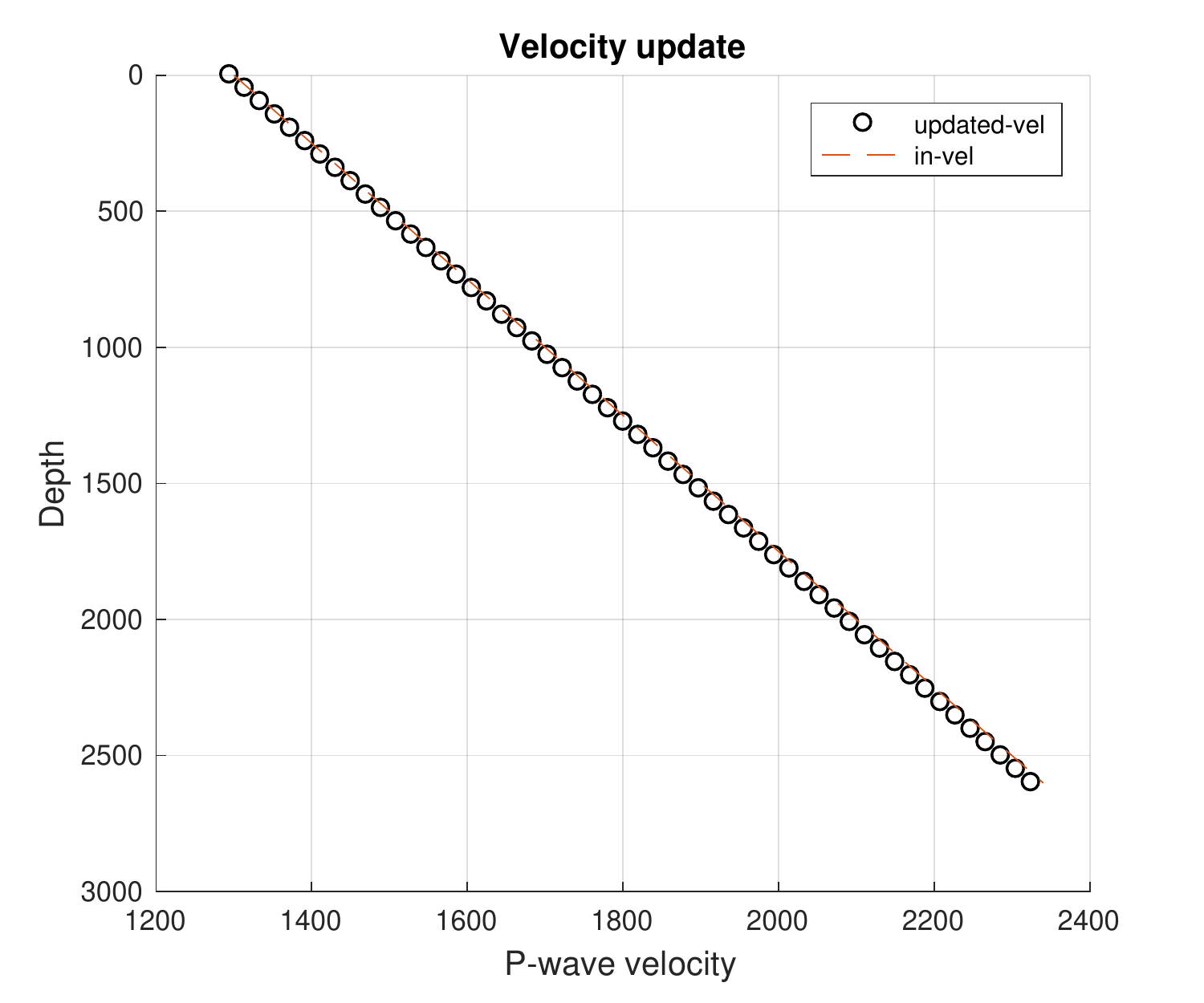}
                \caption{}
                \label{vel_a_1300_b_40}
        \end{subfigure}    
         \begin{subfigure}[b]{.45\textwidth}
                \includegraphics[width=\textwidth]{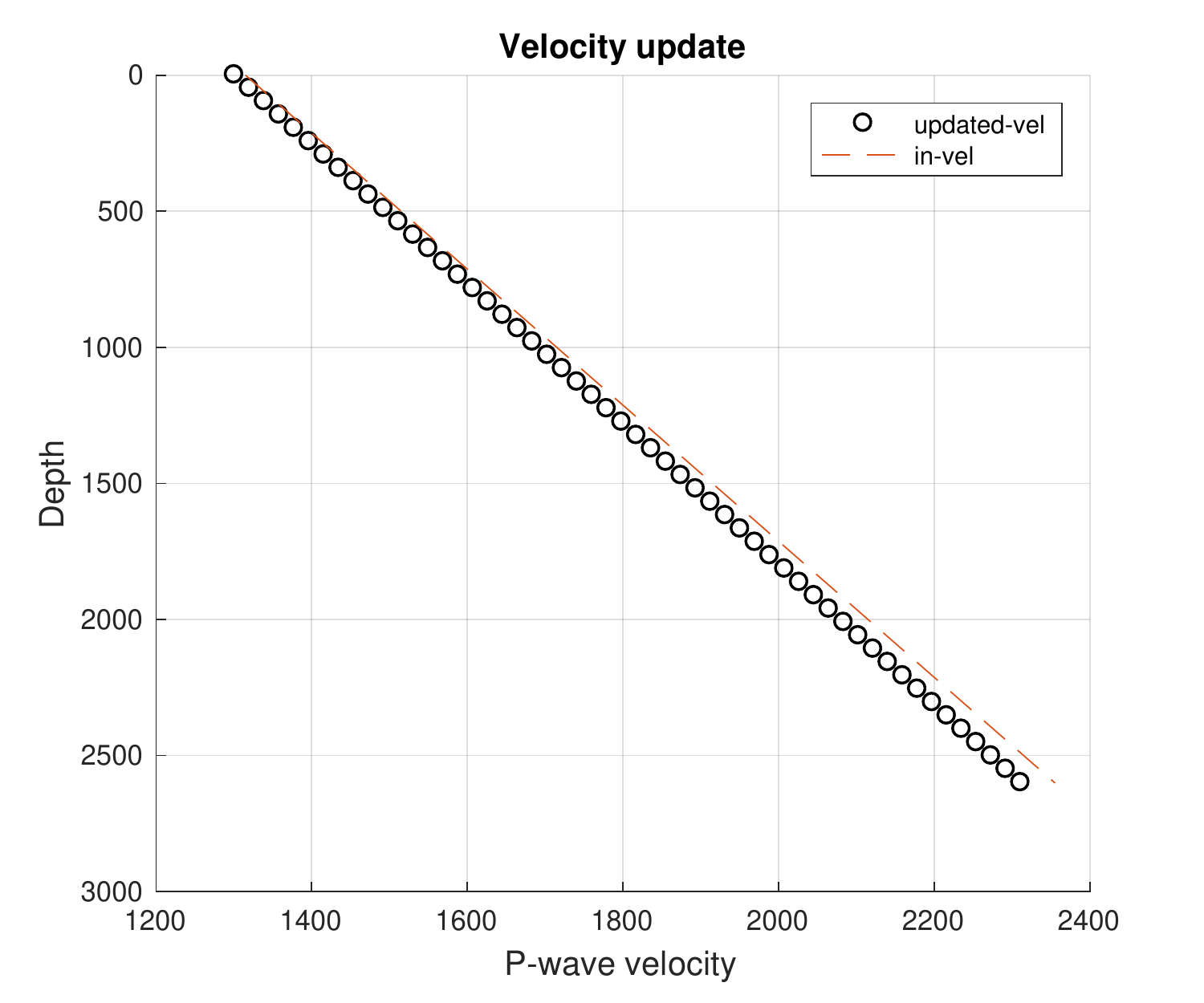}
                \caption{}
                \label{vel_a_1315_b_40}
        \end{subfigure}    
         \begin{subfigure}[b]{.45\textwidth}
                \includegraphics[width=\textwidth]{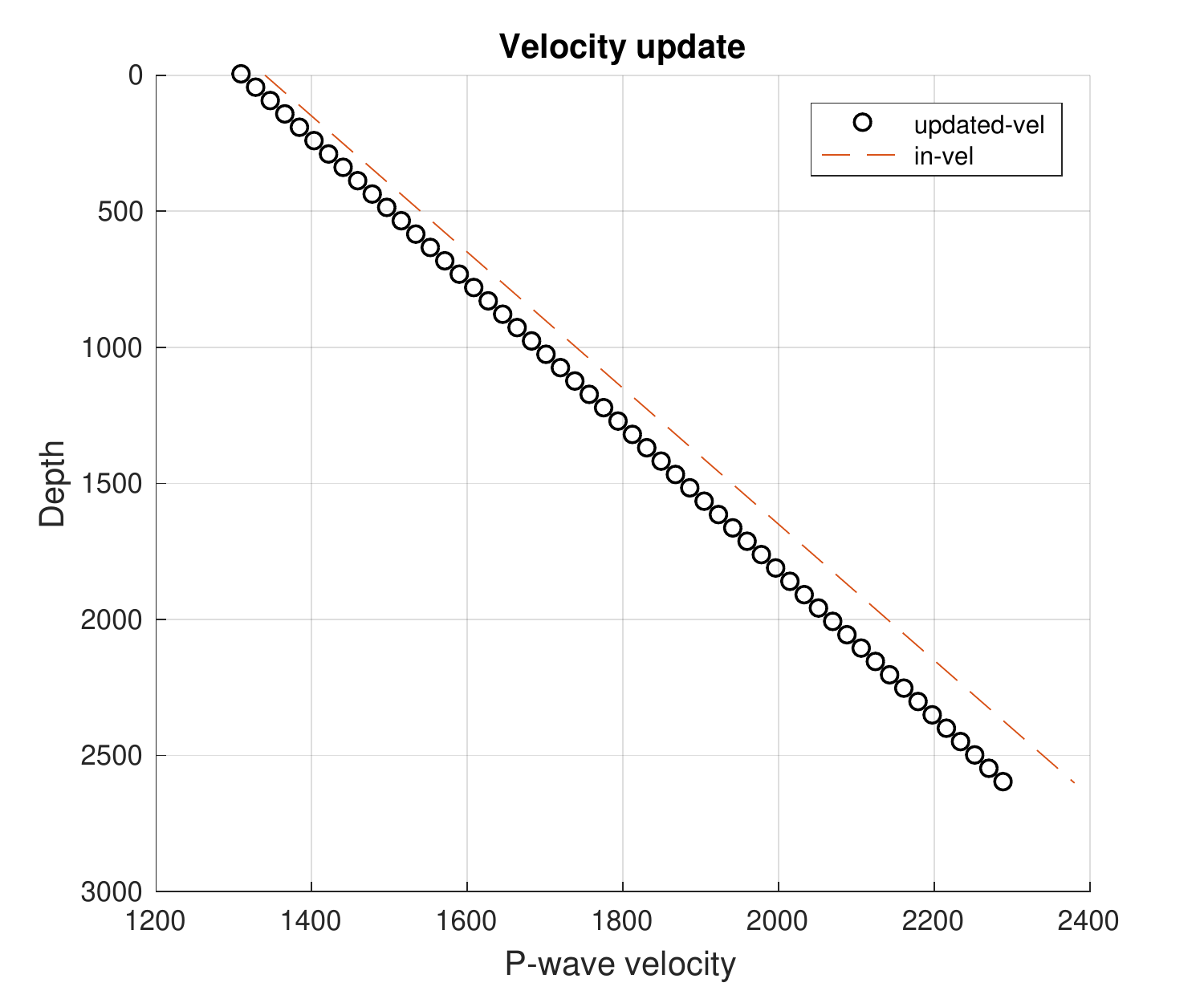}
                \caption{}
                \label{vel_a_1340_b_40}
        \end{subfigure}       
        \caption{Velocity model for different velocity gradients, $v_{in} = v_{true}\pm60  \, (ms^{-1})$.}
        \label{fig:real_vel} 
\end{figure}

The traveltime convergence results are shown in Figure~\ref{fig:real_tt}.
The velocity misfits of the inverted velocity to the reference velocity are shown in Figure~\ref{fig:real_vel}.
Based on the synthetic experiments, we know that the inverted velocity reproduces the reference velocity with less error if the traveltime convergence occurs faster. 
Therefore, we perform several tests with a range of startup values and show that tests 3 and 4 have the best startup values out of the six tests.
Based on the results of experiments 3 and 4, we intuit that the inhomogeneity of the medium
ranges from $0.3960\, s^{-1}$ to $0.4037\,s^{-1}$.
The inhomogeneity results can be improved by increasing the number of data points.

\begin{table}[h]
\centering
\begin{tabular}{c*{9}{c}}
\toprule
Test &
$a_{in}$ &
$b_{in}$ &
$a_{inv}$ &
$b_{inv}$ &
$a_{t_{ab}}$ &
$b_{t_{ab}}$ &
$f({\bf M})$ &
Figure
\\[2pt]
\toprule
1 & 1225 & 0.40 & 1258.66 & 0.4373 & 1247.07 & 0.4384 & 52.01 & \ref{t_t_a_1225_b_40},\ref{vel_a_1250_b_40} \\
2 & 1250 & 0.40 & 1271.63 & 0.4228 & 1247.07 & 0.4384 & 52.72 &  \ref{t_t_a_1250_b_40},\ref{vel_a_1225_b_40} \\
3 & 1285 & 0.40 & 1288.79 & 0.4037 & 1247.07 & 0.4384 & 51.55 &  \ref{t_t_a_1285_b_40},\ref{vel_a_1285_b_40} \\
4 & 1300 & 0.40 & 1295.85 & 0.3960 & 1247.07 & 0.4384 & 49.48 &   \ref{t_t_a_1300_b_40},\ref{vel_a_1300_b_40} \\
5 & 1315 & 0.40 & 1302.69 & 0.3885 & 1247.07 & 0.4384 & 53.79 &  \ref{t_t_a_1315_b_40},\ref{vel_a_1315_b_40} \\
6 & 1340 & 0.40 & 1313.67 & 0.3765 & 1247.07 & 0.4384 & 50.59 &  \ref{t_t_a_1340_b_40},\ref{vel_a_1340_b_40} \\
\bottomrule
\end{tabular}
\caption[Comparison of inversion methods]{Results of 1-D tomography and two-parameter method using real data (units of $a$ and $b$ are ${\rm ms^{-1}}$ and ${\rm s^{-1}}$)}
\label{tab:real}
\end{table}

\captionsetup[figure]{aboveskip=0cm}

\captionsetup[figure]{aboveskip=0cm}
\FloatBarrier
\section{Conclusion}
The synthetic experiments show that the tomography method can reproduce the reference velocity with some misfits. 
The misfit gets higher when there is more noise, and fewer data points.

From the two-parameter method,
we find that the inhomogeneity parameter, $b$, is higher in comparison to the 1-D tomography.

Since, from the traveltime data, the 1-D tomography calculates $m$ parameters and the $ab$ method computes only two parameters to obtain velocity, therefore,
we intuit that, for finding the local inhomogeneity of a segment, the 1-D tomography method is more reliable.

In practical seismology, the velocities are measured in the well log after a few hundred meters of depth from the surface. 
The VSP method can be used as a proxy to obtain the inhomogeneity parameters above the well log region. 

For a common region of interest, we state that the study allows us to obtain linear inhomogeneity of a medium using two different seismic methods.  
To examine that statement, as a future project, we plan to do a comparison study
by applying the developed methods on different sites.

\section*{Acknowledgments}
We acknowledge discussions with Michael A. Slawinski and proof reading of David R. Dalton.
This research was performed in the context of The Geomechanics Project supported by Husky Energy. 
Also, this research was partially supported by the Natural Sciences and Engineering Research Council of Canada, grant~202259.

\bibliographystyle{apa}
\bibliography{article_2_arXiv}


\setcounter{section}{0}
\renewcommand{\thesection}{Appendix~\Alph{section}}
\renewcommand{\thesubsection}{\Alph{section}.\arabic{subsection}}

\section{}
\label{app:A}
\subsection{1-D tomography : synthetic data}
\label{sec:code_syn}
\begin{lstlisting}[language=Matlab]
close all; clear all ;clc

%% Variables from travel time data
rec_depth = 1985:15:2000; % Creating data array for Geophones
offset_fict = 0:20:2000; % Creating data array for Sources 
                         % 201 sources at the surface with 20m apart from
N = length(rec_depth)*length(offset_fict);% Number of Layers (= Number of model parameters)
z_p = 0:(rec_depth(end)/N):rec_depth(end); 
M = length(rec_depth); % Number of Geophone
S_N = length(offset_fict); % Number of sources

for i=1:S_N
    for k=1:M
       x_obs(k,i) = offset_fict(i);
    end
end
% 

%%%%%%%%% Similar_to_previous %%%%%%%%%%%%%
for i = 1:N
d_z_p(i)= z_p(i+1)-z_p(i); % layer thickness
end


for k = 1:M
    for j = 1:N
        H_n(k,j)=d_z_p(1);
    end
end

for k = 1:M
    for j = 1:N-1
        H_n(k,j)=d_z_p(1);
    if sum(d_z_p(1:(j))) <= rec_depth(k)  
        H(k,j)=d_z_p(j);
        H(k,j+1)= rec_depth(k)-sum(d_z_p(1:(j)));
    end
    end
end


%% Variables for initial velocity and ray parameter
a = 1000;
b = .12;

%% From here complication starts
for j=1:N
    v_o_true(j) = a+(b.*z_p(j));
    v_o(j) = (a+20)+((b).*z_p(j));
end


v_true = v_o_true;
v_in = v_o;


%%%%%%%%%%%%%%%%%%% Newton's Gradient %%%%%%%%%%%%%%%5%%
% Initial guess for theta_in to use in Newton method
for i=1:S_N
    for k=1:M
       theta_true(k,i) = atan(x_obs(k,i)./ rec_depth(k));
    end
end

% Initialize the iteration for Newton-Raphson method
for i=1:S_N
    for k=1:M
    dx_true(k,i) = 2500; % Initialize the iteration with higher values
    end
end

myCoordList_true=[];Ite_true = 0;
lim_dx_true = 1e-6;
while (abs(dx_true(:,end)) > lim_dx_true)

for i=1:S_N
    for k=1:M
       ray_p_o_i_true(k,i)  = sin(theta_true(k,i))./ v_o_true(1);
    end
end
    
for i=1:S_N
    for k=1:M
        for j = 1:N
            B_kji_i_true(k,j,i) = ray_p_o_i_true(k,i)*v_o_true(j);
            
            x_jki_true(j,k,i) = H(k,j) .* B_kji_i_true(k,j,i) ./ (1 - B_kji_i_true(k,j,i).^2).^(.5);
            
            
            dx_prime_t1_true(j,k,i) = (H(k,j).* v_o_true(j).*cos(theta_true(k,i)))...
                ./(v_o_true(1).*(1-B_kji_i_true(k,j,i).^2).^(.5));
            
            dx_prime_t2_true(j,k,i) = (H(k,j).* v_o_true(j).*cos(theta_true(k,i)).*B_kji_i_true(k,j,i))...
                ./(v_o_true(1).*(1-B_kji_i_true(k,j,i).^2).^(1.5));
            
            dx_true_all_prime(j,k,i) = dx_prime_t1_true(j,k,i)+dx_prime_t2_true(j,k,i);
        end
    end
end

for i = 1:S_N
    for k = 1:M
        x_ki_true(k,i)=sum(x_jki_true(1:N,k,i)); % in m
        dx_true_prime(k,i)= sum(dx_true_all_prime(1:N,k,i));
    end
end
dx_true = abs(x_obs - x_ki_true);
theta_true = theta_true - (dx_true./dx_true_prime);
Ite_true = Ite_true+1;
myCoordList_true=[myCoordList_true; [Ite_true]];
end

for i=1:S_N
    for k=1:M
       ray_p_true(k,i)  = sin(theta_true(k,i))./ v_o_true(1);
    end 
end

for i=1:S_N
    for k=1:M
        for j = 1:N
            t_jki_true(j,k,i) = H(k,j)./(v_o_true(j).*(1-(ray_p_true(k,i).*v_o_true(j)).^2).^(.5)); % in s
        end
    end
end

for i = 1:S_N
    for k = 1:M
        t_ki_true(k,i)=sum(t_jki_true(1:N,k,i)); % in s
    end
end

%%%%%%% Optimization Starts %%%%%%%%%

lembda = 1;misfit_fn = 10^4;Ite_m = 0;myCoordList_m=[]; 
myCoordList_t_ki = [];
while (misfit_fn >= 202) 
    
%%%%%%%%%%%%%%%%%%% Newton-Raphson method %%%%%%%%%%%%%%%5%%
% Initial guess for theta_in to use in Newton-Raphson method
for i=1:S_N
    for k=1:M
       theta_in(k,i) = atan(x_obs(k,i)./ rec_depth(k));
    end
end
% Initialize the iteration for Newton method
myCoordList_in=[];Ite_in = 0;
for i=1:S_N
    for k=1:M
    dx_in(k,i) = 2500; % Initialize the iteration with higher values
    end
end


myCoordList_in=[];Ite_in = 0;
lim_dx_in = 1e-6;
while (abs(dx_in(:,end)) > lim_dx_in)

for i=1:S_N
    for k=1:M
       ray_p_o_i_in(k,i)  = sin(theta_in(k,i))./ v_o(1);
    end
end
    
for i=1:S_N
    for k=1:M
        for j = 1:N
            B_kji_i_in(k,j,i) = ray_p_o_i_in(k,i)*v_o(j);
            
            x_jki_in(j,k,i) = H(k,j) .* B_kji_i_in(k,j,i) ./ (1 - B_kji_i_in(k,j,i).^2).^(.5);
            
            
            dx_prime_t1_in(j,k,i) = (H(k,j).* v_o(j).*cos(theta_in(k,i)))...
                ./(v_o(1).*(1-B_kji_i_in(k,j,i).^2).^(.5));
            
            dx_prime_t2_in(j,k,i) = (H(k,j).* v_o(j).*cos(theta_in(k,i)).*B_kji_i_in(k,j,i))...
                ./(v_o(1).*(1-B_kji_i_in(k,j,i).^2).^(1.5));
            
            dx_in_all_prime(j,k,i) = dx_prime_t1_in(j,k,i)+dx_prime_t2_in(j,k,i);
        end
    end
end

for i = 1:S_N
    for k = 1:M
        x_ki_in(k,i)=sum(x_jki_in(1:N,k,i)); % in m
        dx_in_prime(k,i)= sum(dx_in_all_prime(1:N,k,i));
    end
end
dx_in = abs(x_obs - x_ki_in);
theta_in = theta_in - (dx_in./dx_in_prime);
Ite_in = Ite_in+1;
myCoordList_true=[myCoordList_in; [Ite_in]];
end

for i=1:S_N
    for k=1:M
       ray_p_in(k,i)  = sin(theta_in(k,i))./ v_o(1);
    end 
end

for i=1:S_N
    for k=1:M
        for j = 1:N
            B_kji(k,j,i) = ray_p_in(k,i).*v_o(j);
            t_jki_in(j,k,i) = H(k,j)./(v_o(j).*(1-(B_kji(k,j,i)).^2).^(.5)); % in s
        end
    end
end

for i = 1:S_N
    for k = 1:M
        t_ki_in(k,i)=sum(t_jki_in(1:N,k,i)); % in s
    end
end


%Calculation of the derivatives from initial estimates 
for i=1:S_N
    for k=1:M
        for j = 1:N
            ddvj_tk(k,j,i) = -H_n(k,j)./(v_o(j).^2.*(1-B_kji(k,j,i).^2).^(.5))...
                + (ray_p_in(k,i).^2.*H_n(k,j))./(1-B_kji(k,j,i).^2).^(1.5);
        end
    end
end


%%%%%%%%%%%%%%%%% Adding_Noise %%%%%%%%%%%%%%%
t = t_ki_in(:);
t_true = t_ki_true(:);
noiseSigma = 0.01 * t_true; % standard deviation = noiseSigma
noise = noiseSigma .* randn(length(t_true),1); % considering mean = 0
noisySignal = t_true + noise;
Y_c = noisySignal;
dt = (Y_c - t);
dt = reshape(dt,[M,S_N]);
noise = reshape(noise,[M,S_N]);
Y_c = reshape(Y_c,[M,S_N]);
%%%%%%%%%%%%%%%%% Adding_Noise %%%%%%%%%%%%%%%
 
for i=1:S_N
    for k=1:M
        for j = 1:N
            term_A1(k,j,i) = ddvj_tk(k,j,i);
        end
    end
end

term_A = [term_A1];
%%
%%%%%%%%%%%%%%%%%%%%%%%%%%%%%%%%%%%%%%%%%%%%%%%%%%%%%%%%%%%%%%%%%%%%%%%%%%%%%%%%%%%%%%%%%%%%%
%%%%%%%%%%%%%%%%%%%%%%%%%%%%%%% Conversion to vectors and matrix%%%%%%%%%%%%%%%%%%%%%%%%%%%%%%%

C = dt(:);

A_c_terms = [];
for ll = 1:size(term_A,3)
  A_c_terms = cat(1,A_c_terms,term_A(:,:,ll));
end
A = A_c_terms;

A_T_A = transpose(A)*A;
A_T_C = transpose(A)*C;

for i = 1:(N)
for j = 1:(N)
    if i==j
        s_ij(i,j)=1./sqrt(A_T_A(i,j));
    elseif i ~= j
        s_ij(i,j)=0;
    end
end
end
S = s_ij;
A_T_A_ast = S.'*A_T_A*S;
A_T_C_ast = S.'*A_T_C;

I = eye(N,N);
lembda = I*1000;

inv_t_ast = A_T_A_ast+lembda;
X_ast = inv(inv_t_ast)*A_T_C_ast;
X = S.'* X_ast;


dt_up = dt(:);
misfit_fn = (sum((dt_up(:)./noise(:)).^2));

dv = X.';
v_o = v_o + dv;
v_f = v_o;
lembda = lembda * 0.1;  
Ite_m = Ite_m+1;
myCoordList_m=[myCoordList_m; [Ite_m, misfit_fn]];
myCoordList_t_ki=[myCoordList_t_ki; [t_ki_in]];
%%%%%%%%%%%%%%%%%%%%%%%%%%%end%%%%%%%%%%%%%%%%%%%%%%%%%
end

%%%%%%%%%%%%%%%%%%%%%%%%%%%end%%%%%%%%%%%%%%%%%%%%%%%%%
figure(1) % Traveltime plot
%%%%%%%%%%%%%%%%%%%
subplot(2,1,1)
time_k1_in = myCoordList_t_ki(1,:);
time_k1_true = t_ki_true(1,:);
time_k1_f = t_ki_in(1,:);
NT = Y_c(1,:); %Noisy Traveltime
err = noise(1,:);
err = std(err).*ones(size(err));

indx = 1:1:length(NT);
scatter(indx,NT);
hold on;
errorbar(indx, NT, err, 'LineStyle','none');
hold on;

plot(time_k1_f,'r')
hold on

plot(time_k1_true,'b')
hold on 

plot(time_k1_in,'--')
legend({'noisy data','errorbar','updated-time','true-time','in-time'},'Location','southeast','FontSize',8)
xlabel('No. of source', 'FontSize', 12)
ylabel('Traveltime (in s)', 'FontSize', 12)
title('Traveltime misfit_fn at R_{1}', 'FontSize', 10)
grid on


% When more than one receiver
subplot(2,1,2)
time_k1_in = myCoordList_t_ki(2,:);
time_k1_true = t_ki_true(2,:);
time_k1_f = t_ki_in(2,:);
NT = Y_c(2,:); %Noisy Traveltime
err = noise(2,:);
err = std(err).*ones(size(err));

indx = 1:1:length(NT);
scatter(indx,NT);
hold on;
errorbar(indx, NT, err, 'LineStyle','none');
hold on;

plot(time_k1_f,'r')
hold on

plot(time_k1_true,'b')
hold on 

plot(time_k1_in,'--')
legend({'noisy data','errorbar','updated-time','true-time','in-time'},'Location','southeast','FontSize',8)
xlabel('No. of source', 'FontSize', 12)
ylabel('Traveltime (in s)', 'FontSize', 12)
title('Traveltime misfit_fn at R_{2}', 'FontSize', 10)
grid on



%%%%%%%%%%%%%%%%%%%%%%%%%%%%%%%%%%%%%%%%%%%%%%%%%%%%
figure(2) % Velocity plot
%%%%%%%%%%%%%%%%%%%%%%%%%%%%%%%%%%%%%%%%%%%%%%%%%%%%


z_p_plot = z_p(1:end-1);
scatter(v_f,z_p_plot,'r')
hold on 
scatter(v_true,z_p_plot,'b')
hold on 
plot(v_in,z_p_plot,'--')
legend({'updated-vel','true-vel','in-vel'},'Location','northeast','FontSize',12)
set(gca,'Ydir','reverse')
xlabel('P-wave velocity', 'FontSize', 12)
ylabel('Depth', 'FontSize', 12)
title('Velocity misfit_fn', 'FontSize', 12)
grid on


\end{lstlisting}
\newpage
\subsection{1-D tomography : real data}
\label{sec:code_real}
\begin{lstlisting}[language=Matlab]
close all; clear all ;clc
%% Data upload
[num,txt,raw] = xlsread('mizzen_o_16_cs.xlsm'); % Checkshot data
vert_depth = num(:,2);
vert_depth = rmmissing(vert_depth)-5;  % True vertical depth from source)
traveltime = num(:,3);
traveltime = rmmissing(traveltime);  % Measured travetime from source to receivers 
traveltime = traveltime(1:54);
offset = 26.5;   % Source offset 26.5 m
sigma = .0003; % from time picking  

%% Variables from travel time data
rec_depth = vert_depth(1:54).';
z_p = 0:(rec_depth(end)/54):rec_depth(end);
N = length(z_p)-1;
M = length(rec_depth); % Number of Geophone
S_N = length(offset); % Number of sources


for i=1:S_N
    for k=1:M
       x_obs(k,i) = offset(i);
    end
end
% 

%%%%%%%%% Similar_to_previous %%%%%%%%%%%%%
for i = 1:N
d_z_p(i)= z_p(i+1)-z_p(i); % layer thickness
end


% Not working this H?
for k = 1:M
    for j = 1:N
        H_n(k,j)=d_z_p(1);
    end
end

for k = 1:M
    for j = 1:N-1
        H_n(k,j)=d_z_p(1);
    if sum(d_z_p(1:(j))) <= rec_depth(k)  
        H(k,j)=d_z_p(j);
        H(k,j+1)= rec_depth(k)-sum(d_z_p(1:(j)));
    end
    end
end


%% Variables for initial velocity and ray parameter
a_in =  1250;
b_in =   0.40;

%% Initial Velocity
for j=1:N
    v_o(j) = a_in+(b_in.*z_p(j));
end
v_in = v_o;



%%%%%%% Optimization Starts %%%%%%%%%
%%%%%%% Initialize Levenberg-Marquardt Method %%%%%%%%% 
lembda = 1e3;misfit_fn = 10^4;Ite_m = 0;myCoordList_m=[]; 
myCoordList_t_ki = [];
while (misfit_fn >= 54) 

%%%%%%%%%%%%%%%%%%% Newton's Gradient %%%%%%%%%%%%%%%5%%
% Initial guess for theta_in to use in Newton method
for i=1:S_N
    for k=1:M
       theta_in(k,i) = atan(x_obs(k,i)./ rec_depth(k));
    end
end
% Initialize the iteration for Newton method
myCoordList_in=[];Ite_in = 0;
for i=1:S_N
    for k=1:M
    dx_in(k,i) = 2500; 
    end
end


myCoordList_in=[];Ite_in = 0;
lim_dx_in = 1e-10;
while (abs(dx_in(:,end)) > lim_dx_in)

for i=1:S_N
    for k=1:M
       ray_p_o_i_in(k,i)  = sin(theta_in(k,i))./ v_o(1);
    end
end
    
for i=1:S_N
    for k=1:M
        for j = 1:N
            B_kji_i_in(k,j,i) = ray_p_o_i_in(k,i)*v_o(j);
            
            x_jki_in(j,k,i) = H(k,j) .* B_kji_i_in(k,j,i) ./ (1 - B_kji_i_in(k,j,i).^2).^(.5);
            
            
            dx_prime_t1_in(j,k,i) = (H(k,j).* v_o(j).*cos(theta_in(k,i)))...
                ./(v_o(1).*(1-B_kji_i_in(k,j,i).^2).^(.5));
            
            dx_prime_t2_in(j,k,i) = (H(k,j).* v_o(j).*cos(theta_in(k,i)).*B_kji_i_in(k,j,i))...
                ./(v_o(1).*(1-B_kji_i_in(k,j,i).^2).^(1.5));
            
            dx_in_all_prime(j,k,i) = dx_prime_t1_in(j,k,i)+dx_prime_t2_in(j,k,i);
        end
    end
end

for i = 1:S_N
    for k = 1:M
        x_ki_in(k,i)=sum(x_jki_in(1:N,k,i)); % in m
        dx_in_prime(k,i)= sum(dx_in_all_prime(1:N,k,i));
    end
end
dx_in = abs(x_obs - x_ki_in);
theta_in = theta_in - (dx_in./dx_in_prime);
Ite_in = Ite_in+1;
myCoordList_true=[myCoordList_in; [Ite_in]];
end

for i=1:S_N
    for k=1:M
       ray_p_in(k,i)  = sin(theta_in(k,i))./ v_o(1);
    end 
end

for i=1:S_N
    for k=1:M
        for j = 1:N
            B_kji(k,j,i) = ray_p_in(k,i).*v_o(j);
            t_jki_in(j,k,i) = H(k,j)./(v_o(j).*(1-(B_kji(k,j,i)).^2).^(.5)); % in s
        end
    end
end

for i = 1:S_N
    for k = 1:M
        t_ki_in(k,i)=sum(t_jki_in(1:N,k,i)); % in s
    end
end


%Calculation of the derivatives from initial estimates 
for i=1:S_N
    for k=1:M
        for j = 1:N
            ddvj_tk(k,j,i) = -H_n(k,j)./(v_o(j).^2.*(1-B_kji(k,j,i).^2).^(.5))...
                + (ray_p_in(k,i).^2.*H_n(k,j))./(1-B_kji(k,j,i).^2).^(1.5);
        end
    end
end


%%%%%%%%%%%%%%%%% Adding_Noise %%%%%%%%%%%%%%%
t = t_ki_in(:);
t_true = traveltime(:);
dt = (t_true - t);
dt = reshape(dt,[M,S_N]);
t_true = reshape(t_true,[M,S_N]);
%%%%%%%%%%%%%%%%% Adding_Noise %%%%%%%%%%%%%%%
 
for i=1:S_N
    for k=1:M
        for j = 1:N
            term_A1(k,j,i) = ddvj_tk(k,j,i);
        end
    end
end

term_A = [term_A1];
%%
%%%%%%%%%%%%%%%%%%%%%%%%%%%%%%% Conversion to vectors and matrix%%%%%%%%%%%%%%%%%%%%%%%%%%%%%%%

C = dt(:);

A_c_terms = [];
for ll = 1:size(term_A,3)
  A_c_terms = cat(1,A_c_terms,term_A(:,:,ll));
end
A = A_c_terms;

A_T_A = transpose(A)*A;
A_T_C = transpose(A)*C;

for i = 1:(N)
for j = 1:(N)
    if i==j
        s_ij(i,j)=1./sqrt(A_T_A(i,j));
    elseif i ~= j
        s_ij(i,j)=0;
    end
end
end
S = s_ij;
A_T_A_ast = S.'*A_T_A*S;
A_T_C_ast = S.'*A_T_C;

I = eye(N,N);
lembda = I*1000;

inv_t_ast = A_T_A_ast+lembda;
X_ast = inv(inv_t_ast)*A_T_C_ast;
X = S.'* X_ast;


dt_up = dt(:);
misfit_fn = sum(dt_up(:)./sigma).^2; 

dv = X.';
v_o = v_o + dv;
v_f = v_o;
lembda = lembda * 0.1;  
Ite_m = Ite_m+1;
myCoordList_m=[myCoordList_m; [Ite_m, misfit_fn]];
myCoordList_t_ki=[myCoordList_t_ki; [t_ki_in]];
end

%%%%%%%%%%%%%%%%%%%%%%%%%%%%%%%%%%%%%%%%%%%%%%%%%%%%%%%
figure(1)
%%%%%%%%%%%%%%%%%%%%%%%%%%%%%%%%%%%%%%%%%%%%%%%%%%%%%%%
time_k1_in = myCoordList_t_ki(1:54,1);
time_k1_true = t_true;
time_k1_f = t_ki_in;
time_k1_f = myCoordList_t_ki((end-53):end,1);

plot(time_k1_f,'r')
hold on
plot(time_k1_true,'b')
hold on 
plot(time_k1_in,'--')
legend({'updated-time','true-time','in-time'},'Location','southeast','FontSize',10)
xlabel('No. of receiver', 'FontSize', 12)
ylabel('Traveltime (in s)', 'FontSize', 12)
title('Source at offset 26.5 m', 'FontSize', 12)
grid on 

z_p_plot = z_p(1:end-1);
p_n = polyfit(z_p_plot,v_f,1);
yfit = polyval(p_n,z_p_plot);
p_nn = polyfit(z_p_plot(39:end),v_f(39:end),1);
yfitn = polyval(p_nn,z_p_plot(39:end));


%%%%%%%%%%%%%%%%%%%%%%%%%%%%%%%%%%%%%%%%%%%%%%%%%%%%%%%
figure(2)
%%%%%%%%%%%%%%%%%%%%%%%%%%%%%%%%%%%%%%%%%%%%%%%%%%%%%%%
scatter(v_f,z_p_plot,'k')
hold on
plot(v_in,z_p_plot,'--')
legend('updated-vel','in-vel')
set(gca,'Ydir','reverse')
xlabel('P-wave velocity', 'FontSize', 12)
ylabel('Depth', 'FontSize', 12)
title('Velocity update', 'FontSize', 12)
grid on
 
upd = [a_in b_in p_n(1,2) p_n(1,1) p_nn(1,2) p_nn(1,1) misfit_fn]


rec_depth_t = rec_depth.';
ind_t = 1:length(rec_depth_t);
ind_t = ind_t.';

upd_t = [ind_t rec_depth_t traveltime];
\end{lstlisting}
\newpage
\subsection{$ab$ model inversion : real data}
\label{sec:code_ab}
\begin{lstlisting}[language=Matlab]
close all; clear all ;clc
[num,txt,raw] = xlsread('mizzen_o_16_cs.xlsm'); % Checkshot data
vert_depth = num(:,2);
vert_depth = rmmissing(vert_depth)-5;  % True vertical depth from source (-5 beacause MSL-SP=5m)
traveltime = num(:,3);
traveltime = rmmissing(traveltime);  % Measured travetime from source to receiver 
traveltime = traveltime(1:54);
offset = 26.5;   % Source offset 26.5 m
 

%% Variables from travel time data
rec_depth = vert_depth(1:54);
d_p =length(rec_depth);



Depth_of_layer = 0; % in m
Depth_of_receiver = rec_depth; % in m
z = Depth_of_receiver - Depth_of_layer; 
x_o = offset;
t_d = traveltime(1:d_p,1);
a_in = 1285;
b_in = .40;

lb=[eps eps];
ub=[inf inf];
x0=[a_in b_in];


fun = @(X)fn_ab(X(1),X(2), z, x_o, t_d);
x = lsqnonlin(fun,x0,lb,ub)

a_p = x(1);
b_p = x(2);

t_test = fn_ab(a_p,b_p, z, x_o, t_d);
mis_fit = sqrt(t_test.^2);
t_mean = mean(t_test)


function fun=fn_ab(a, b, z, x, t_d)
p_b_1 = (b.^2.*x.^2) + a.^2 + (a+b.*z).^2;
p_b_2 = 2.*a.*(a+b.*z);
p = (2.*b.*x)./sqrt(p_b_1.^2-p_b_2.^2);
 
term_3 = (a + (b.*z))./a;
term_4 = sqrt(1-(a.^2).*(p.^2));
term_5 = sqrt(1-((term_3.*a).^2).*(p.^2));

t_time = (1./b).*log( (term_3).*((1+term_4)./(1+term_5)));  

fun = abs(t_time - t_d);
end  




\end{lstlisting}

\end{document}